\documentclass{aa}  
\bibpunct{(}{)}{;}{a}{}{,}

\usepackage{verbatim}
\usepackage{graphicx}
\usepackage{lscape}
\usepackage{txfonts}
\begin{document}

   \title{Physical Properties of Galactic Planck Cold Cores revealed by the Hi-GAL survey \thanks{{\it Herschel} is an ESA space observatory with science instruments provided by European-led Principal Investigator consortia and with important participation from NASA.}}
   \author{S. Zahorecz
	  \inst{1,2}
          \and
          I. Jimenez-Serra
          \inst{3,1}
          \and
	  K. Wang
	  \inst{1}
          \and 
          L. Testi
	  \inst{1}
          \and
	  L. V. T\'oth
	  \inst{2}
      	   \and
      S. Molinari
      \inst{4}
}

   \institute{European Southern Observatory, Karl-Schwarzschild-Str. 2, D-85748, Garching bei M\"unchen, Germany\\
              \email{szahorec@eso.org}
         \and
	E\"otv\"os Lor\'and University, Department of Astronomy, P\'azm\'any P\'eter s\'et\'any 1/A, 1117, Budapest, Hungary
         \and
	University College London, Department of Physics and Astronomy, Hampstead Road 132, NW1 2PS, London, United Kingdom
    	\and
    INAF-Istituto Fisica Spazio Interplanetario, Via Fosso del Cavaliere 100, 00133 Roma, Italy 
           }

   \date{}

 
  \abstract
   {Previous studies of the initial conditions of massive star and star cluster formation have mainly targeted Infrared-Dark Clouds (or IRDCs) toward the inner Galaxy. This is due to the fact that IRDCs were first detected in absorption against the bright mid-IR background of the inner Galaxy, requiring a favourable location to be observed. By selection, IRDCs therefore represent only a fraction of the Galactic clouds capable of forming massive stars and star clusters. Due to their low dust temperatures, IRDCs are however bright in the far-IR and millimeter and thus, observations at these wavelengths have the potential to provide a complete sample of star-forming massive clouds across the Galaxy.}
   {Our aim is to identify the clouds at the initial conditions of massive star and star cluster formation across the Galaxy and compare their physical properties as a function of Galactic longitude and galactocentric distance.}
   {We have examined the physical properties of a homogeneous galactic cold core sample obtained with the {\it Planck} satellite across the Galactic Plane. With the use of {\it Herschel} Hi-GAL observations, we have characterized the internal structure of the most reliable Galactic cold clumps within the Early Cold Core (ECC) {\it Planck} catalog. By using background-subtracted Herschel images, we have derived the H$_2$ column density and dust temperature maps for 48 {\it Planck} clumps covered by the {\it Herschel} Hi-GAL survey. Their basic physical parameters (size, mass, and average dust temperature) have been calculated and analyzed as a function of location within the Galaxy. These properties have also been compared with the empirical relation for massive star formation derived by \citet{2010ApJ...723L...7K}, and with the clump evolutionary tracks calculated by \citet{2008A&A...481..345M}.}
   {Most of the Planck clumps contain signs of star formation. About 25\% of the clumps are massive enough to form high mass stars / star clusters since they exceed the empirical threshold for massive star formation of \citet{2010ApJ...723L...7K}. Planck clumps toward the Galactic center region show higher peak column densities and higher average dust temperatures than those of the clumps in the outer Galaxy. Although we only have seven clumps without associated YSOs, the Hi-GAL data show no apparent differences in the properties of Planck cold clumps with and without star formation.
   Most of the clumps are found close to the pre-main sequence evolutionary stage, with some undergoing an efficient clearing of their envelopes.}
   {}
\titlerunning{Physical Properties of Galactic Planck Cold Cores}
   \maketitle
%

\section{Introduction}
In contrast to low-mass stars, the process by which massive stars form remains poorly understood. The characterization of the initial conditions of massive star and star cluster formation is indeed challenging: massive stars are less common and have much shorter lifetimes than low mass stars.
The formation of massive stars and star clusters is believed to start in cold and dense molecular structures. When viewed against the bright Galactic mid-IR background, these clouds are called Infrared Dark Clouds (IRDCs). While some show evidence for ongoing star formation, others appear to be starless. Thus, IRDCs represent the best objects where to study the initial conditions for massive star and star cluster formation. This has been verified by single-dish observations in low-resolution \citep[e.g.][]{2006A&A...450..569P, chambers09} and interferometric observations in high-resolution \citep[e.g.][]{2011ApJ...735...64W,2012ApJ...745L..30W,2014MNRAS.439.3275W,2013MNRAS.433L..15L,2013ApJ...779...96T, zhang15}.  

Some of the early studies of IRDCs focused on the characterization of the global physical properties of IRDCs by using \textit{Spitzer} 8 $\mu$m data, i.e. based on extinction maps \citep{2006ApJ...641..389R,2008ApJS..174..396R,2010A&A...518L..98P,2009ApJ...696..484B}. However, such studies have a statistical bias because IRDCs, by selection, only represent the population of dense clouds projected onto the bright background of the inner Galaxy, as they need a favorable location to absorb background infrared radiation. Not all the dark patches are real star-forming objects. IRDC catalogues based only on mid-infrared data indeed overestimate the number of real sources by a factor of $\sim$2 \citep{2008ApJ...680..349J}. In addition, \citet{2012MNRAS.422.1071W} studied 3171 IRDC candidates within the l = 300-330$^\circ$ region of the Hi-GAL survey. Only 38 $\%$ of them were bright at the Herschel wavelengths, and therefore associated with cold cloud structures. The other mid-infrared IRDC candidates are simply minima in the mid-infrared background.

An unbiased selection criterion to identify the whole Galactic population of massive pre-star/pre-cluster forming clouds is based on the emission properties of dust in the far-IR and sub-mm. The APEX Telescope Large Area Survey of the Galaxy \citep[ATLASGAL,][]{2009A&A...504..415S} and the CSO Bolocam Galactic Plane Survey \citep[BGPS,][]{Aguirre11} have provided catalogues of submillimeter sources throughout the inner Galaxy. The outer Galaxy is however out of the coverage for these systematic surveys.

For the first time, the {\it Planck} satellite has provided an inventory of the cold condensations throughout the Galaxy. The {\it Planck} survey covered the submm-to-mm wavelength range with unprecedented sensitivity, furnishing the first all sky catalogue of cold objects. This is thus a perfect database where to identify the coldest structures in the whole Galaxy, offering the opportunity to search for massive star forming clumps also in the outer Galaxy and investigate the dependence of massive star formation with location in the Galactic disk. While in extragalactic studies only the average star formation properties can be studied \citep{2012ARA&A..50..531K}, a more detailed analysis can be performed in the Galaxy toward different environments (e.g. the Galactic Centre, the spiral arms, and the outer Galaxy). 
The angular resolution of the Planck observations is however scarce for a thorough study of these objects, and higher angular resolution observations are needed to obtain detailed information about the temperature, column density structure, and star-formation content of these cold objects.

In this paper we present the analysis of the Herschel observations of 48 Planck sources that fall within the Hi-GAL \citep{2010PASP..122..314M,2011hers.prop.1899M,2012hers.prop.2454M} survey area. We use the Planck ECC catalogue, which provides continuum fluxes for over 900 Planck sources across a broad range of wavelengths (from 350 to 850 $\mu$m), to systematically analyze the distribution of cold clumps in the Galactic Plane. The higher spatial resolution of Herschel PACS and SPIRE \citep{2010A&A...518L...2P,2010A&A...518L...3G} makes it possible to examine the internal structure of these Planck sources and allows us to identify their high column density peaks and sample their deeply embedded sources with angular resolutions of $\sim$5$\arcsec$-36$\arcsec$. In Section 2 we present the sample and datasets used in this paper, including the distance estimate for our sample. In Section 3 we describe the analysis methods and results based on the Hi-GAL and Planck data. Finally, in Section 4 we discuss our results.

\section{Data and sample selection}
\subsection{Planck catalogues of Galactic Cold Core Objects}
The Planck satellite \citep{2010A&A...520A...1T} has provided an all-sky submillimetre / millimetre survey covering the wavelengths around and longwards of the intensity maximum of cold dust emission. While $\nu^2$B$_\nu$(T=10K) peaks close to 300 $\mu$m, the coldest dust with a temperature of T$\sim$6 K shows its maximum emission close to 500 $\mu$m. Combined with far-infrared data such as the IRAS survey, the data enable accurate determination of both the dust temperature and the spectral index. 
The Early Cold Cores \footnote{Note that these ``cores'' are loosely defined as compact sources seen by the \textit{Planck}, thus \textit{Planck} ``cores'' have a wide range of physical sizes depending on distances. They are different than the 0.1 pc ``cores'' usually adopted by the star formation community.}
Catalogue \citep[ECC,][]{2011A&A...536A..23P} is part of the Planck Early Release Compact Source Catalogue (ERCSC) and the ECC forms a subset of the full Cold Core Catalogue of Planck Objects \citep[C3PO; more than 10000 objects have been detected; ][]{2011A&A...536A..23P}. The ECC, with a total of 915 objects over the sky, contains only the most secure detections (SNR > 15) with colour temperatures below 14 K.

The Planck Catalogue of Galactic Cold Clumps (PGCC, Planck Collaboration 2015) is the last release of Planck clump data tables. It contains 13188 Galactic sources with a temperature ranging from 5.8 K to 20 K. 42 \% of these sources have available distance estimates. A few ECC sources are not part of the PGCC catalogue because they do not satisfy the compactness criterion of the PGCC catalogue, i.e. they are slightly more extended compared to the other sources. We note however that the old ECC and the new PGCC data are in good agreement for the selected 48 clumps. Indeed, the derived dust temperatures, spectral indices and clump sizes differ by less than 6\%, 1.5\% and 2\%, respectively. 

\subsection{Planck ECC sources in the Hi-GAL survey}
To build up our sample, we have selected the ECC sources that were also covered by the Hi-GAL survey. Hi-GAL was one of the Herschel Open Time Key-Projects \citep{2010PASP..122..314M} and with its extensions \citep[in OT1 and Hi-GAL2pi; ][]{2012hers.prop.2454M,2011hers.prop.1899M} it mapped the entire Galactic Plane of the Milky Way (-1$^\circ$ < b < +1$^\circ$, following the Galactic warp) in 5 bands with the PACS instrument at 70 and 160 $\mu$m, and with the SPIRE instrument at 250, 350 and 500 $\mu$m with spatial resolutions of 5$\arcsec$, 13$\arcsec$, 18$\arcsec$, 25$\arcsec$ and 36$\arcsec$, respectively. 

We have selected the 48 ECC sources that were also covered by the Hi-GAL survey (see Table \ref{table:calculated_values}). These are all the ECC objects that fall within the Herschel Hi-GAL survey area and all the 5 Herschel band images are available for each source. Three sources from our sample are not part of the latest PGCC catalogue due to their lower level of compactness (G001.64-00.07, G201.13+00.31 and G296.52-01.19), but we included them in our analysis. Based on their physical properties, there are no significant differences between these 3 rejected ECC clumps and the rest of the ECCs within the Hi-GAL region (see later in Section \ref{clumpdef}).

The main parameters associated with the Planck clumps are shown in Table \ref{table:calculated_values}. The mass calculation based on Planck data is described in Section \ref{planck_mass_calculation}. 


\subsection{Distance determination of ECC objects}
\begin{figure}[h]
   \centering
   \includegraphics[width=0.45\columnwidth]{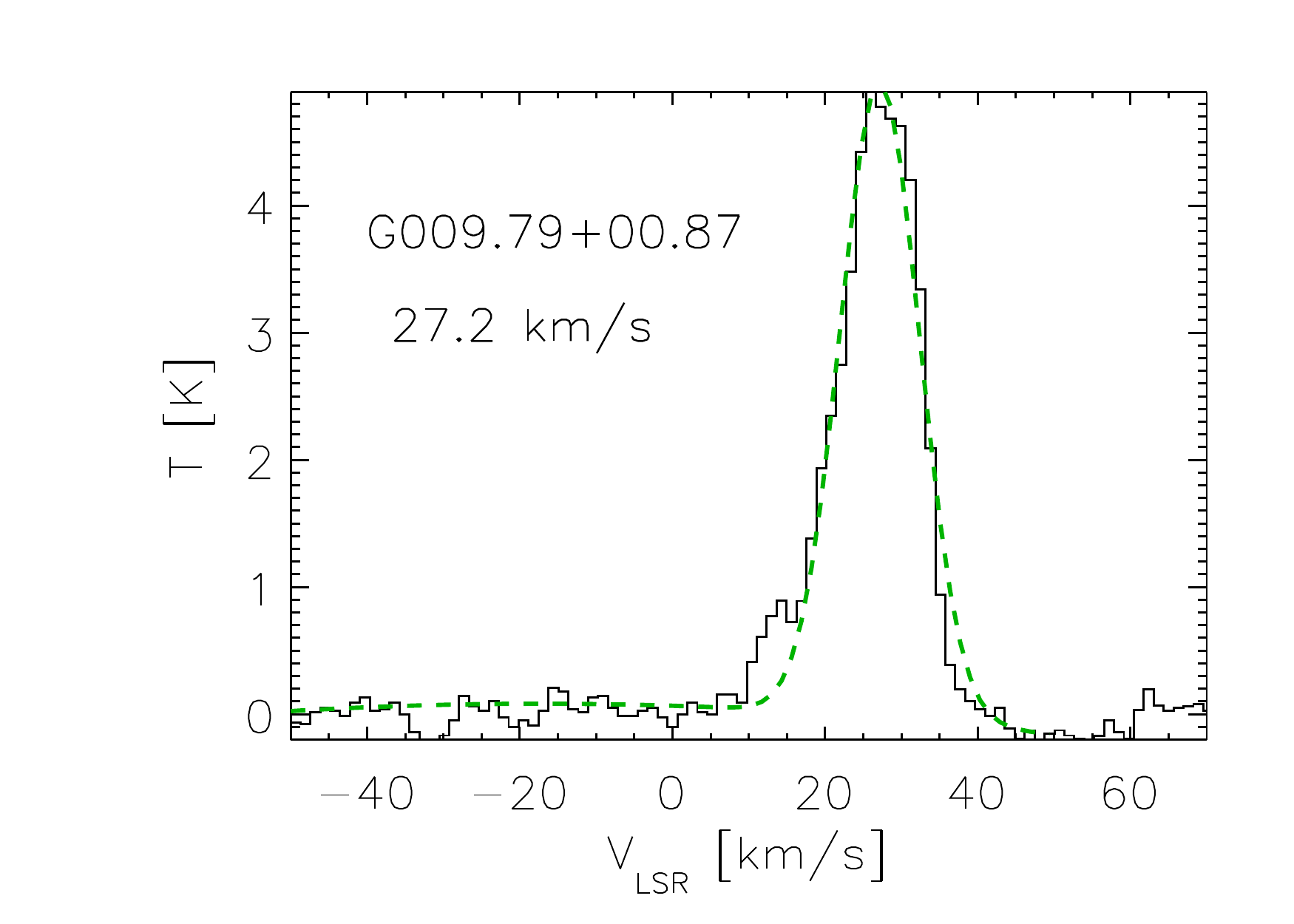}
   \includegraphics[width=0.45\columnwidth]{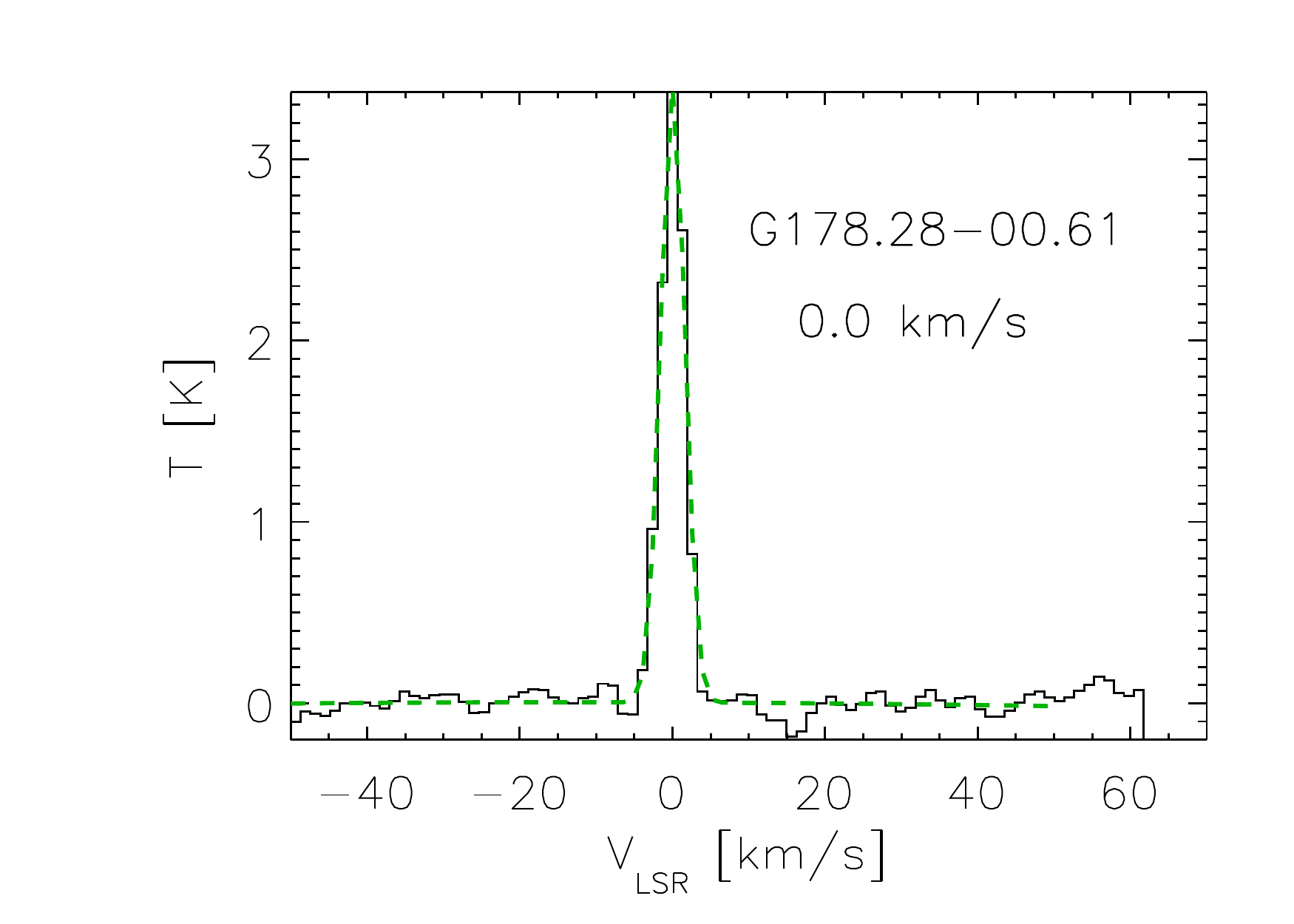}
   \includegraphics[width=0.45\columnwidth]{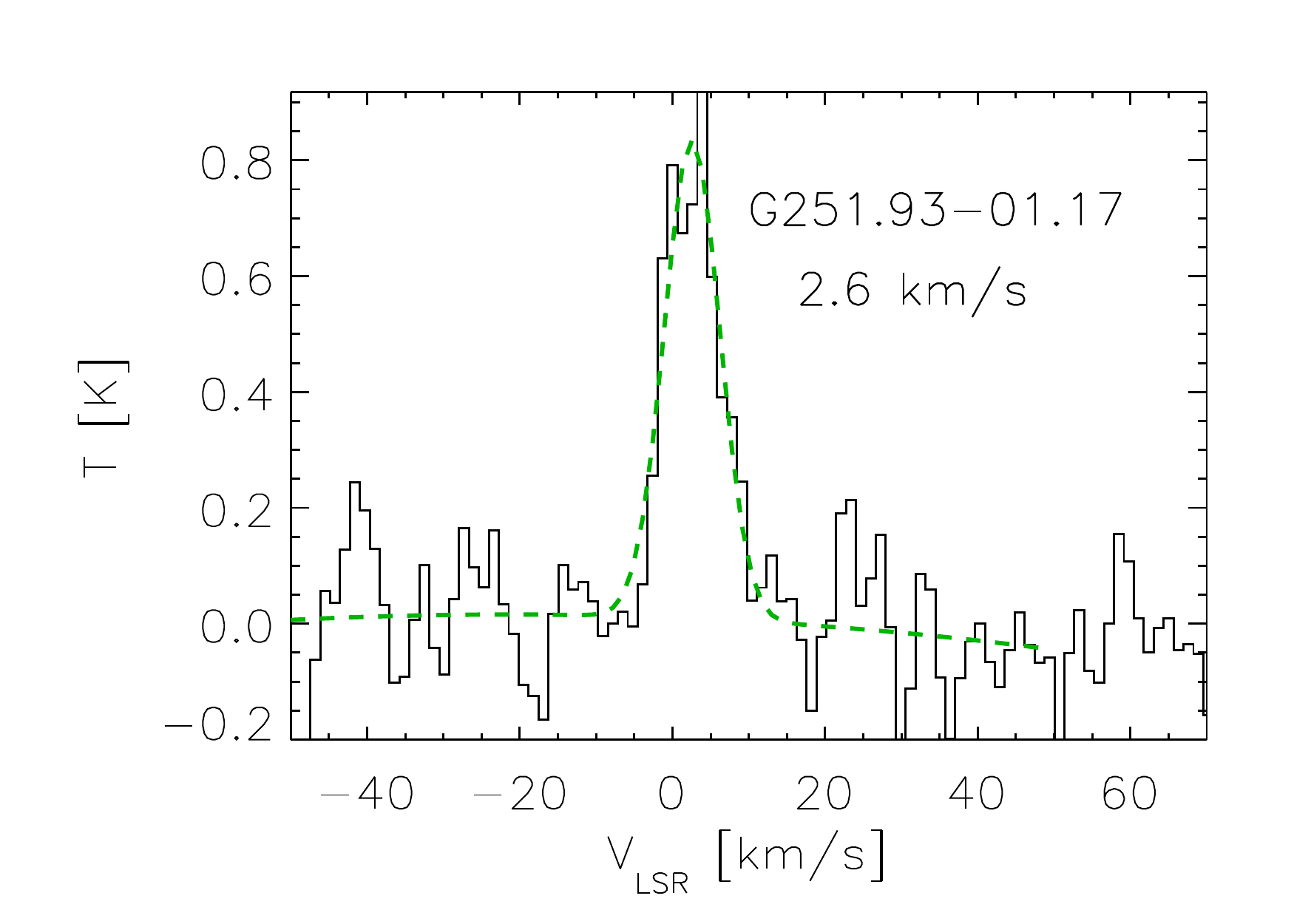}
   \includegraphics[width=0.45\columnwidth]{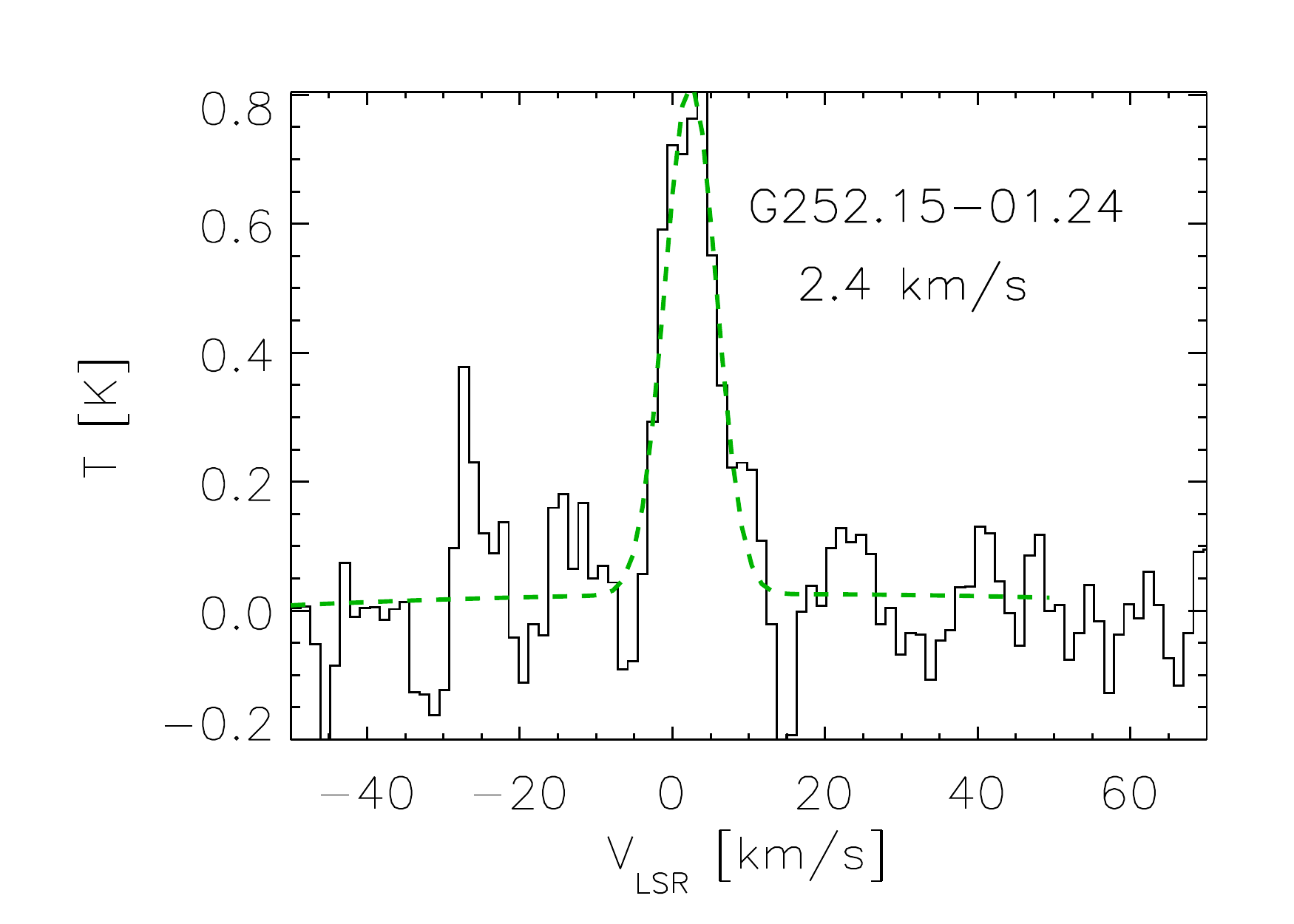}
   \includegraphics[width=0.45\columnwidth]{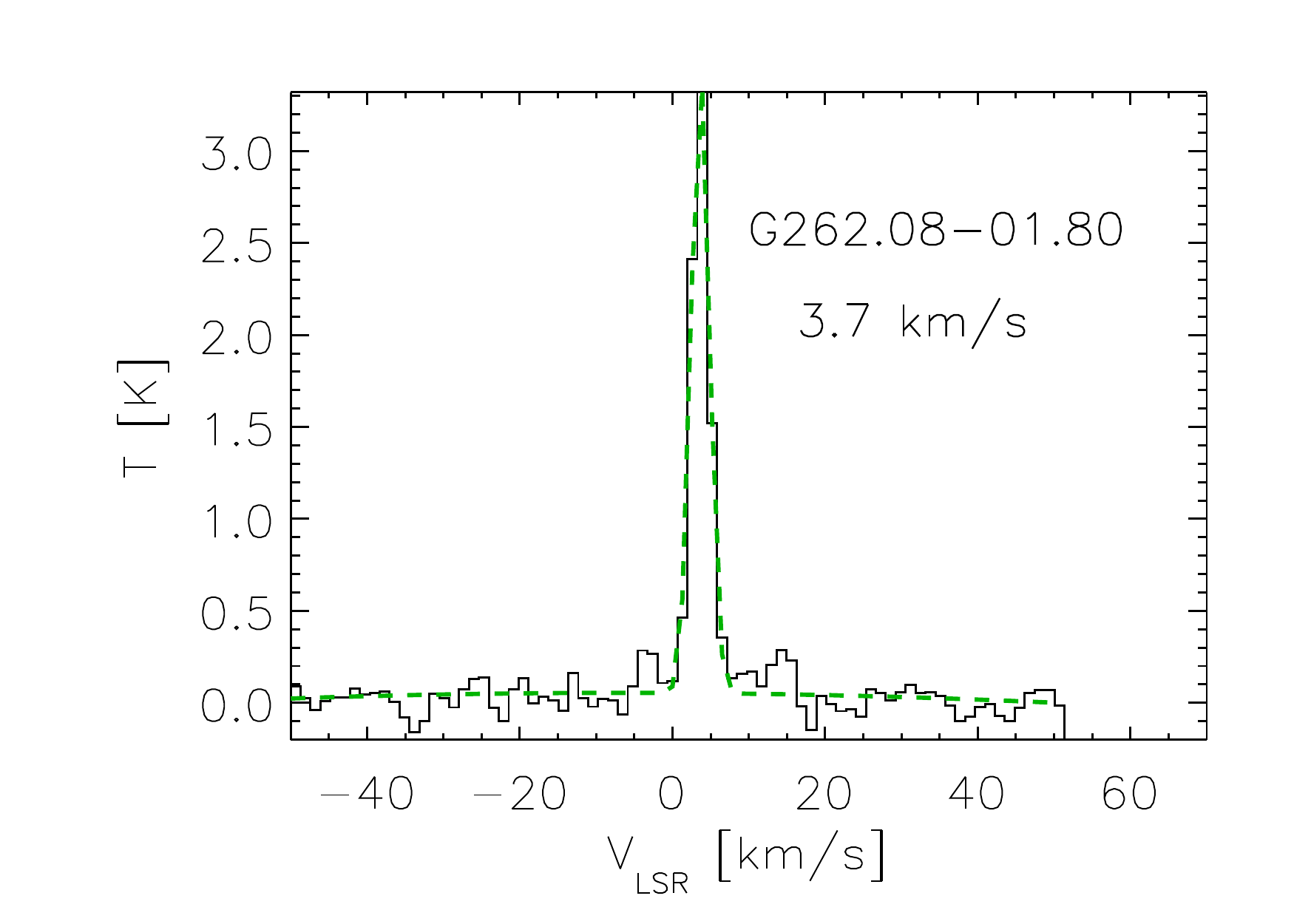}
   \includegraphics[width=0.45\columnwidth]{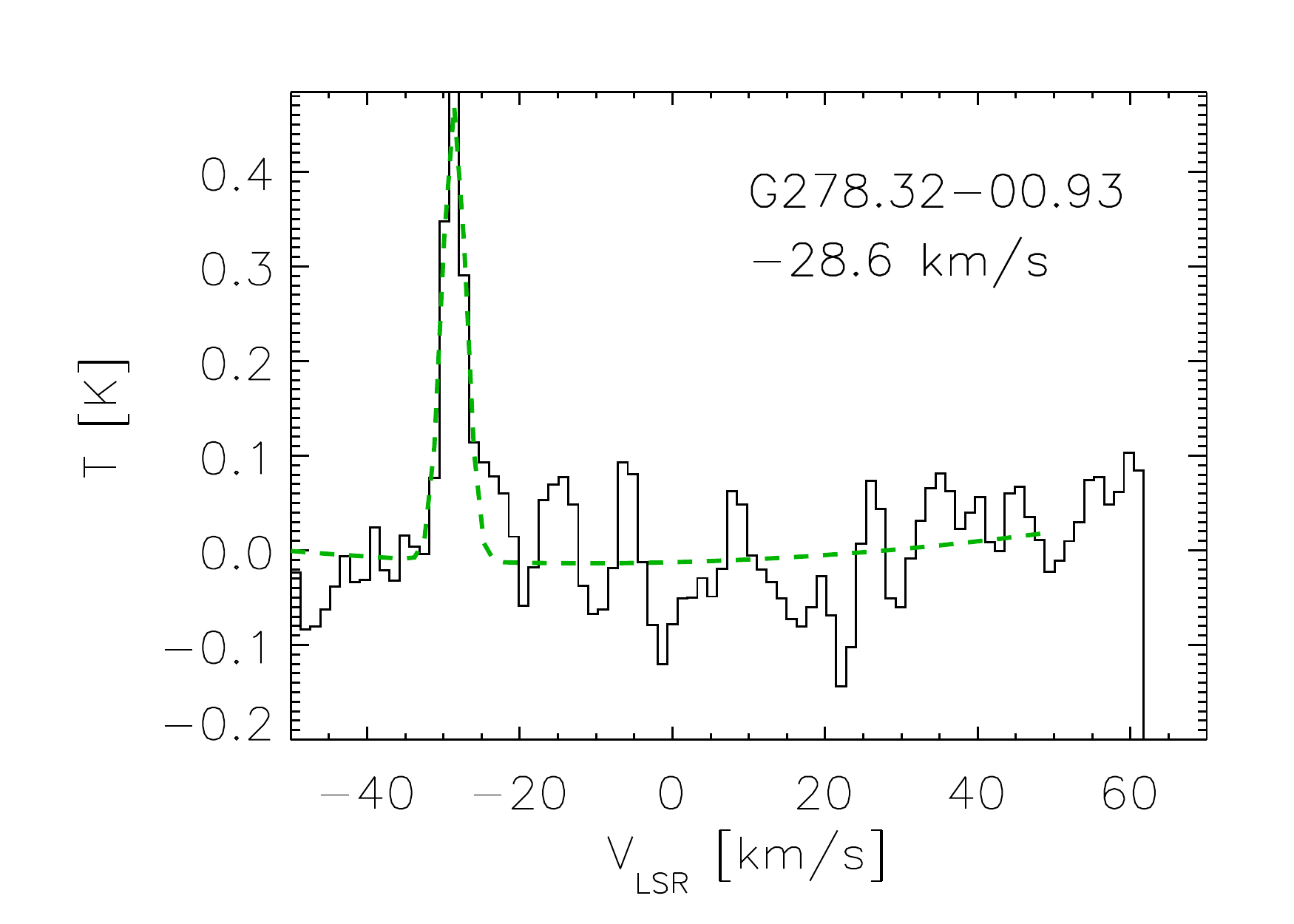}
      \caption{Sample of CfA spectra toward the ECC clumps outside the \citet{2012ApJ...756...76W} survey. Green line shows the fitted Gaussian component centered at the velocity shown in the upper left of the panels.}
      \label{fig:cfa_part1}
\end{figure}

Distance determination is needed for the mass and size calculation of the ECC clumps. For this purpose, association with known IRDCs and molecular line follow up survey data were used. Table 1 in the Appendix shows all the available estimated distances for the ECC objects and the final adopted distances. The data are coming from different sources: 
\begin{enumerate}
\item{Kinematic distance based on the Purple Mountain Observatory’s CO J = 1-0 survey data by \citet{2012ApJ...756...76W}. CO J = 1-0 emission is detected in all ECC clumps except one. Among these sources, 28 clumps belong to our Hi-GAL ECC sample and have velocity and distance estimates in the \citet{2012ApJ...756...76W} survey. Half of them (15 sources) show more than one velocity components. In those cases, we used the velocity component with the brightest line intensity to estimate the distance of the ECC clump.}

\item{Kinematic distance based on the MALT90 (Millimetre Astronomy Legacy Team 90 GHz, ATNF Mopra 22-m telescope, \citealt{2013PASA...30...57J}) survey data. Six ECC clumps were covered by this survey. For the clumps with more than one molecular line detected (i.e. N$_2$H$^+$, $^{13}$CO and $^{12}$CO), we have adopted the kinematic distance derived from the species with the higher critical density.}

\item{Kinematic distance based on our own APEX observations (Zahorecz et al. in prep). APEX observations were performed towards 3 ECC clumps as part of the E-093.C-0866A-2014 project. Again, the kinematic distance adopted for these clumps was the one associated with the emission from the species with the higher critical density.}

\item{Kinematic distance based on the CfA CO survey \citep{2001ApJ...547..792D} data. 24 objects out of our sample of 48 Planck clumps show a single / dominating velocity component in the CfA $^{12}$CO survey data. In Figure \ref{fig:cfa_part1}, we show some examples of the CfA CO survey spectra measured for the ECC sources that do not have any distance estimate from the Purple Mountain Observatory survey.}

\item{Associations with known IRDCs. 6 ECC clumps are associated with known IRDC objects. Their kinematic distances were obtained by \citet{2009MNRAS.399.1506P} based on the Galactic Ring Survey data.}

 \item{Distance from the Planck PGCC catalogue. This distance determination available in the PGCC catalogue was obtained by using several methods: kinematic distance estimates, optical extinction based on SDSS DR7, NIR extinction towards IRDCs, and NIR extinction. In case of the NIR extinction methods, negative values indicated upper limits.}

\end{enumerate}

The distance flags for the ECC clumps considered in our sample are indicated in Table \ref{table:calculated_values}. Once the molecular line velocity information was collected for our sample, we estimated the kinematic distance of the ECC clumps by using the rotation curve of the Galaxy from \citet{2014ApJ...783..130R}. We employed the publicly available fortran code from \citet{2009MNRAS.399.1506P} to relate the radial velocity to the source's Galactrocentric radius. In the inner Galaxy, this Galactocentric radius corresponds to two distances along the line of sight, the near and the far kinematic distances. The near kinematic distance was adopted if only kinematic distance estimation was available for the source. For the sources in the outer Galaxy, the distance solution is unique.

We note that in twelve cases we have both kinematic- and extinction-based distance estimates available. The average difference between these two methods is 10\%, although for two objects they differ by more than 50\%. When an extinction-based distance estimate was available, we used the latter one as the adopted distance (see Table \ref{table:calculated_values}) since this method reproduces maser parallax distances (with very low level of uncertainties) better than kinematics distances (see \citealt{2012ApJ...751..157F}). Otherwise, we adopted the kinematic distance estimates obtained for the remaining ECC clumps.

Calculations of mass, size and H$_2$ column density were performed only for the 40 clumps with available distance estimations. The distances are between 0.1 kpc and 8.1 kpc, the angular sizes lie between 4$\arcmin$ - 15$\arcmin$, which give us spatial scales in the range of 0.5 pc - 29 pc.

We note that the distance estimates for the majority (92\%) of our ECC clumps are likely correct within a factor of 2, and so our mass estimates also lie within a factor of 4. Only 3 sources (i.e. 8\% of the sample) may have distance discrepancies as large as a factor of 5, implying a factor of 25 uncertainty in the calculated mass. Therefore, our results on the mass distribution of ECC clumps as a function of Galactic longitude and Galactocentric distance are not expected to change substantially due to the errors in the distance determination.

\subsection{Location within the Galaxy}
\begin{figure*}[ht]
   \centering
   \includegraphics[width=1.0\columnwidth]{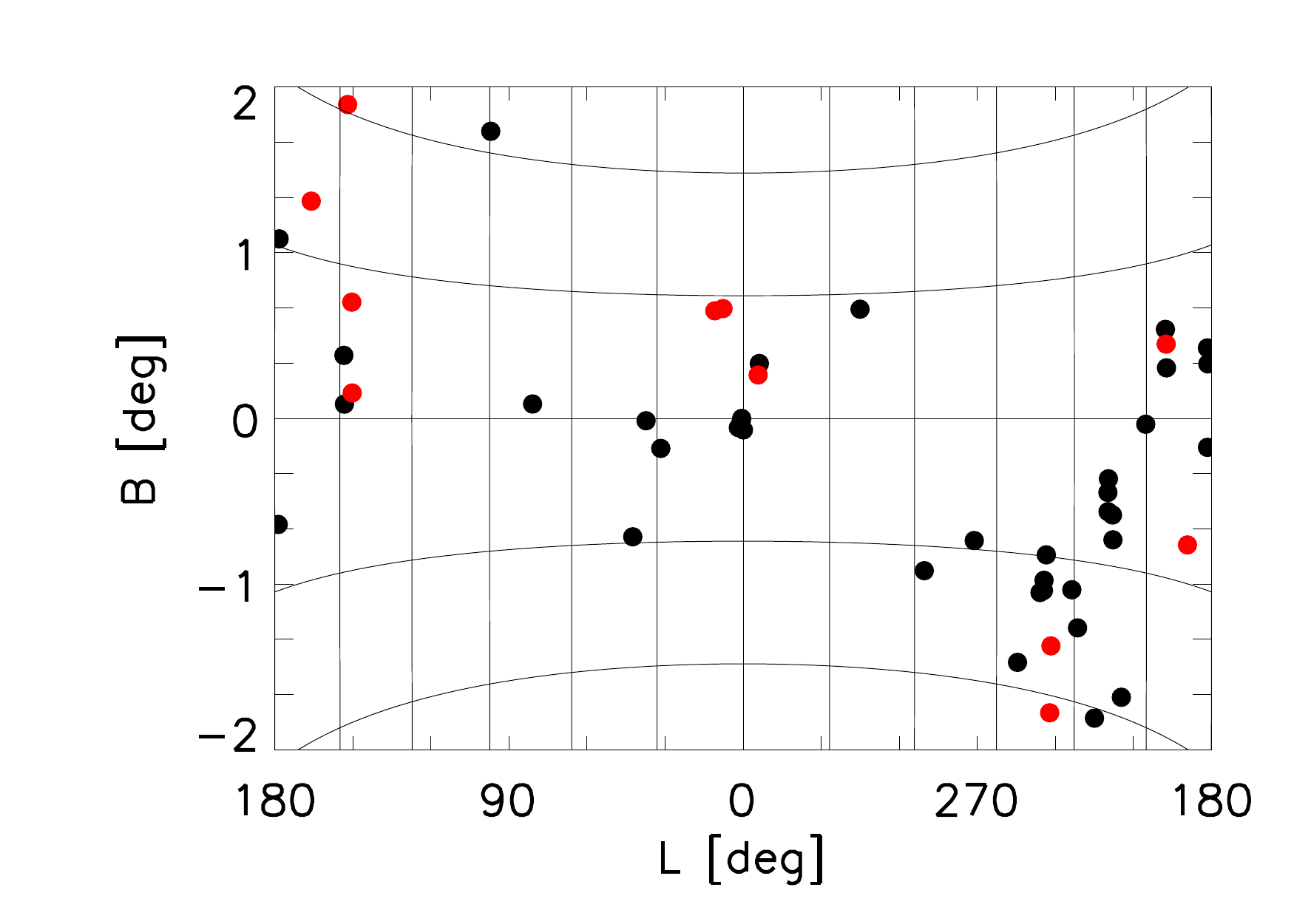}
   \includegraphics[width=0.7\columnwidth]{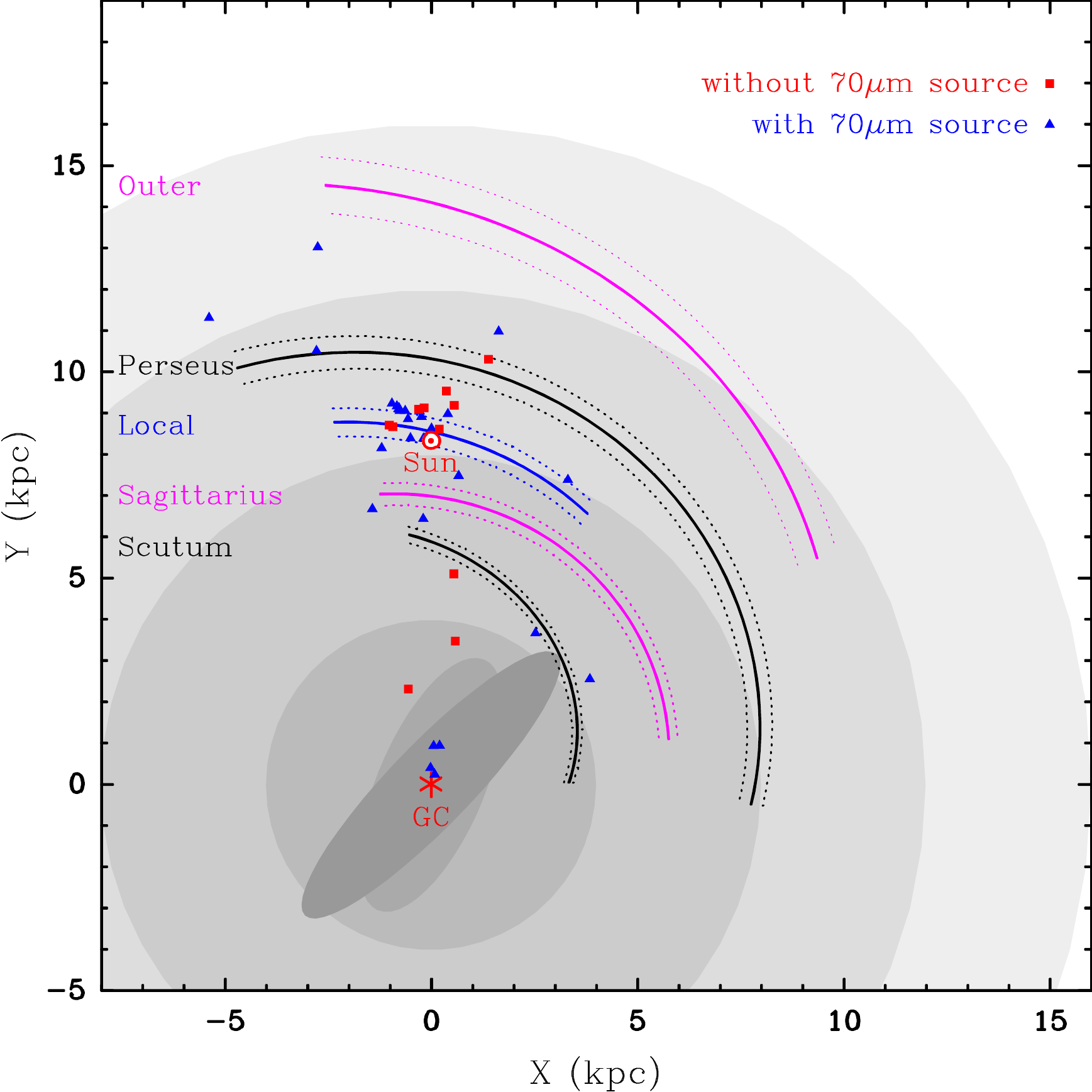}
      \caption{Left: Galactic distribution of the selected 48 ECC clumps observed by the Hi-GAL project in galactic latitude and longitude. Right: Plan view of the Milky Way \citep{2014ApJ...783..130R} with the 40 ECC clumps with available distance estimates overplotted. Red and blue symbols indicate the clumps from Category I and II, respectively (see Section \ref{section:categories} for the definition of these two categories). The position and 1$\sigma$ width of spiral arms \citep{2014ApJ...783..130R} are indicated with solid and dotted lines, respectively. The background gray disks with radii of $\sim$4 kpc, $\sim$8 kpc, $\sim$12 kpc and $\sim$16 kpc indicate the Galactic bar, the solar circle, co-rotation of the spiral pattern and the end of major star formation regions, respectively. }
      \label{fig:distribution}
\end{figure*}

Once the distance of the selected ECC clumps has been estimated, we can determine the location where these objects fall within the Galaxy. Figure \ref{fig:distribution} shows the galactic distribution of the selected clumps. 1/3 and 2/3 of them are located in the inner and outer part of the Galaxy, respectively, between Galactic longitudes of 278$^\circ$ to 89$^\circ$ and 144$^\circ$ to 262$^\circ$. Based on our estimated distances, sixty percent of them fall on top of the Galactic spiral arms as defined in  \citet{2014ApJ...783..130R}.

\section{Methods and Results}

\subsection{Mass determination based on Planck data \label{planck_mass_calculation}}
From the fluxes measured by Planck for every cold clump and provided in the ECC catalogue, we can estimate their individual masses by using \begin{equation}
M=\frac{S_\nu D^2}{\kappa_\nu B_\nu(T)}
\end{equation}
where $S_\nu$ is the integrated flux density, D is the distance, $\kappa_\nu$ is the dust opacity, and $B_\nu(T)$ is the Planck function for dust temperature at T. The dust opacity was adopted by \citet{2011A&A...536A..23P} as:
\begin{equation}
\kappa_\nu=0.1\times \left( \frac {\nu}{1 THz} \right) ^\beta cm^2/g
\label{eq_kappa}
\end{equation}

For this calculation, we used the integrated flux density at the frequency $\nu$=857GHz from the ECC catalogue, the derived core temperature, T$_{core}$, and the emissivity spectral index, $\beta_{core}$.
We note that the Planck fluxes at $\nu$= 857GHz were used for the computation because this frequency is the closest one to the frequency reference (i.e. 1THz) for the formula of the dust opacity (see Eq.\ref{eq_kappa}), so that the impact on the spectral index $\beta$ remains small compared to the uncertainty in $\kappa_\nu$. In addition the 857GHz band has also the best SNR of all Planck observations and the dust emission at this frequency is optically thin \citep[see details in ][]{2011A&A...536A..23P}.

The average calculated mass based on the Planck data is 1.2$\times10^6$ M$_\odot$, with a minimum and maximum value of 10 M$_\odot$ for G089.62+02.16 and 2.7$\times$10$^7$ M$_\odot$ for G000.65-00.01, respectively. This average value is biased by a few very massive sources within the sample. The 25th, 50th and 75th percentile of the data are 1.6$\times10^2$ M$_\odot$, 1.7$\times10^3$ M$_\odot$ and 2.2$\times10^4$ M$_\odot$, respectively.

\subsection{Clump categories based on Hi-Gal 70$\mu$m data \label{section:categories}}

In this Section, we evaluate the level of star and star cluster formation within the Planck cold clumps observed by Herschel using the 70$\mu$m Hi-GAL images. We note that the 70 $\mu$m images have the best angular resolution within the Hi-GAL data and therefore provide information about the point-like sources found within the clumps and likely associated with protostellar objects. Also it is an excellent probe of the population of intermediate- and high-mass young stellar objects in star forming regions \citep[see eg. ][]{2012A&A...547A..49R,2013A&A...549A.130V}. Planck clumps without 70 $\mu$m sources are the best candidates to study the earliest phases of star formation, since the absence of 70 $\mu$m emission likely indicates that protostellar activity has affected very little its surrounding environment.

In order to investigate whether there are significant differences in the physical properties (i.e. dust temperature, mass, maximum H$_2$ column density) between clumps with and without star formation, we divided the sample of Planck cold clumps into two different categories: clumps with no sign of star formation (Category I, 11 sources) and clumps with active star formation (Category II, 37 sources).

\begin{figure}[!h]
   \centering
   \includegraphics[width=0.95\columnwidth]{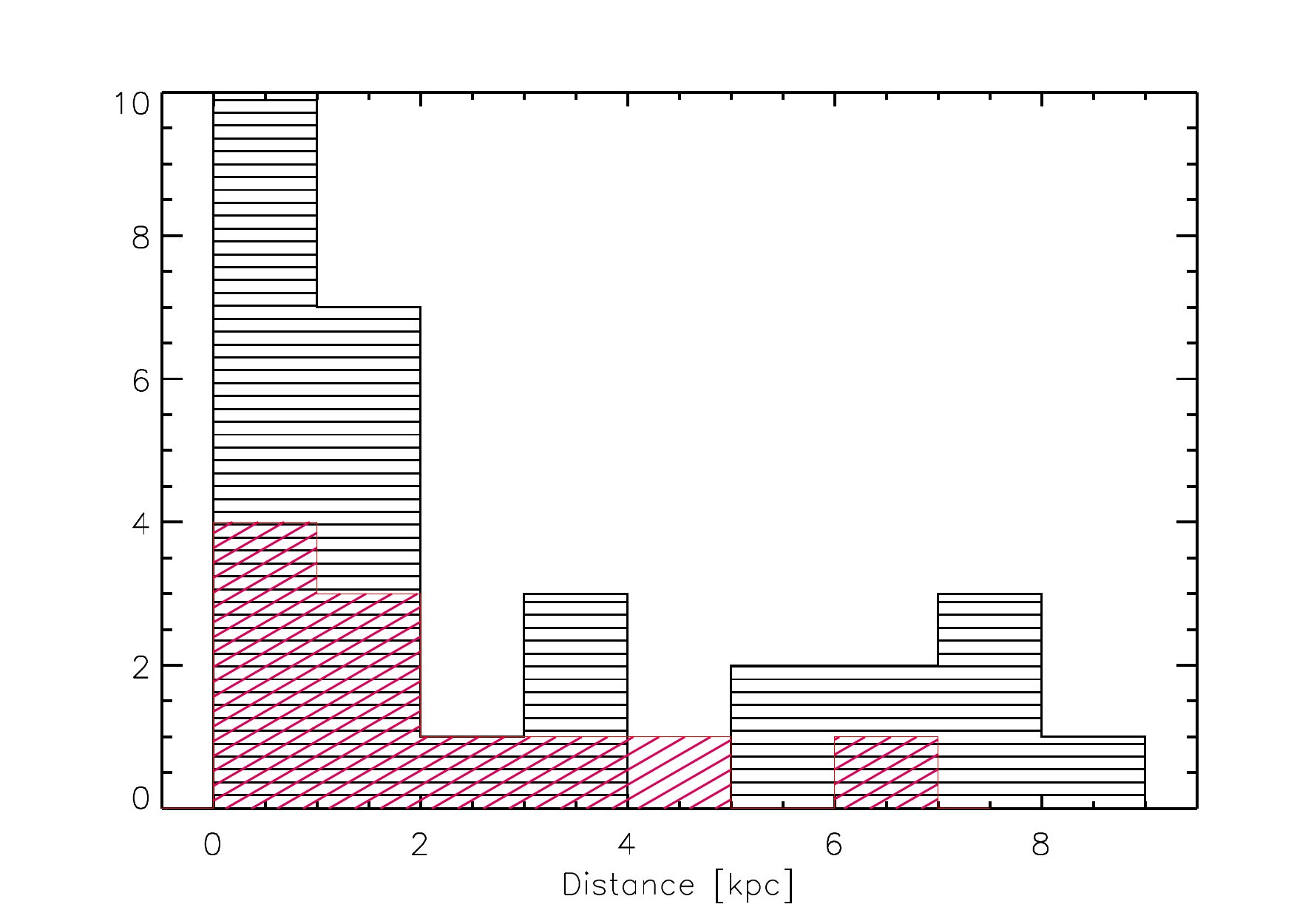}
      \caption{Distribution of the ECC clumps as a function of distance. Red and black histograms indicate the clumps from Category I (clumps without 70 $\mu$m sources) and Category II (clumps with 70 $\mu$m sources), respectively.}
      \label{fig:distances_category}
\end{figure}

We probe the star formation activity by checking the presence/absence of 70 $\mu$m point sources within the defined clump boundaries (see Section \ref{clumpdef}). From our classification of clumps with/without 70$\mu$m point-sources, we find that 11 clumps (23\% of the sample of 48 clumps) do not show any star formation activity. The criterion may be distant-dependent due to the sensitivity of the 70 $\mu$m observations. To check for this possible bias, we plotted the distribution of ECC clumps with and without 70 $\mu$m sources as a function of distance (see Figure \ref{fig:distances_category}). We find that sources without 70 $\mu$m sources are not, in average, further away than sources with 70 $\mu$m emission. On the contrary, most of the cores with no 70 $\mu$m sources lie within $\sim$4.5 kpc from the Sun, with only one exception, G354.81+00.35 at $\sim$6kpc.

\begin{figure*}[ht]
   \centering
   \includegraphics[width=1.8\columnwidth]{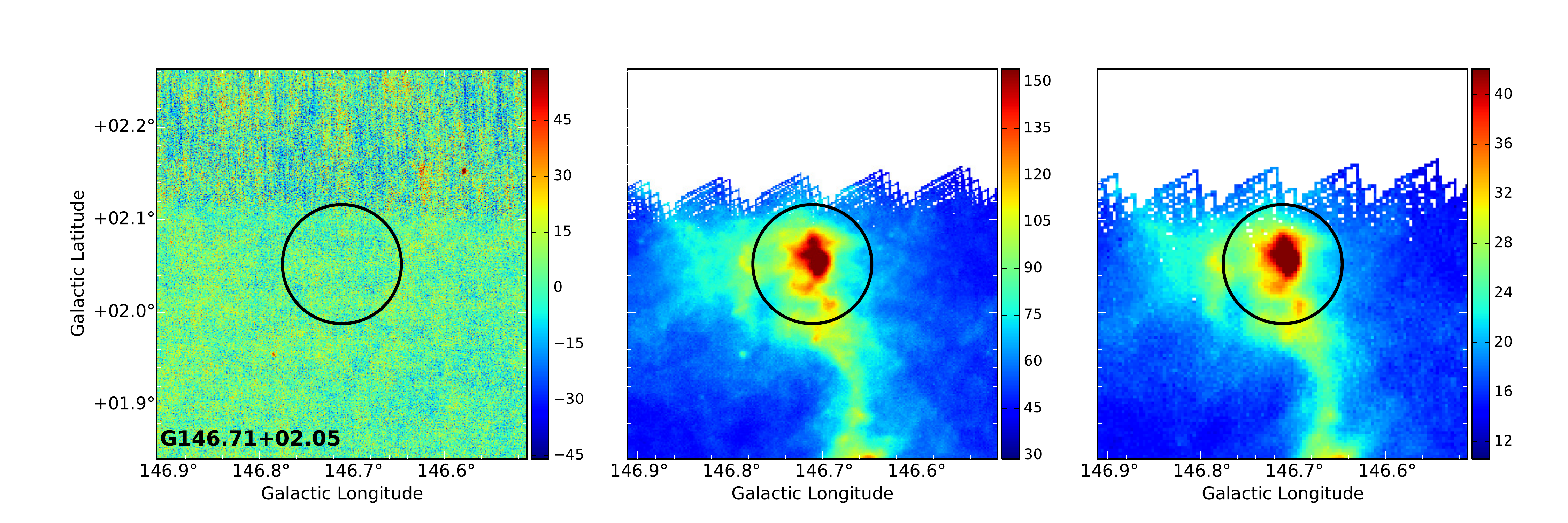}
   \includegraphics[width=1.8\columnwidth]{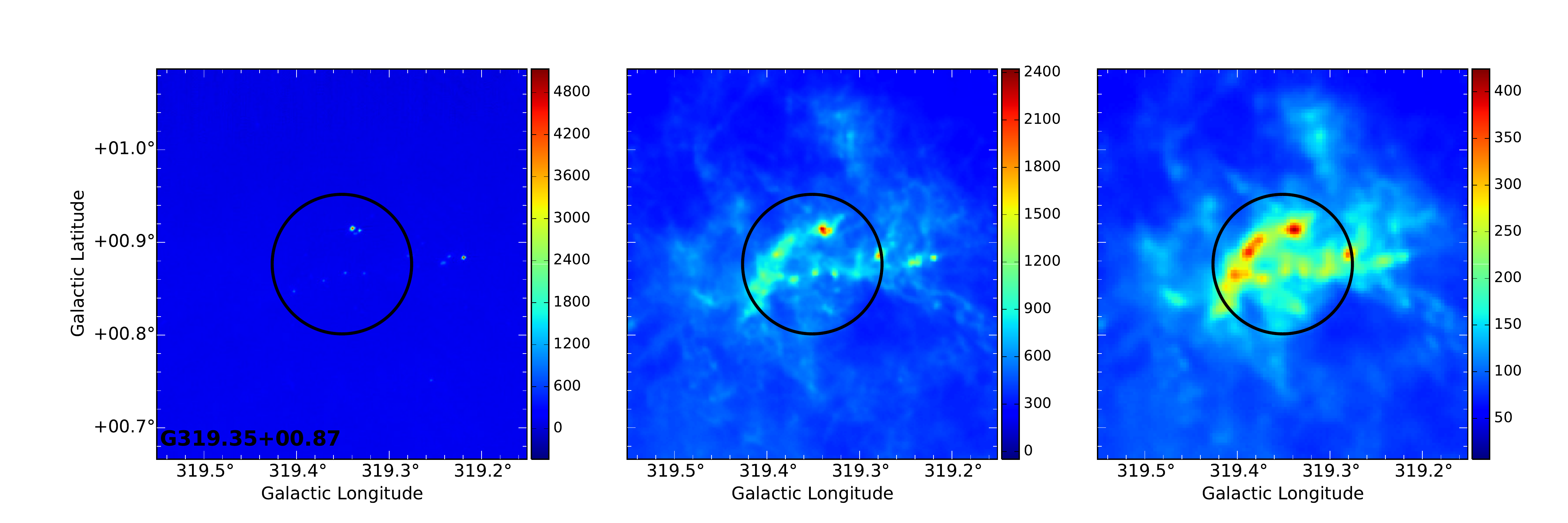}
      \caption{Herschel 70, 250 and 500 $\mu$m images in MJy/sr of G146.71+2.05 and G319.35+0.87 from Category I and II, respectively, with a resolution of 5'', 18'' and 36''. Black circle shows the Planck clump's position and major axis as shown in the ECC catalog.}
      \label{figure:clump_classes}
\end{figure*}

Figure \ref{figure:clump_classes} shows examples of Category I (G146.71+2.05) and Category II (G319.35+0.87) clumps. Herschel 70, 250 and 500 $\mu$m images are shown with angular resolution of 5$\arcsec$, 18$\arcsec$ and 36$\arcsec$, respectively. Similar set of figures are available for all the 48 sources in the Appendix.

To investigate the possibility of chance alignments, we estimated the expected number of point sources within the clump regions. For this, we identified the sources present within 2$\times$2 square degree Hi-GAL 70 $\mu$m tiles using SExtractor \citep{1996A&AS..117..393B}. Except for the ECC clumps located in the inner 20 degree of the Galaxy, the expected number of point sources due to chance alignment is less than 2 (median values of 0.02 for |l-180$^\circ$| $<$ 90$^\circ$ and of 1.8 for 90$^\circ$ < |l-180$^\circ$| $<$ 170$^\circ$), which implies that the 70 $\mu$m sources detected in the ECC clumps genuinely trace the star formation activity within the clumps. For the ECC clumps in the Galactic center, we note however that the number of expected sources by chance alignment is higher than for the ECC clumps in the Galactic disk (median value of 78 for |l-180$^\circ$| $>$ 170$^\circ$). This is related to the H$_2$ column density value chosen to define the clump boundary (3$\times$10$^{21}$ cm$^{-2}$, see Section 3.3), which covers larger areas. In any case, even for these Galactic Center ECC clumps, the number of 70 $\mu$m sources detected in the Hi-GAL images is, in average, higher than the expected number of sources due to chance alignment.

From the integrated flux of the faintest detected 70$\mu$m point sources, we can estimate the luminosity (and mass) limit below which no source is detected in the Hi-GAL images. The faintest detected sources in the ECC clumps have fluxes F$_{70}\sim$45mJy. In order to make the conversion from 70$\mu$m flux to luminosity, we use the following correlation (see \citealt{2008ApJS..179..249D}):
\begin{equation}
L_{int}=3.3 \times 10^8 F_{70}^{0.94} \left(\frac{d}{0.14 kpc}\right)^2 L_{\odot}
\label{eq_dunham}
\end{equation}
where F$_{70}$ is in [ergs cm$^{-2}$ s$^{-1}$] and it is normalised to 0.14 kpc, and $d$ is the distance in kpc.
The faintest sources in our ECC clumps (with F$_{70}$=45mJy) have luminosities 0.2 - 0.7 - 1.5 - 2.6 L$_\odot$ at 1 - 2 - 3 - 4 kpc. For distances larger than 3 kpc, only high mass protostars are luminous enough to be visible.

The all-sky YSO candidate catalogues of \citet{2014PASJ...66...17T} and \citet{2016arXiv160205777M} can be used to search for low and intermediate mass star formation. We find that four of our ECC clumps without 70$\mu$m point-sources show AKARI/WISE YSOs. This implies that, although these clumps have not formed massive protostars yet, they show a certain level of low-mass star formation. From this, the total number of clumps without any star formation activity (low-mass and high-mass) within their clump boundaries is reduced to 7, which represents 15\% of the whole sample. Since we are mostly interested in the potential of the ECC clumps to form massive protostars, we will hereafter focus on the clump classification based on the presence/absence of 70$\mu$m sources only. 

\subsection{Physical properties derived from Hi-GAL data \label{clumpdef}}
The higher-angular resolution of the Herschel Hi-GAL observations allows us to investigate not only the internal physical structure (compactness and fragmentation) of the Planck cold clumps, but also the small-scale variations in the dust temperature and H$_2$ column density distribution of these clumps. These higher angular resolution data also provide more accurate estimates of the fraction of the mass locked into the densest parts of the clumps, i.e. into the regions that will form massive stars and star clusters.

\begin{figure*}[ht]
    \centering
   \includegraphics[width=1.8\columnwidth]{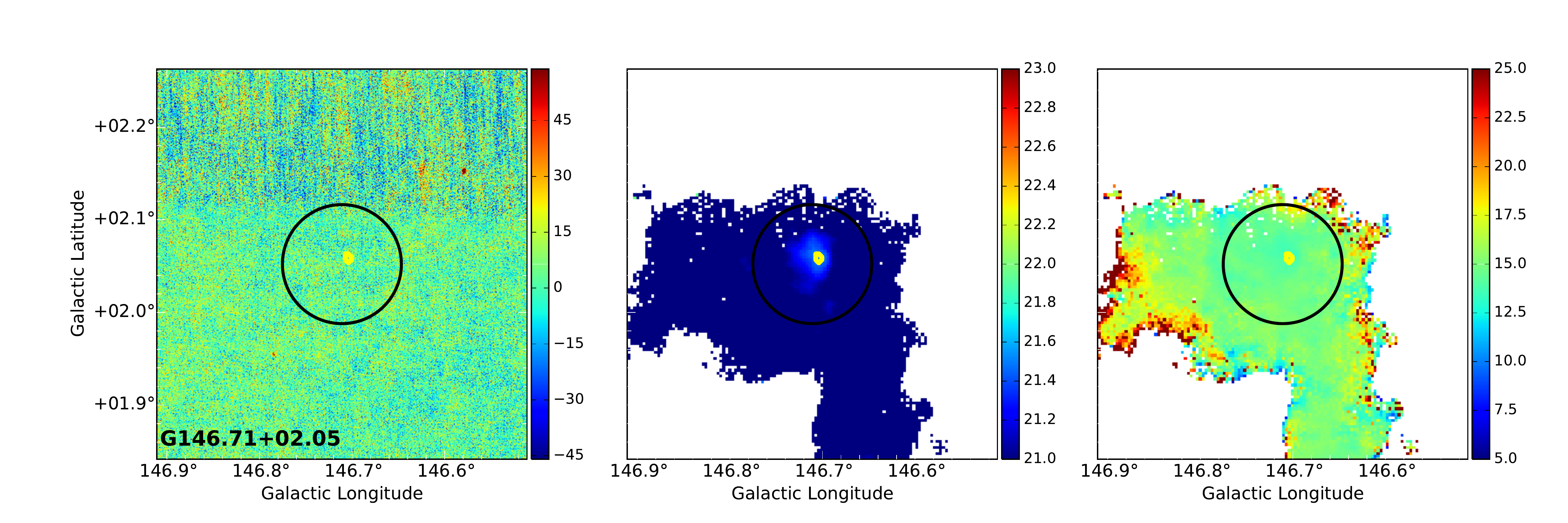}
	\includegraphics[width=1.8\columnwidth]{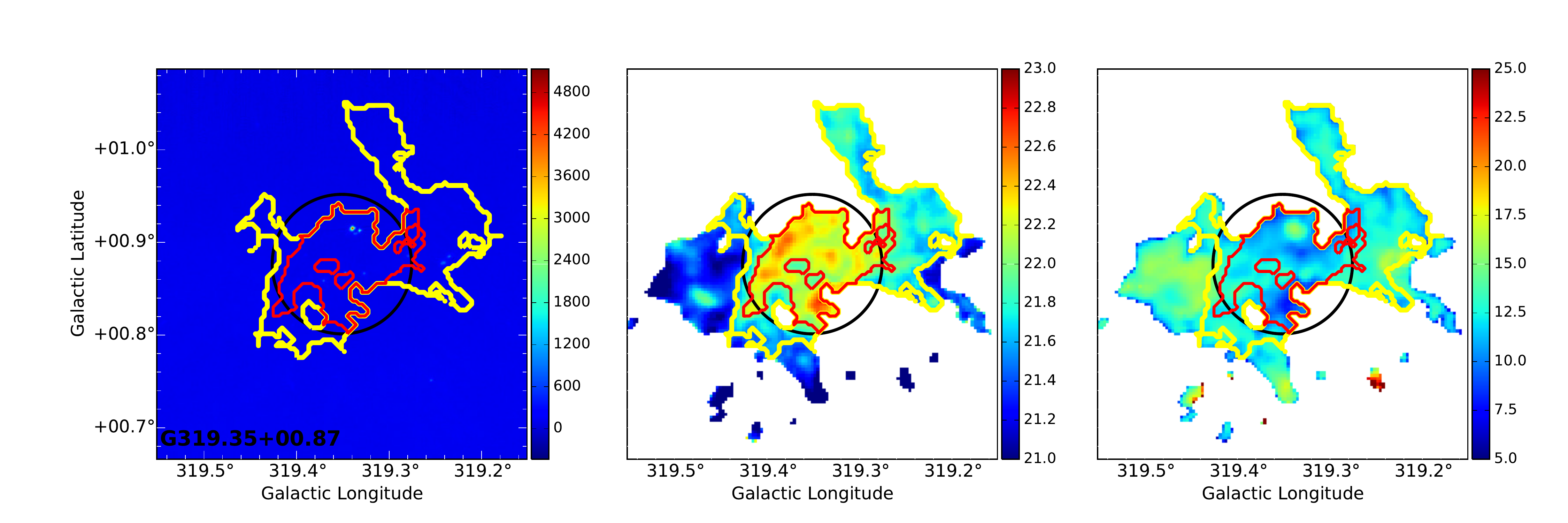}
    \caption{70 $\mu$m images ([MJy/sr], left) and calculated column density ([cm$^{-2}$], middle) in logarithmic scale and dust temperature ([K], right) maps for the same clumps show in Fig. \ref{figure:clump_classes} ECC G146.71+02.05 (top) and G319.35+0.87 (bottom), from Category I and II with a resolution of 36''. Yellow and red contour levels refer to the H$_2$ column density thresholds 3$\times10^{21}$ and 10$^{22}$ cm$^{-2}$, respectively. Black circles are centered at the position of the Planck clump and the circles size corresponds to the major axis given in the ECC catalog.}
    \label{figure:example_source}
\end{figure*}

Dust temperature (T$_{dust}$) and H$_2$ column density maps (N(H$_2$)) were derived by fitting the SED constructed based on the Herschel PACS 160 $\mu$m and SPIRE 250-500 $\mu$m data pixel-by-pixel (see the detailed description of the method in \citealt{wangke15}). As a first step, we smoothed the maps to the SPIRE 500$\mu$m data resolution (i.e. to 36'') and we performed background subtraction. See the Appendix for a detailed description of the background subtraction methods we tested.
As an example, Figure \ref{figure:example_source} shows the 70 $\mu$m images (left) and calculated column density (middle) and dust temperature (right) maps for ECC G146.71+02.05 and G319.35+0.87. 

To derive the mass, average temperature and size of the clumps, we defined the clumps' boundaries 
on the background-subtracted H$_2$ column density maps. The peak column density values span 2 order of magnitudes, so we have used two different column density thresholds for the source definition: 3$\times$10$^{21}$ cm$^{-2}$ and 10$^{22}$ cm$^{-2}$. These values correspond to extinctions of $\sim$3 mag and $\sim$9 mag, respectively (see \citealt{2009MNRAS.400.2050G}). These thresholds are shown in Figure \ref{figure:example_source} in red (for N(H$_2$)=10$^{22}$ cm$^{-2}$) and yellow contours (for N(H$_2$)=3$\times10^{21}$ cm$^{-2}$).
The different column density thresholds give us information about the fraction of mass associated with the diffuse structure of the clump and with the densest parts of the clumps, where star formation is expected to occur.

We note that for one clump, G249.23-01.64, the derived peak column density is smaller than 3$\times$10$^{21}$ cm$^{-2}$, and therefore its average dust temperature, mass and size within the 3$\times$10$^{21}$ cm$^{-2}$ and 10$^{22}$ cm$^{-2}$ contours could not be calculated (see Table \ref{table:calculated_values}). We do not detect 70 $\mu$m sources in it. The peak column density is smaller than 10$^{22}$ cm$^{-2}$ for 24 clumps; for 12 of them we have multiple subclumps. In this case, the total mass of the clump was calculated as the addition of the subclump masses. For four sources in the Galactic Centre region, the peak column density is larger than 10$^{22}$ cm$^{-2}$, but their masses, sizes and temperatures within the used contours were not calculated, because the size within these column density limits is not comparable with the ECC's size. The region within the column density limit of 5-10$\times$10$^{22}$ cm$^{-2}$ is comparable with the ECC objects' size.

In order to provide a characteristic dust temperature for the ECC clumps based on the Hi-GAL data, a column density weighted temperature was calculated within the clump boundary for every object. In this way, the derived temperature is dominated by the temperature of the dense material. The average temperature among all ECC clumps is 13.9 K based on the Herschel images, while the average dust temperature based on the Planck data differs by 1.5-3.0 K with respect to the Herschel one. This difference will be discussed in more detail in the next section.

In Table \ref{table:calculated_values}, we report all the calculated average temperatures, sizes and masses values based on the two column density limits of 10$^{22}$ cm$^{-2}$ and 3$\times$10$^{21}$ cm$^{-2}$.

\begin{longtab}
\begin{landscape}
\begin{longtable}{l r r r r r r r r r r r r c}
\caption{Calculated T, M, R for the ECC clumps. Flag in the distance column: (1) kinematic distance from Wu et al. (2012), (2) kinematic distance based on the MALT90 survey data, (3) kinematic distance based on APEX observations, (4) kinematic distance based on CfA survey data, (5) associations with known IRDCs and molecular cloud cores, (6) distance estimate from the Planck PGCC catalogue \label{table:calculated_values}}\\
\hline \hline
Name	&	Distance	&	T$_{ECC}$	&	major$_{ECC}$	&	R$_{ECC}$ & M$_{ECC}$	&	Max N(H$_2$)	&	T$_{N\geq10^{22}}$	&	M$_{N\geq10^{22}}$	&	R$_{N\geq10^{22}}$	&	T$_{N\geq3\times10^{21}}$	&	M$_{N\geq3\times10^{21}}$	&	R$_{N\geq3\times10^{21}}$	& Category\\
& [kpc] & [K] & ['] & [pc] & [M$\odot$] & [cm$^{-2}$] & [k] & [M$_\odot$] & [pc] & [K] & [M$_\odot$] & [pc] & \\ 
\hline
\endfirsthead
\caption{continued.}\\
\hline \hline
Name	&	Distance	&	T$_{ECC}$	&	major$_{ECC}$	&	R$_{ECC}$ & M$_{ECC}$	&	Max N(H$_2$)	&	T$_{N\geq10^{22}}$	&	M$_{N\geq10^{22}}$	&	R$_{N\geq10^{22}}$	&	T$_{N\geq3\times10^{21}}$	&	M$_{N\geq3\times10^{21}}$	&	R$_{N\geq3\times10^{21}}$	& Category\\
& [kpc] & [K] & ['] & [pc] & [M$\odot$] & [cm$^{-2}$] & [k] & [M$_\odot$] & [pc] & [K] & [M$_\odot$] & [pc] & \\ 
\hline
\endhead
\hline
\endfoot
G000.50+00.00	&	7.40	$^	6	$	&	9.49	&	7.48	&	8.05	&	5.45E+06	&	2.03E+24	&		&		&		&		&		&		&	II	\\
G000.65-00.01	&	8.09	$^	2	$	&	8.91	&	6.73	&	7.91	&	2.73E+07	&	7.40E+24	&		&		&		&		&		&		&	II	\\
G001.64-00.07	&	7.39	$^	2	$	&	9.59	&	4.51	&	4.85	&	1.21E+06	&	2.65E+23	&		&		&		&		&		&		&	II	\\
G006.96+00.89	&	4.89	$^	1	$	&	9.29	&	7.27	&	5.17	&	1.14E+04	&	1.22E+23	&	12.95	&	6.30E+03	&	2.40	&	12.91	&	1.66E+04	&	5.81	&	I	\\
G009.79+00.87	&	3.27	$^	4	$	&	9.24	&	10.02	&	4.76	&	8.31E+04	&	2.62E+23	&	12.20	&	8.85E+03	&	2.97	&	12.74	&	1.72E+04	&	5.61	&	I	\\
G028.56-00.24	&	5.30	$^	6	$	&	12.37	&	10.1	&	7.78	&	4.85E+04	&	1.05E+23	&	13.76	&	1.75E+04	&	3.29	&	15.10	&	3.44E+04	&	7.63	&	II	\\
G033.70-00.01	&	6.94	$^	1	$	&	12.11	&	14.07	&	14.19	&	1.53E+05	&	6.22E+22	&	14.77	&	3.02E+04	&	4.94	&	19.08	&	1.51E+05	&	18.46	&	II	\\
G038.36-00.95	&	1.08	$^	1	$	&	9.28	&	8.47	&	1.33	&	4.28E+03	&	1.70E+23	&	10.98	&	7.76E+01	&	0.24	&	10.98	&	7.85E+01	&	0.25	&	II	\\
G074.11+00.11	&	3.45	$^	5	$	&	9.14	&	9.15	&	4.59	&	2.80E+04	&	3.05E+22	&	12.39	&	1.39E+03	&	1.24	&	12.77	&	5.13E+03	&	3.70	&	II	\\
G089.62+02.16	&	0.06	$^	1	$	&	8.35	&	8.5	&	0.07	&	1.06E+01	&	2.61E+22	&	11.82	&	2.69E-01	&	0.02	&	12.18	&	1.30E+00	&	0.06	&	II	\\
G144.66+00.16	&	0.35	$^	1	$	&	13.94	&	10.39	&	0.53	&	2.05E+01	&	8.58E+21	&		&		&		&	12.01	&	3.39E+00	&	0.11	&	I	\\
G144.84+00.76	&	2.42	$^	1	$	&	8.23	&	9.82	&	3.45	&	6.08E+03	&	8.42E+21	&		&		&		&	10.97	&	2.73E+02	&	1.09	&	I	\\
G146.71+02.05	&	1.03	$^	6	$	&	11.14	&	7.71	&	1.15	&	3.67E+02	&	3.21E+21	&		&		&		&	13.70	&	1.74E+00	&	0.10	&	I	\\
G148.00+00.09	&	0.77	$^	6	$	&	13.13	&	10.16	&	1.14	&	1.48E+02	&	6.33E+21	&		&		&		&	13.46	&	4.73E+01	&	0.45	&	II	\\
G148.24+00.41	&	3.12	$^	1	$	&	11.97	&	7.88	&	3.57	&	2.13E+03	&	2.74E+22	&	6.61	&	9.60E+01	&	0.33	&	9.29	&	7.21E+02	&	1.44	&	II	\\
G162.79+01.34	&	1.26	$^	6	$	&	11.82	&	13.1	&	2.40	&	6.74E+02	&	4.57E+21	&		&		&		&	14.12	&	1.74E+01	&	0.29	&	I	\\
G177.86+01.04	&	0.30	$^	6	$	&	12.52	&	7.97	&	0.35	&	1.32E+01	&	8.75E+21	&		&		&		&	11.07	&	3.42E+00	&	0.11	&	II	\\
G178.28-00.61	&					&	9.32	&	10.16	&		&		&	1.16E+22	&	12.84	&		&		&	13.32	&		&		&	II	\\
G181.84+00.31	&					&	10.80	&	8.57	&		&		&	1.16E+22	&	13.36	&		&		&	12.33	&		&		&	II	\\
G182.02-00.16	&					&	12.55	&	7.27	&		&		&	6.50E+21	&		&		&		&	13.40	&		&		&	II	\\
G182.04+00.41	&					&	11.19	&	6.14	&		&		&	5.87E+21	&		&		&		&	12.36	&		&		&	II	\\
G191.51-00.76	&	0.82	$^	6	$	&	11.54	&	12.39	&	1.48	&	3.55E+02	&	6.96E+21	&		&		&		&	13.21	&	4.70E+01	&	0.44	&	I	\\
G201.13+00.31	&	0.76	$^	1	$	&	15.08	&	7.85	&	0.87	&	1.37E+02	&	8.45E+21	&		&		&		&	13.21	&	1.71E+01	&	0.25	&	II	\\
G201.26+00.46	&	0.82	$^	1	$	&	18.39	&	7.18	&	0.86	&	4.93E+01	&	8.46E+21	&		&		&		&	12.55	&	5.65E+01	&	0.43	&	I	\\
G201.59+00.55	&	0.63	$^	6	$	&	10.31	&	10.54	&	0.97	&	5.71E+02	&	1.85E+22	&	12.18	&	5.29E+01	&	0.26	&	12.70	&	2.68E+02	&	0.85	&	II	\\
G210.30-00.03	&	5.44	$^	6	$	&	10.78	&	8.16	&	6.45	&	3.47E+03	&	4.00E+21	&		&		&		&	11.50	&	1.04E+02	&	0.71	&	II	\\
G220.67-01.86	&	0.96	$^	1	$	&	14.39	&	8.4	&	1.17	&	1.60E+02	&	1.85E+22	&	11.92	&	3.39E+01	&	0.20	&	11.87	&	1.01E+02	&	0.50	&	II	\\
G224.27-00.82	&	1.11	$^	1	$	&	11.72	&	7.82	&	1.26	&	2.06E+03	&	7.80E+22	&	12.09	&	8.26E+02	&	0.88	&	11.68	&	2.86E+03	&	2.37	&	II	\\
G224.47-00.65	&	1.18	$^	1	$	&	9.32	&	6.86	&	1.18	&	3.15E+03	&	6.20E+22	&	12.10	&	4.46E+02	&	0.72	&	11.73	&	3.20E+03	&	2.51	&	II	\\
G226.16-00.41	&	1.05	$^	6	$	&	11.29	&	9.08	&	1.39	&	3.29E+02	&	5.55E+21	&		&		&		&	13.03	&	3.97E+01	&	0.43	&	II	\\
G226.29-00.63	&	1.31	$^	1	$	&	11.65	&	6.29	&	1.20	&	1.41E+02	&	5.98E+21	&		&		&		&	12.07	&	5.72E+01	&	0.49	&	II	\\
G226.36-00.50	&	0.77	$^	6	$	&	11.28	&	5.76	&	0.64	&	1.29E+02	&	8.20E+21	&		&		&		&	12.25	&	2.21E+01	&	0.30	&	II	\\
G231.81-02.07	&	3.53	$^	6	$	&	13.97	&	6.91	&	3.55	&	1.72E+03	&	5.79E+21	&		&		&		&	12.64	&	3.08E+02	&	1.17	&	II	\\
G238.71-01.47	&					&	11.48	&	8.28	&		&		&	9.49E+21	&		&		&		&	12.82	&		&		&	II	\\
G240.99-01.21	&	6.15	$^	6	$	&	10.78	&	15.24	&	13.63	&	1.20E+04	&	6.84E+21	&		&		&		&	12.97	&	7.95E+02	&	1.80	&	II	\\
G249.23-01.64	&	1.08	$^	4	$	&	11.66	&	8.86	&	1.39	&	2.03E+02	&	1.85E+21	&		&		&		&		&		&		&	I	\\
G249.67-02.12	&	0.99	$^	4	$	&	9.25	&	7.71	&	1.11	&	5.12E+02	&	3.67E+21	&		&		&		&	12.31	&	1.15E+00	&	0.08	&	I	\\
G251.05-00.98	&					&	12.81	&	8.01	&		&		&	7.47E+21	&		&		&		&	12.32	&		&		&	II	\\
G251.93-01.17	&	0.19	$^	4	$	&	10.34	&	9.38	&	0.26	&	1.81E+01	&	5.81E+21	&		&		&		&	12.86	&	3.56E-01	&	0.04	&	II	\\
G252.15-01.24	&	0.16	$^	4	$	&	9.92	&	8.74	&	0.20	&	1.30E+01	&	8.64E+21	&		&		&		&	11.94	&	7.43E-01	&	0.05	&	II	\\
G253.49-01.26	&					&	11.23	&	8.39	&		&		&	1.23E+22	&	12.86	&		&		&	13.05	&		&		&	II	\\
G262.08-01.80	&	0.50	$^	4	$	&	8.67	&	14.34	&	1.04	&	8.79E+02	&	2.05E+22	&	12.37	&	2.79E+01	&	0.20	&	13.23	&	1.30E+02	&	0.61	&	II	\\
G278.32-00.93	&	1.21	$^	4	$	&	12.37	&	8.59	&	1.51	&	6.70E+02	&	9.05E+21	&		&		&		&	13.38	&	2.06E+02	&	0.85	&	II	\\
G296.52-01.19	&					&	14.47	&	7.63	&		&		&	2.20E+22	&	13.85	&		&		&	13.58	&		&		&	II	\\
G319.35+00.87	&	2.18	$^	6	$	&	11.59	&	9.04	&	2.86	&	8.09E+03	&	4.32E+23	&	10.47	&	6.31E+03	&	2.30	&	10.85	&	9.40E+03	&	3.72	&	II	\\
G354.39+00.44	&	1.90	$^	6	$	&	12.08	&	8.92	&	2.46	&	2.22E+04	&	5.06E+23	&	12.90	&	1.53E+04	&	3.05	&	13.14	&	1.82E+04	&	4.03	&	II	\\
G354.81+00.35	&	6.04	$^	2	$	&	9.46	&	7.44	&	6.53	&	8.12E+05	&	9.86E+22	&	14.19	&	1.76E+05	&	9.90	&	14.85	&	2.27E+05	&	15.76	&	I	\\
G359.91-00.09	&	7.93	$^	2	$	&	9.04	&	10.69	&	12.32	&	1.15E+07	&	2.03E+24	&		&		&		&		&		&		&	II	\\
\end{longtable}
\end{landscape}
\end{longtab}

\subsection{Distribution of the ECC physical properties with Galactic longitude and Galactocentric distance}

After calculating the average H$_2$ column densities, dust temperatures and masses of the ECC clumps from the Herschel data, we can investigate whether variations in these properties are observed depending on the location of the ECC clumps within the Galaxy.

\begin{figure}[h]
   \centering
   \includegraphics[width=0.80\columnwidth]{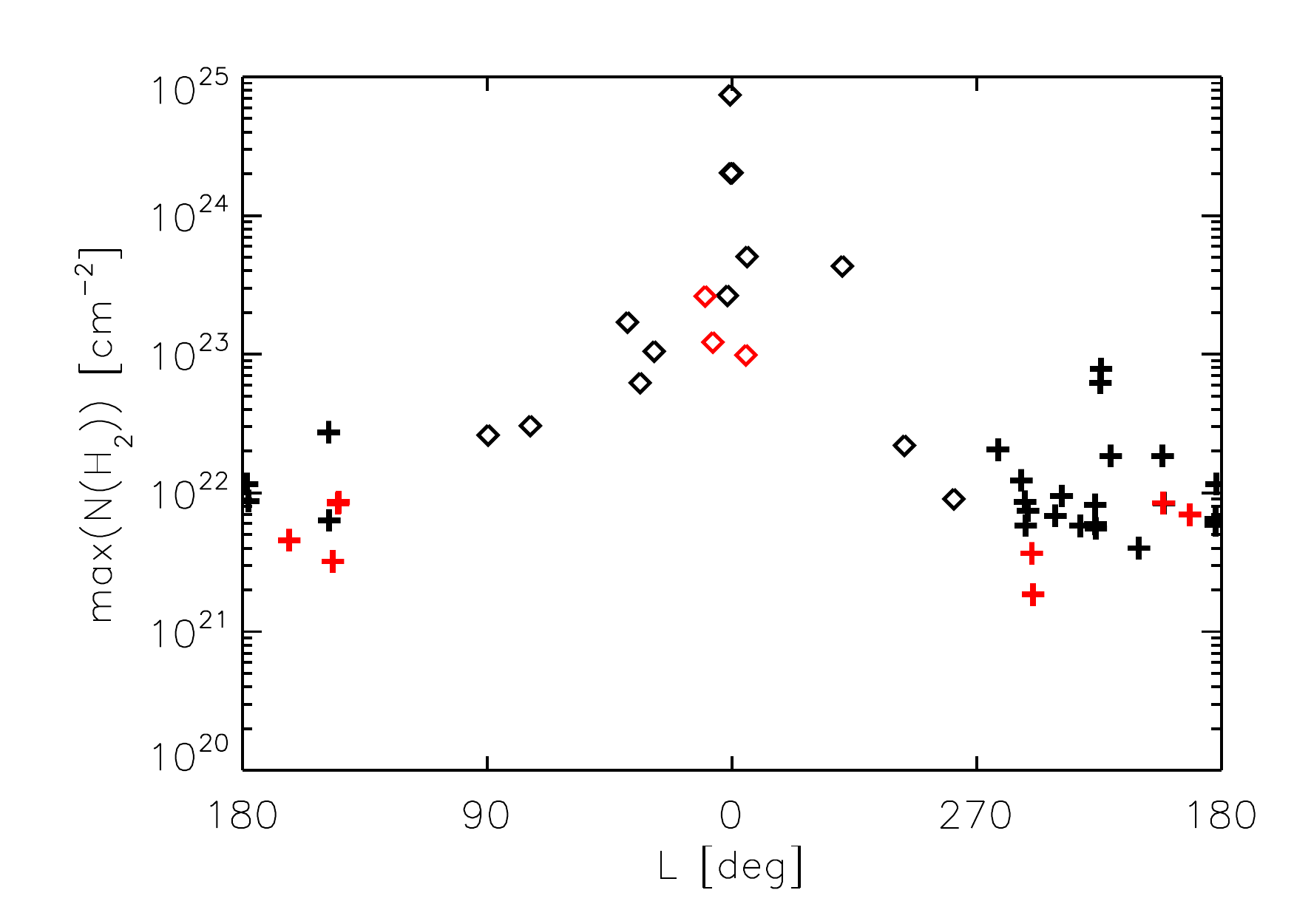}
   \includegraphics[width=0.80\columnwidth]{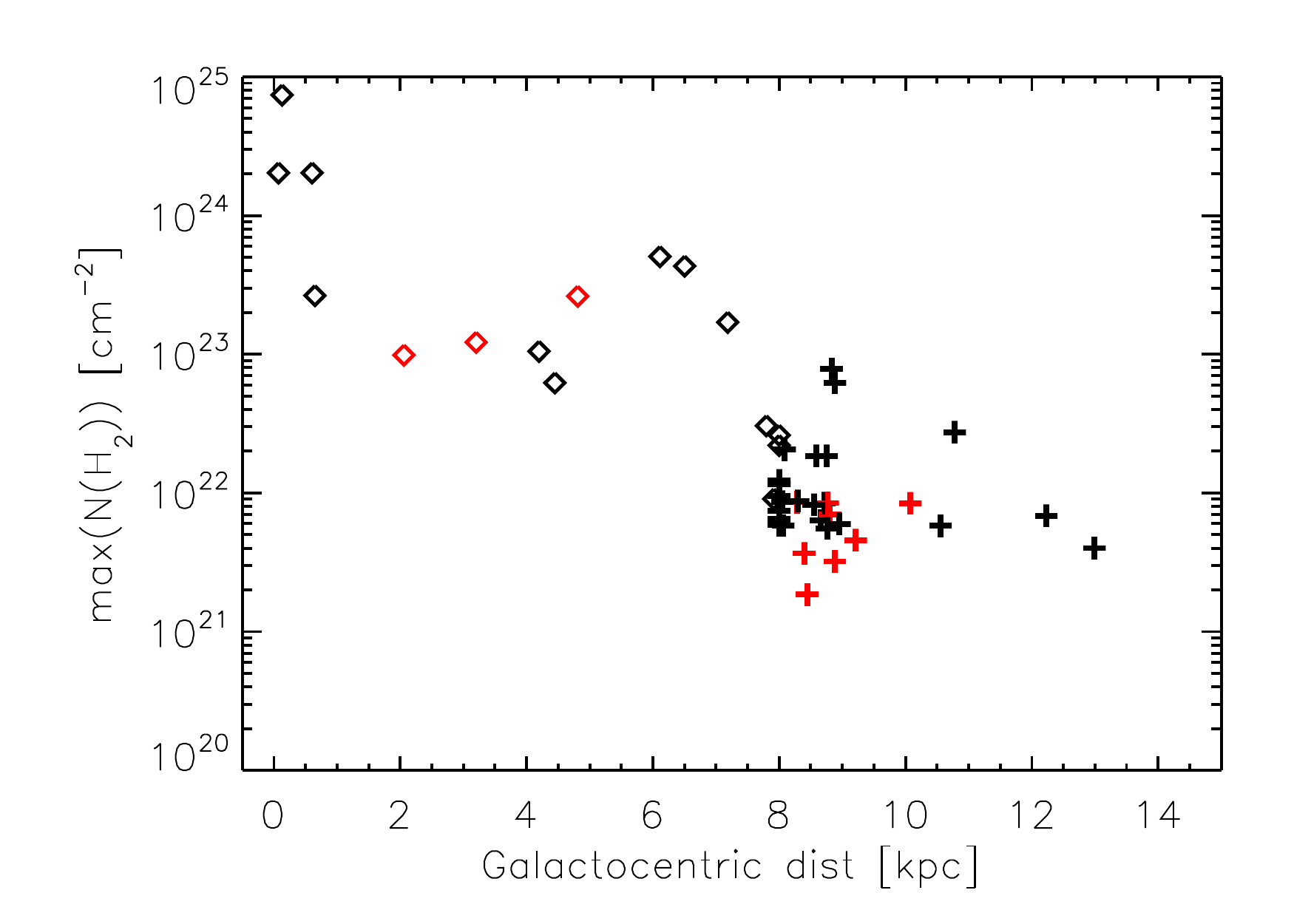}
      \caption{Peak H$_2$ column density distribution of the ECC clumps as a function of Galatic longitude for the 48 objects (top) and Galactocentric distance for the 40 clumps with available distance estimates (bottom). Red symbols indicate the clumps without 70 $\mu$m point sources. 
      }
      \label{figure:nh2_dist}
\end{figure}

In Figure \ref{figure:nh2_dist}, we show the maximum $H_2$ column density distribution of the ECC clumps as a function of Galactic longitude and galactocentric distance. ECC clumps toward the Galactic center region show the highest peak column density values. The peak column density decreases with galactocentric distance. There is a 3 order of magnitude difference between the peak N(H$_2$) values in the Galactic center region and in the outer part of the Galaxy. From Figure \ref{figure:nh2_dist}, we also find that H$_2$ peak column densities N(H$_2$)>9$\times$10$^{23}$ cm$^{-2}$ are detected only toward the Central Molecular Zone region. No high column density clumps with N(H$_2$)>10$^{23}$ cm$^{-2}$ are found in the outer Galaxy. There is no clear trend between Category I and Category II clumps and their location within the Galaxy. These two types of clumps are found indistinctly across the Galaxy.

\begin{figure}[h]
   \centering
   \includegraphics[width=0.8\columnwidth]{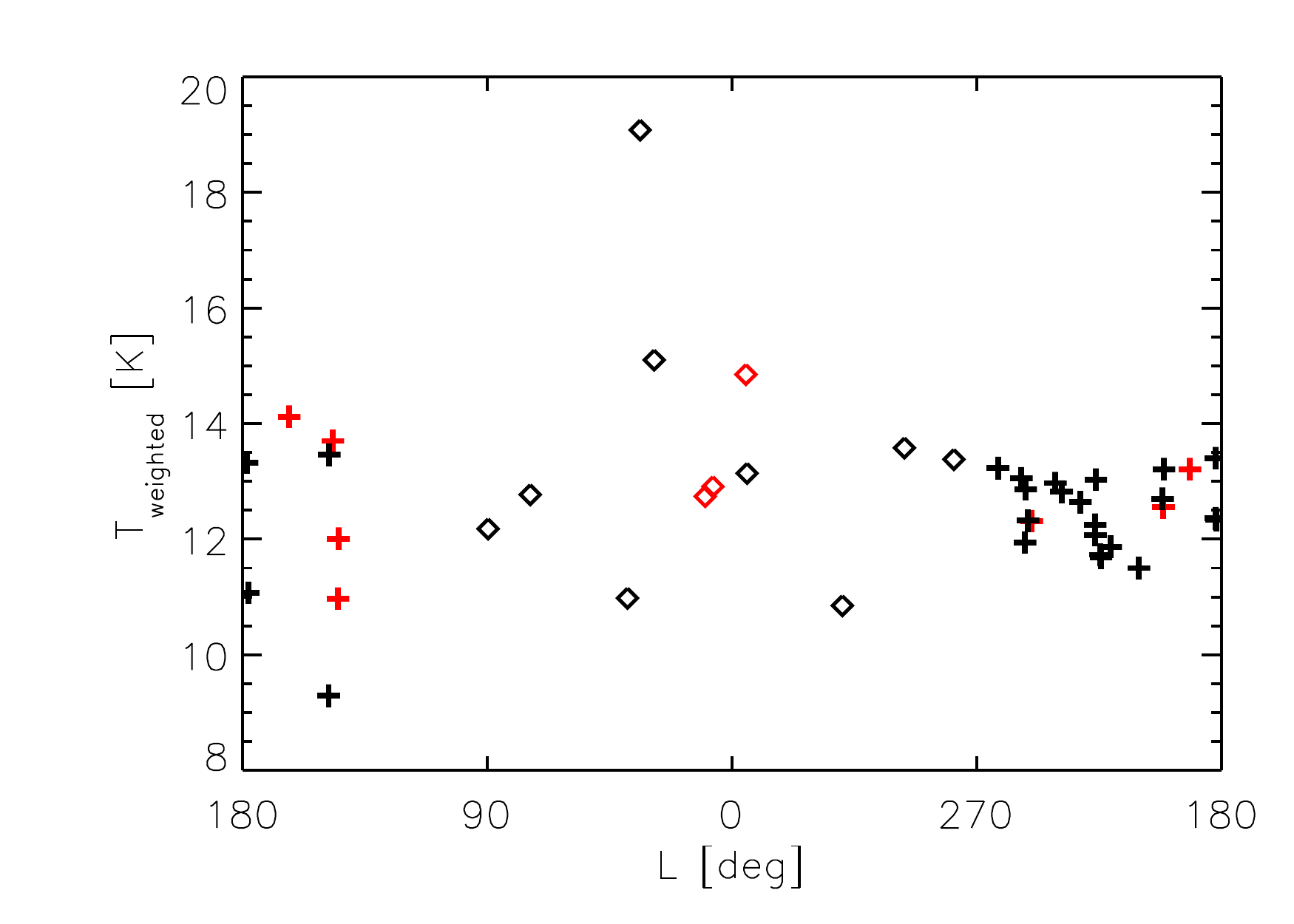}
   \includegraphics[width=0.8\columnwidth]{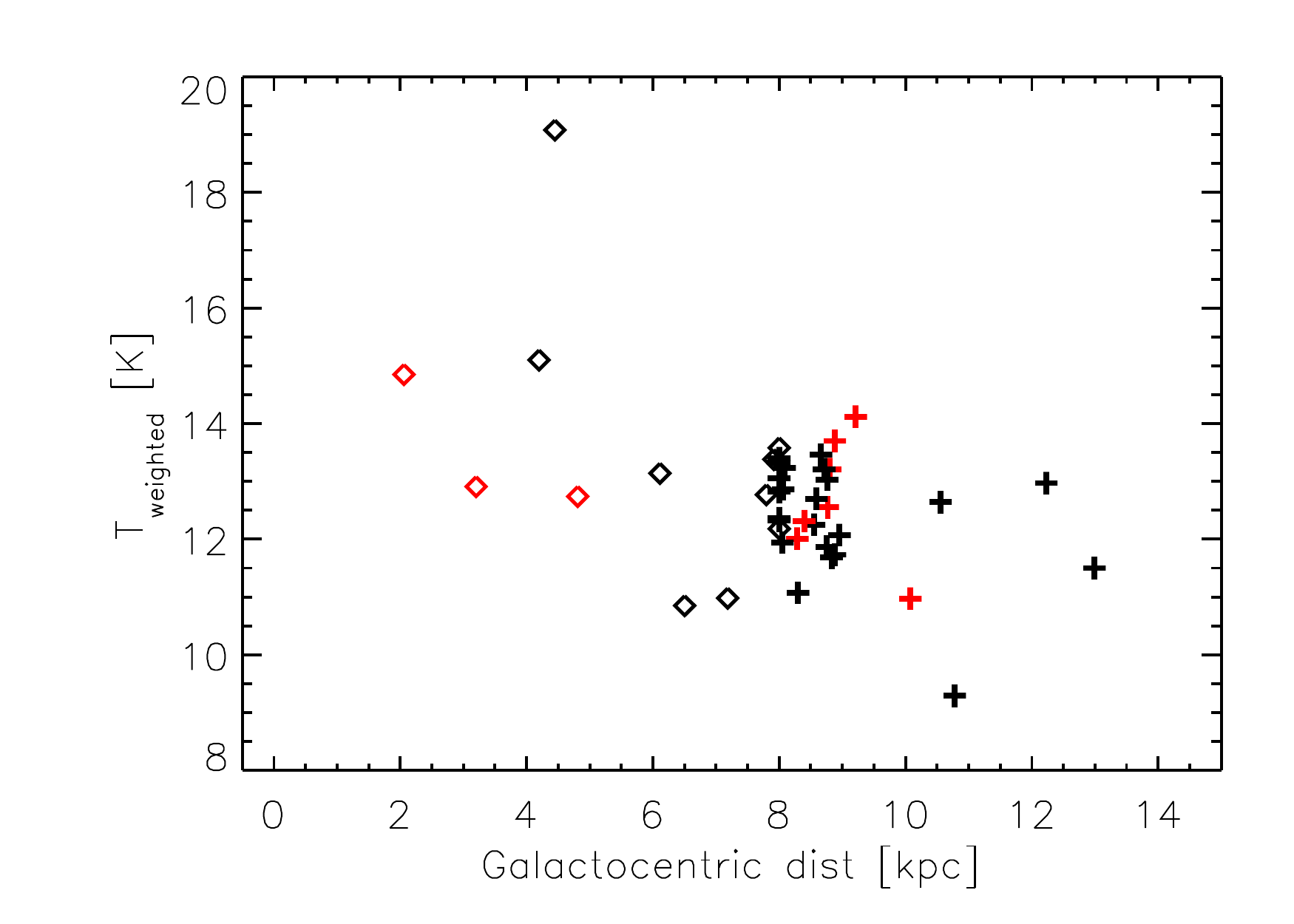}
      \caption{Column density weighted average temperature distribution of the ECC clumps as a function of Galactic longitude for the 43 objects (top, 4 sources in the Galactic center region and a low column density source are excluded; see details in Section \ref{clumpdef}) and as a function of Galactocentric distance for the 35 clumps with available distance and average temperature estimates (bottom). Red symbols indicate the clumps without 70 $\mu$m point sources.}
      \label{figure:temp_dist}
\end{figure}

Figure \ref{figure:temp_dist} shows the column density weighted temperature, T$_{weighted}$, distribution of the ECC clumps as a function of Galactic longitude and Galactocentric distance. It shows similar trends as the distribution of the maximum column density values. ECC objects in the Galactic center show higher temperatures (by more than 5 K difference). 
For the clumps with Galactocentric distances larger than 6 kpc, there is no significant difference in the calculated temperature between clumps with (in red) and without (in black) 70 $\mu$m sources. For Galactocentric distances shorter than 6 kpc, we have too few sources to draw conclusions. The weighted temperature outside the molecular ring is $\sim$13 K, values larger than 14.5 K are detected only within the molecular ring.

\subsection{Comparison of Planck and Hi-GAL temperatures}

\begin{figure}[h]
   \centering
   \includegraphics[width=0.8\columnwidth]{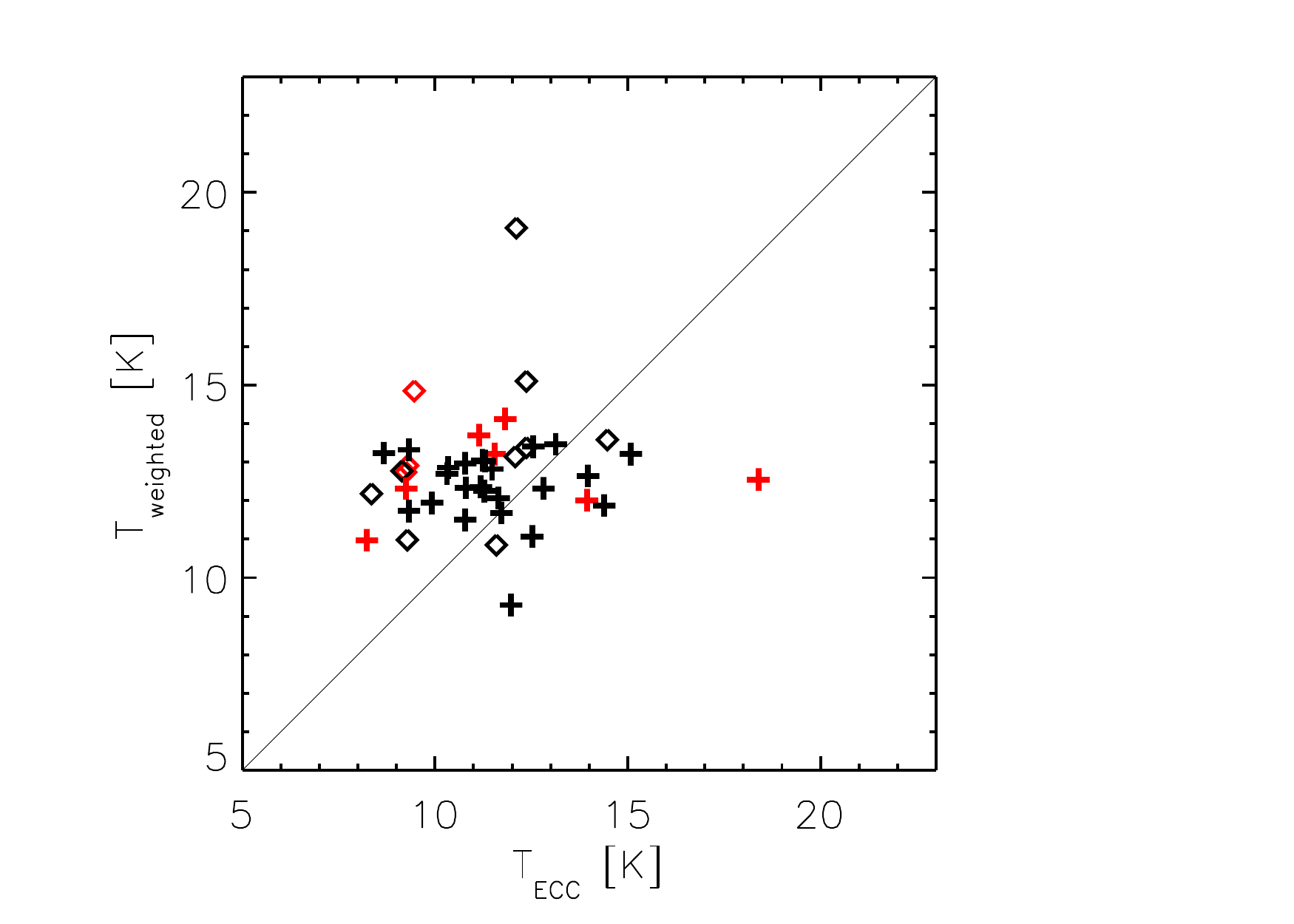}
      \caption{T$_{ECC}$ vs T$_{weighted}$ distribution. Red symbols indicate the clumps without 70 $\mu$m point sources. Plus signs indicate the clumps in the outer part of the Galaxy, diamonds indicate the sources in the inner Galaxy. 
      }
      \label{figure:tecc_twei}
\end{figure}

In Figure \ref{figure:tecc_twei}, we make the comparison between the clump dust temperature calculated from Planck data, T$_{ECC}$, and the column density weighted average dust temperature derived from the Herschel images, T$_{weighted}$. The Planck and Herschel based temperatures agree within $\pm$3 K, except for the warmer sources in the Galactic center region, where our column density criteria select a much larger region than the Planck clump size and includes warmer regions. The average temperatures in the inner Galaxy are 10.4 K and 15.4 K based on the Planck and Herschel data, respectively. The same values are 11.8 K and 12.5 K in the outer part of our Galaxy. We note that there is one outlier clump in the outer Galaxy without any 70 $\mu$m emission (G201.26+00.46), for which the Planck based temperature is higher by $\sim$5 K with respect to the average temperature derived from Herschel data. A bright 70 $\mu$m-emission region with high dust temperatures is visible toward the south-west from the clump. The Planck ECC size is larger than the denser region used in the Herschel-based calculation, which causes lower weighted temperature for the Herschel data.
This comparison shows that the Planck ECC temperatures are, in most cases, a reasonable approximation for the average temperatures of the high column density regions of the cores. Temperatures derived from the integrated Herschel fluxes within the two column density contours that we used in the paper lead to the same conclusion.

\section{Discussion}
In this Section, we evaluate the potential of forming massive stars and star clusters by the selected Planck cold clumps and determine their evolutionary stage. This is done by using the column density maps and bolometric luminosities based on the Herschel observations. In Section 4.1 we investigate the fraction of mass in the densest parts of the clumps compared to their surrounding, more diffuse regions, while in Section 4.2 we identify the clumps with on-going massive star formation as probed by 70 $\mu$m emission. \citet{2010ApJ...723L...7K} showed that regions forming massive stars are, at a given radius, more massive than the limit mass (m$_{lim}$[r]=870M$_\odot$[r/pc]$^{1.33}$) and thus, we evaluate the fraction of Planck cold clumps with potential to form massive stars and star clusters. In Section 4.3 we finally investigate the evolutionary stage of the clumps by comparing their L$_{bol}$ - M$_{env}$ values with the theoretical evolutionary tracks calculated by \citet{2008A&A...481..345M}.

\subsection{Mass distribution}
As mentioned in Section 3.3, the clump masses and sizes were calculated from the Hi-GAL column density maps by considering two different H$_2$ column density thresholds of 3$\times10^{21}$ and 10$^{22}$ cm$^{-2}$. The comparison of the masses contained within these levels gives us information about the fraction of gas mass locked into the densest parts of the clumps and the total mass enclosed within the surrounding region. This parameter may be important since it is possible to increase the dense core's total mass through the accretion of this lower-column density material.

\begin{figure}
   \centering
   \includegraphics[width=0.8\columnwidth]{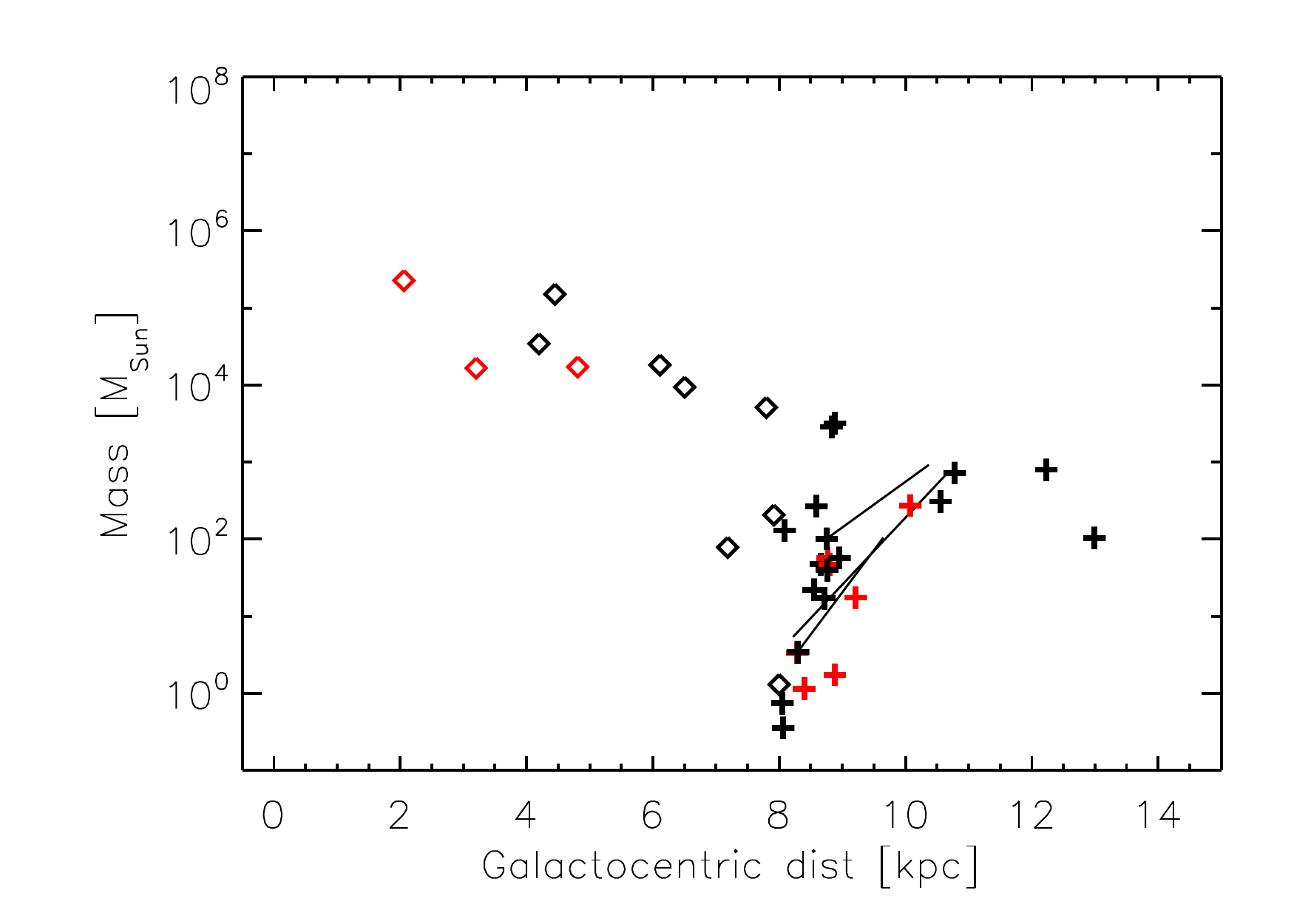}
   \includegraphics[width=0.8\columnwidth]{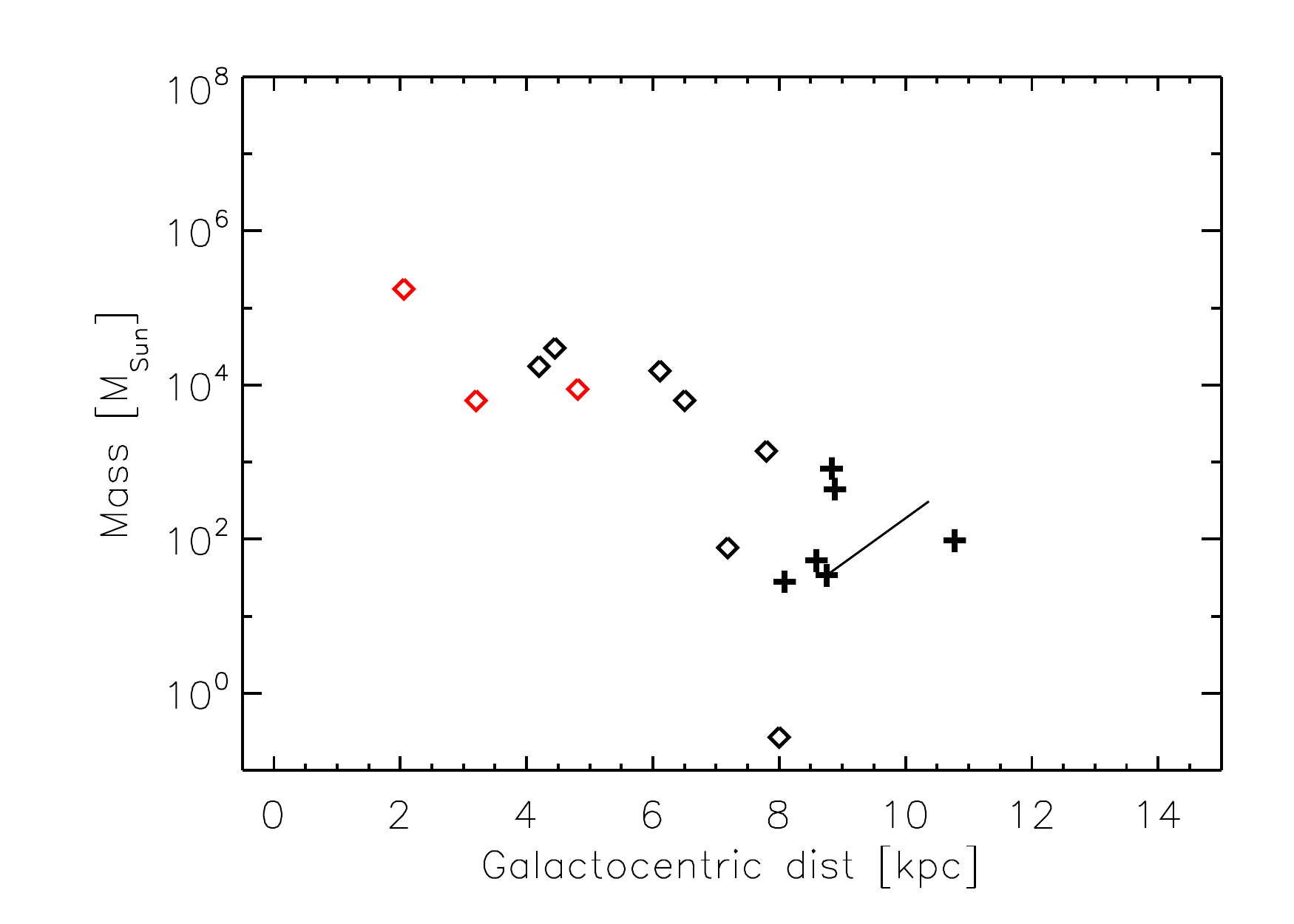}
      \caption{Mass as function of Galactocentric distance for the ECC Hi-GAL objects. Red and black symbols indicate the clumps from Category I and II, respectively. Plus signs indicate the clumps in the outer part of the Galaxy, diamonds indicate the sources in the inner Galaxy. Solid lines show the ranges of masses and Galactocentric distances for sources with large distance discrepancies. Top: calculated mass above 3$\times$10$^{21}$ cm$^{-2}$. Bottom: mass of the high column density gas, N(H$_2$) > 10$^{22}$cm$^{-2}$. 
      }
      \label{figure:mass_distance}
\end{figure}

Figure \ref{figure:mass_distance} shows the mass distribution of the sources as a function of Galactocentric distance. The top panel shows the total mass associated with a column density limit of 3$\times$10$^{21}$ cm$^{-2}$, while the bottom panel shows the mass contained within the high column density gas, N(H$_2$) > 10$^{22}$cm$^{-2}$. Solid lines show the range of masses and Galactocentric distances for the sources with large distance discrepancies. No sources with mass above 10$^4$ M$_\odot$ were detected at a Galactocentric distance larger than 6 kpc. All the densest clumps with N(H$_2$) > 10$^{22}$cm$^{-2}$ are found within 9 kpc to the Galactic Center.

\begin{figure}
   \centering
   \includegraphics[width=0.8\columnwidth]{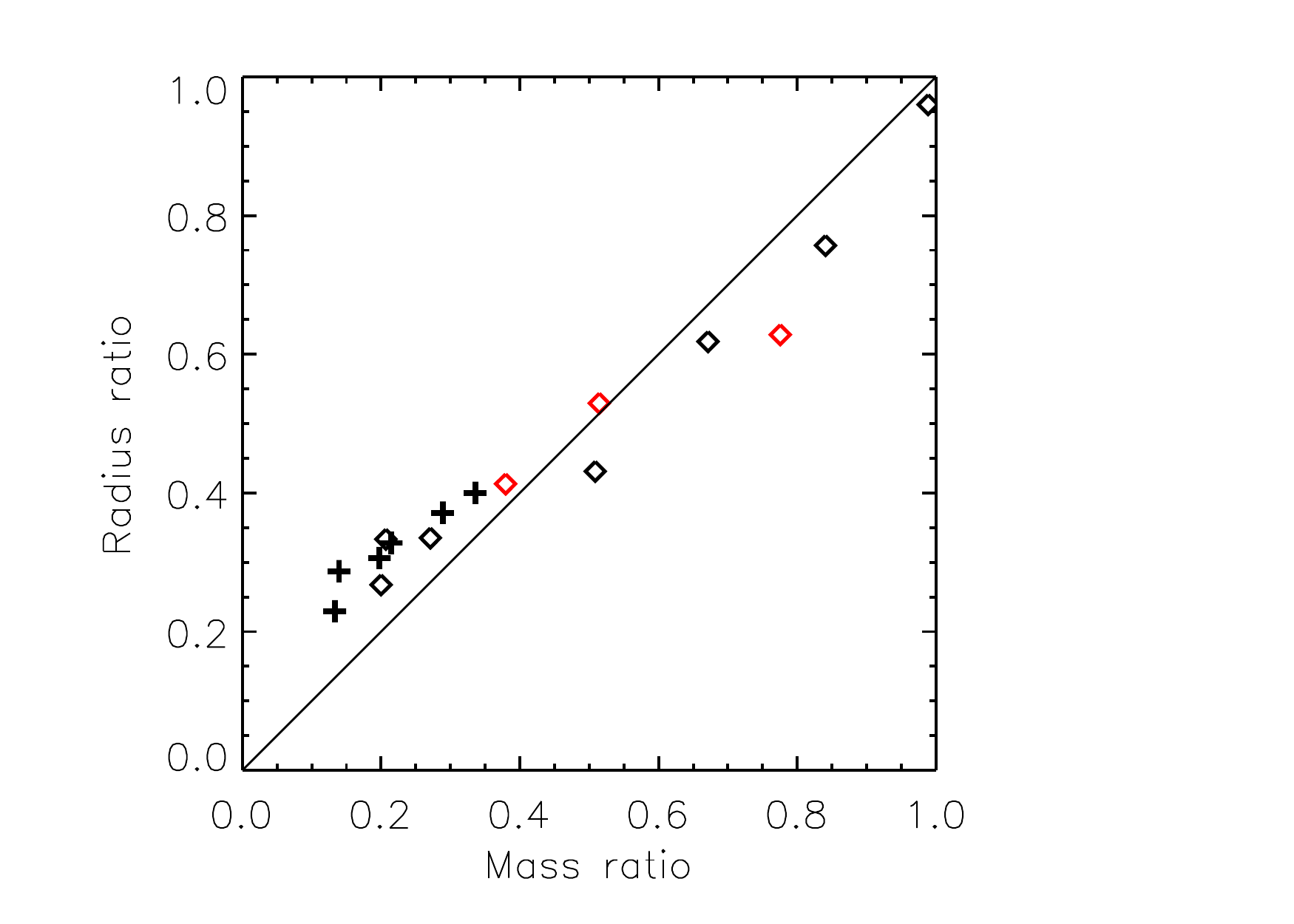}
   \includegraphics[width=0.8\columnwidth]{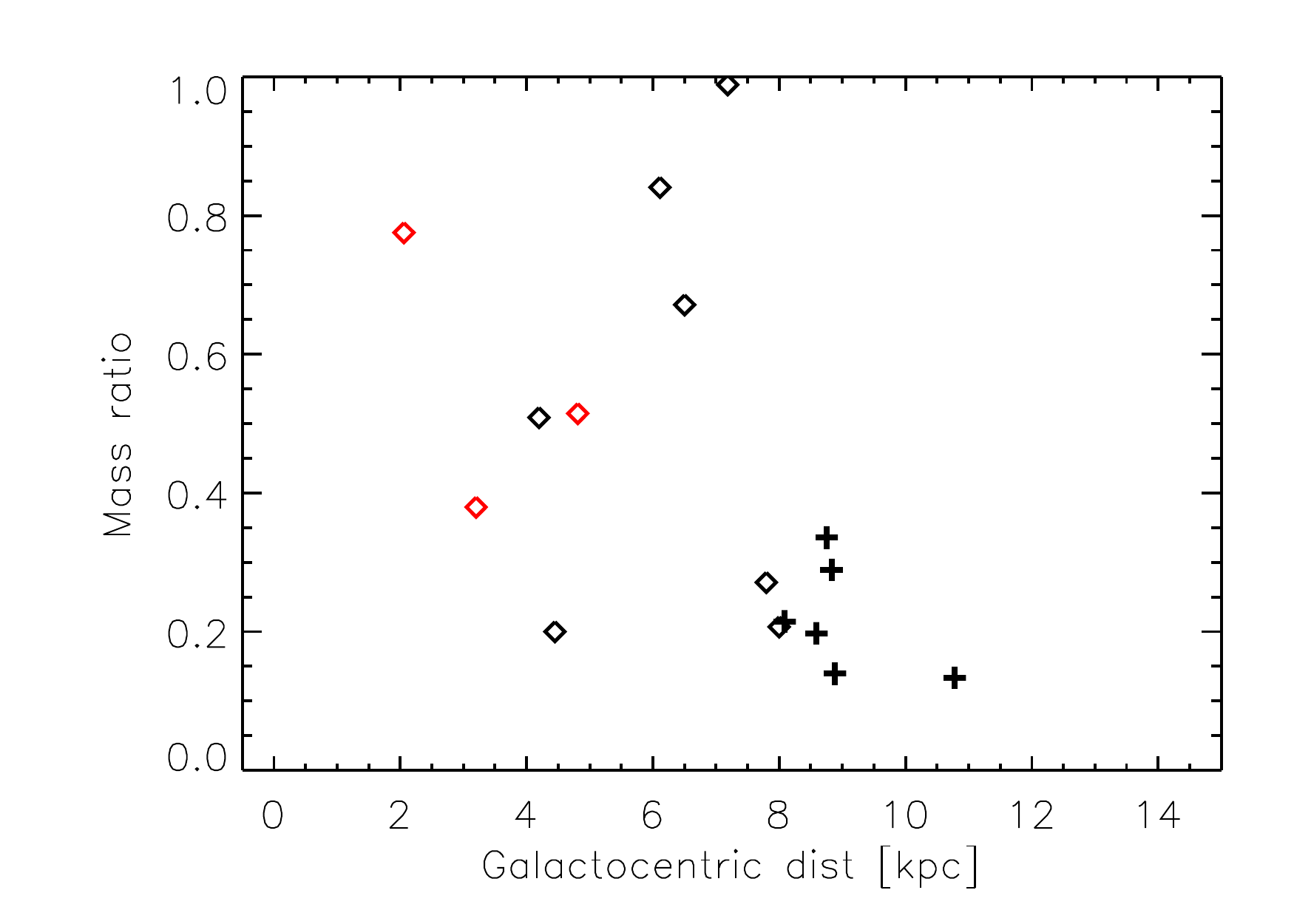}
      \caption{Top: Distribution of the size ratio [Size(10$^{22}$ cm$^{-2}$)/Size(3$\times$10$^{21}$ cm $^{-2}$)] as a function of the mass ratio [Mass(10$^{22}$ cm$^{-2}$)/Mass(3$\times$10$^{21}$ cm $^{-2}$)] for two different column density thresholds: 3$\times$10$^{21}$ cm$^{-2}$ and 10$^{22}$cm$^{-2}$. Bottom: Distribution of the mass ratio as a function of the Galactocentric distance. On both plots, red and black symbols indicate the clumps from Category I and II, respectively. Plus signs indicate the clumps in the outer part of the Galaxy, diamonds indicate the sources in the inner Galaxy.}
      \label{figure:mass_size_ratio}
\end{figure}

Figure \ref{figure:mass_size_ratio} shows the size and mass ratio for the two column density thresholds considered, 3$\times$10$^{21}$ cm$^{-2}$ and 10$^{22}$ cm$^{-2}$, and the mass ratio as a function of Galactocentric distance. The ECC clumps with a Galactocentric distance of > 9 kpc do not have high column density gas in our sample. The fraction of the mass in the densest parts of the nearby Planck clumps is lower than that of the more massive clumps closer to the Galactic Center. We also note that the densest regions tend to be more compact (i.e. smaller in spatial extent) for the clumps in the outer Galaxy than for the clumps in the inner Galaxy. This is likely a consequence of the higher overall H$_2$ column densities found in the inner Galaxy (by more than a factor of 10) with respect to the outer Galaxy.  

\subsection{Mass-size relation}
\begin{figure}
   \centering
   \includegraphics[width=0.8\columnwidth]{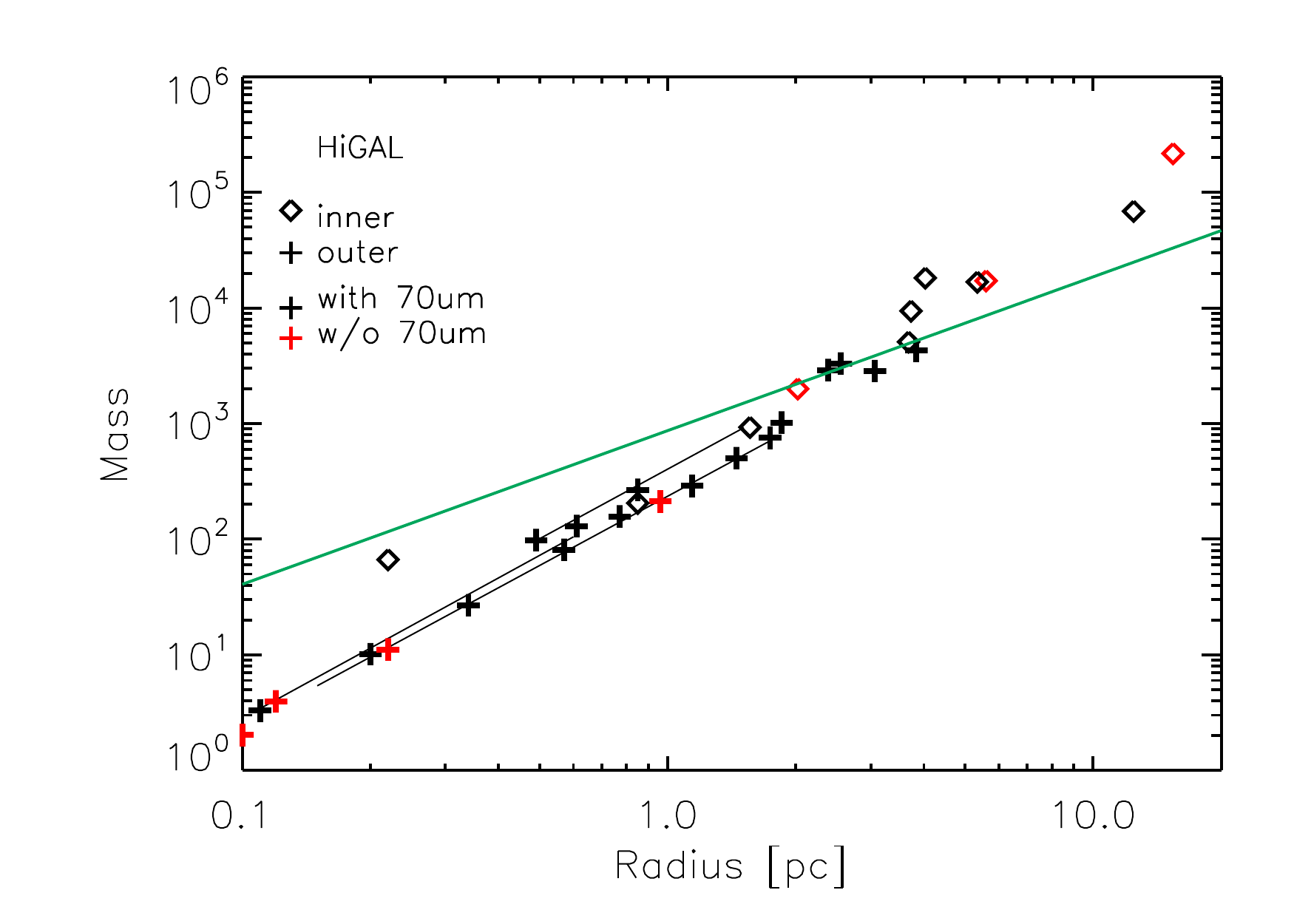}
   \includegraphics[width=0.8\columnwidth]{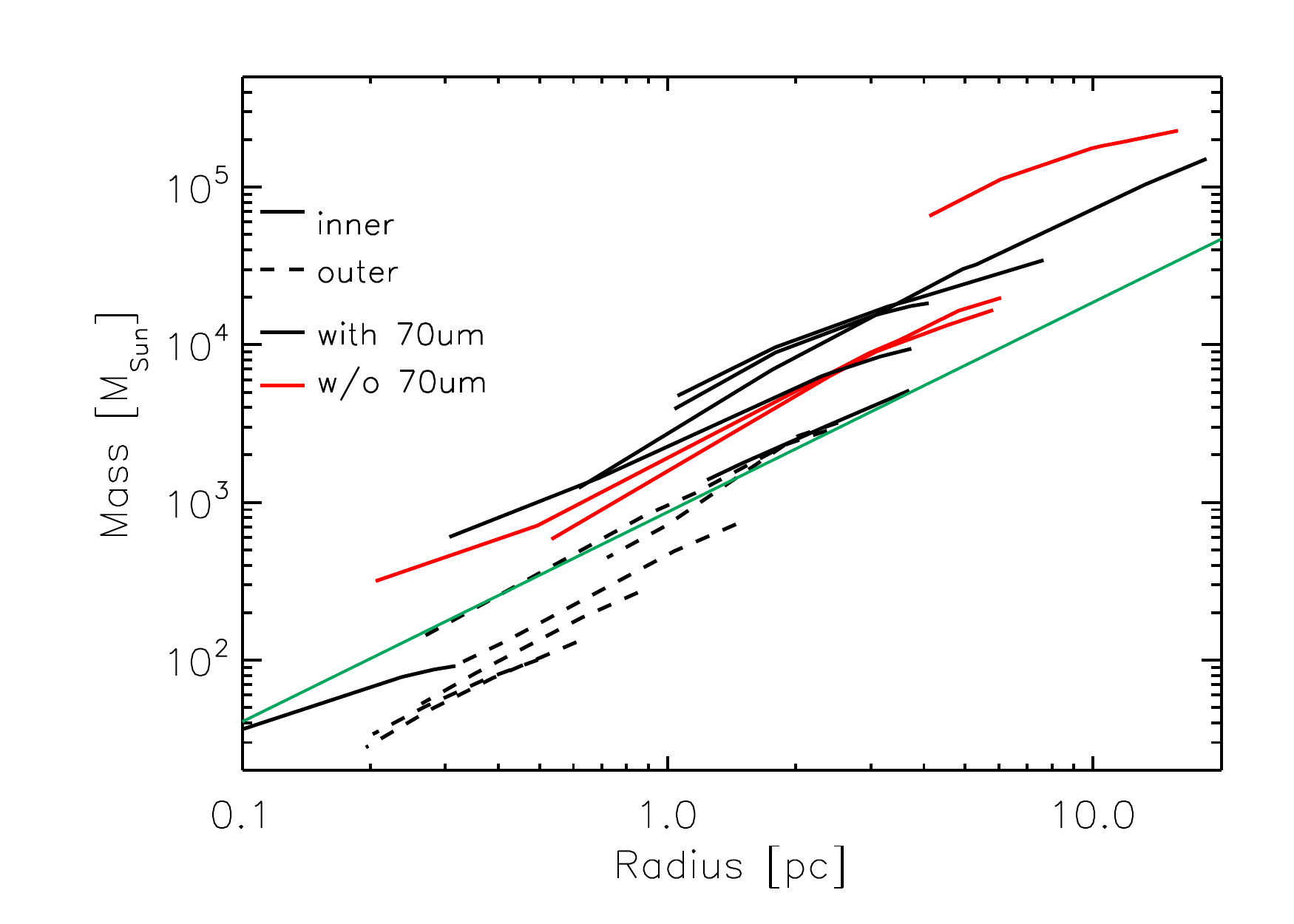}
      \caption{Top: Mass-size diagram for the ECC Hi-GAL objects, based on the Hi-GAL Herschel column density maps. Sources with available distance estimates are shown. Red and black symbols indicate the clumps from Category I and II, respectively. Plus signs indicate the clumps in the outer part of the Galaxy, diamonds indicate the sources in the inner Galaxy. Green line indicates the mass - size threshold for massive star formation: m$_{lim}$=870$M_\odot (r/pc)^{1.33}$ \citep{2010ApJ...723L...7K}. Solid lines show the ranges of masses and radii for sources with large distance discrepancies. Bottom: Mass-size diagram based on the Hi-GAL column density maps. Masses and sizes are calculated within different column density limits, see details in text. Sources with peak column density values larger than 10$^{22}$ cm$^{-2}$ are shown. Dashed lines indicate the sources in the outer Galaxy. Red lines indicate the sources without 70$\mu$m point sources.}
      \label{figure:masssize}
\end{figure}

By using solar neighbourhood clouds devoid of massive star formation, \citet{2010ApJ...723L...7K} found an empirical mass-size threshold above which massive star formation can occur. Regions forming massive stars must thus be, at a given radius, more massive than the limit mass (m$_{lim}$[r]=870M$_\odot$[r/pc]$^{1.33}$). In Figure \ref{figure:masssize}, we show the mass-size diagram for the selected Planck cold clumps based on their Herschel data. Solid lines indicate the range of masses and radii for the sources with large distance discrepancies. They all located below the empirical mass-size threshold of massive star formation. 25\% of the Planck ECC clumps in the Galactic Plane with Herschel-based derived masses have the potential to form massive stars since these masses lie above the mass limit proposed by \citet{2010ApJ...723L...7K}. These ECC clumps are the most massive and also the largest in size among our sample, which is consistent with the idea that massive stars tend to form in the largest, most massive clumps \citep{2009A&A...507..369W}.  Their distances are quite typical for the sample, between 1.1 kpc and 6.04 kpc. We note that three out of the these clumps do not have 70 $\mu$m point sources (G006.96+00.89, G009.79+00.87 and G354.81+00.35), which makes them good candidates to study the earliest stages of star-formation. Two of the sources above the mass limit are in the outer Galaxy (G224.27-00.82 and G224.47-00.65), but they contain 70$\mu$m point sources already, which confirms their ability to form stars.

In addition, we can investigate the possible evolution of the clumps' mass by calculating the ECC clump masses and sizes contained within five different H$_2$ column density thresholds: 10$^{21}$, 3$\times10^{21}$, 5$\times10^{21}$, 8$\times10^{21}$ and 10$^{22}$ cm$^{-2}$. The bottom panel of Figure \ref{figure:masssize} show the mass-size relationship. The lines represent the evolution of the calculated values within the different thresholds. As shown in Figure \ref{figure:masssize}, the lines are almost parallel to the mass-size threshold. This demonstrates that the number of sources with the potential to form high mass stars does not depend on the selected column density threshold.

We stress that two clumps are found in the Outer Galaxy which fulfill the criteria of high mass star formation. Their masses and sizes based on the Planck catalogue information also fulfill this criteria. Three clumps without 70 $\mu$m emission are found above the high mass star formation threshold in the inner Galaxy. These clumps represent good candidates to study the earliest phases of massive star and star cluster formation.

\section{Summary}
In this paper we investigate the basic physical properties of 48 Planck ECC clumps in the Galactic Plane using Herschel Hi-GAL data. Starless ($\sim$22\%) and star-forming ($\sim$78\%) clumps were identified in the inner and outer part of the Galaxy. We have calculated their masses and sizes based on background subtracted {\it Herschel} images and characterized their evolutionary stage based on the presence, or absence, of point-like 70 $\mu$m emission (an excellent probe of massive star formation).
No clear difference was found in the mass and temperature of starless and star forming clumps based on our data. Planck clumps located in the Galactic center show higher column densities and average dust temperatures than those located in the outer Galaxy. 
We identified 5 particularly interesting objects in the Galactic Plane, which are good candidates for higher resolution continuum and molecular line studies. Three of them (G006.96+00.89, G009.79+00.87 and G354.81+00.35) are located in the inner Galaxy, fulfill the criteria of massive star and star cluster formation and show no 70 $\mu$m emission. These objects are thus excellent templates where to study the earliest phases of massive star and star cluster formation. Two objects with 70 $\mu$m  emission (G224.27-00.82 and G224.47-00.65) were found in the outer part of the Galaxy which also fulfill the massive star formation criteria: with their follow-up study, the properties of massive star-forming clumps can be compared directly in the inner and outer part of the Galaxy.

\begin{acknowledgements}
This work has made use of the APLpy plotting package (https://aplpy.github.io/) and the agpy code package (https://pypi.python.org/pypi/agpy/0.1.4). K.W. acknowledges the support from Deutsche Forschungsgemeinschaft (DFG) grant WA3628-1/1 through priority programme 1573 ('Physics of the Interstellar Medium''). I.J.-S. acknowledges the financial support received from the People Programme (Marie Curie Actions) of the European Union's Seventh Framework  Programme (FP7/2007–2013) under REA grant agreement PIIF-GA-2011-301538 and from the STFC through an Ernest Rutherford Fellowship (proposal number ST/L004801/1). L.V.T. and S.Z. acknowledges the support by the OTKA grants NN-111016 and K101393.
\end{acknowledgements}

\bibliographystyle{aa}
\bibliography{references}

\Online
\begin{appendix}

\section{Herschel images of the clumps}
\begin{figure}[!h]
   \centering
\includegraphics[width=\columnwidth]{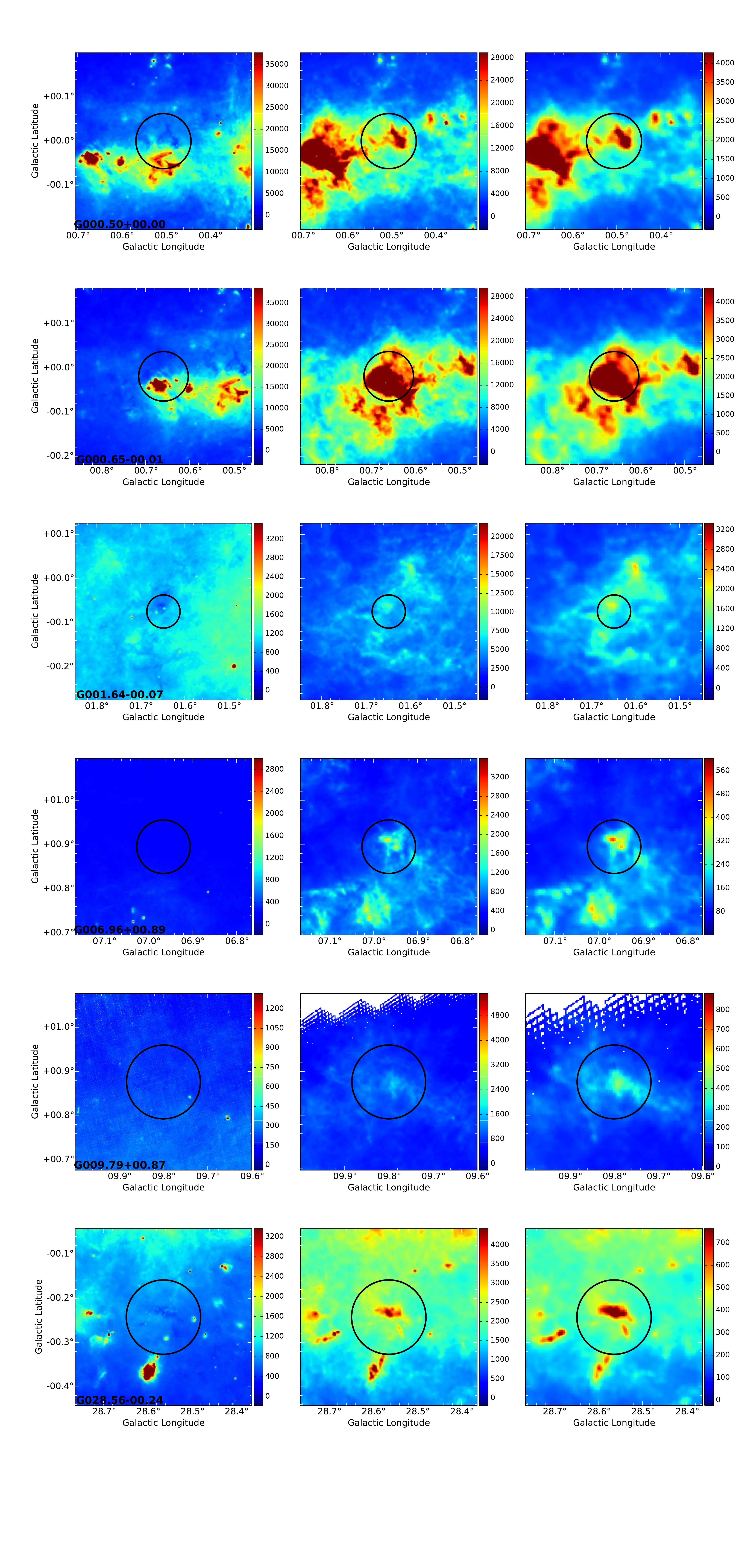}
\caption{Herschel 70, 250 and 500 $\mu$m images of the 48 selected Planck cold clumps given in units of MJy/sr with a resolution of 5$\arcsec$, 18$\arcsec$ and 36$\arcsec$. Black circles are centered at the position of the Planck clumps' positions and the circle sizes correspond to the derived major axis of the clumps.}
      \label{fig:herschel_all_maps}
\end{figure}

\begin{figure}[!h]
   \centering
\includegraphics[width=\columnwidth]{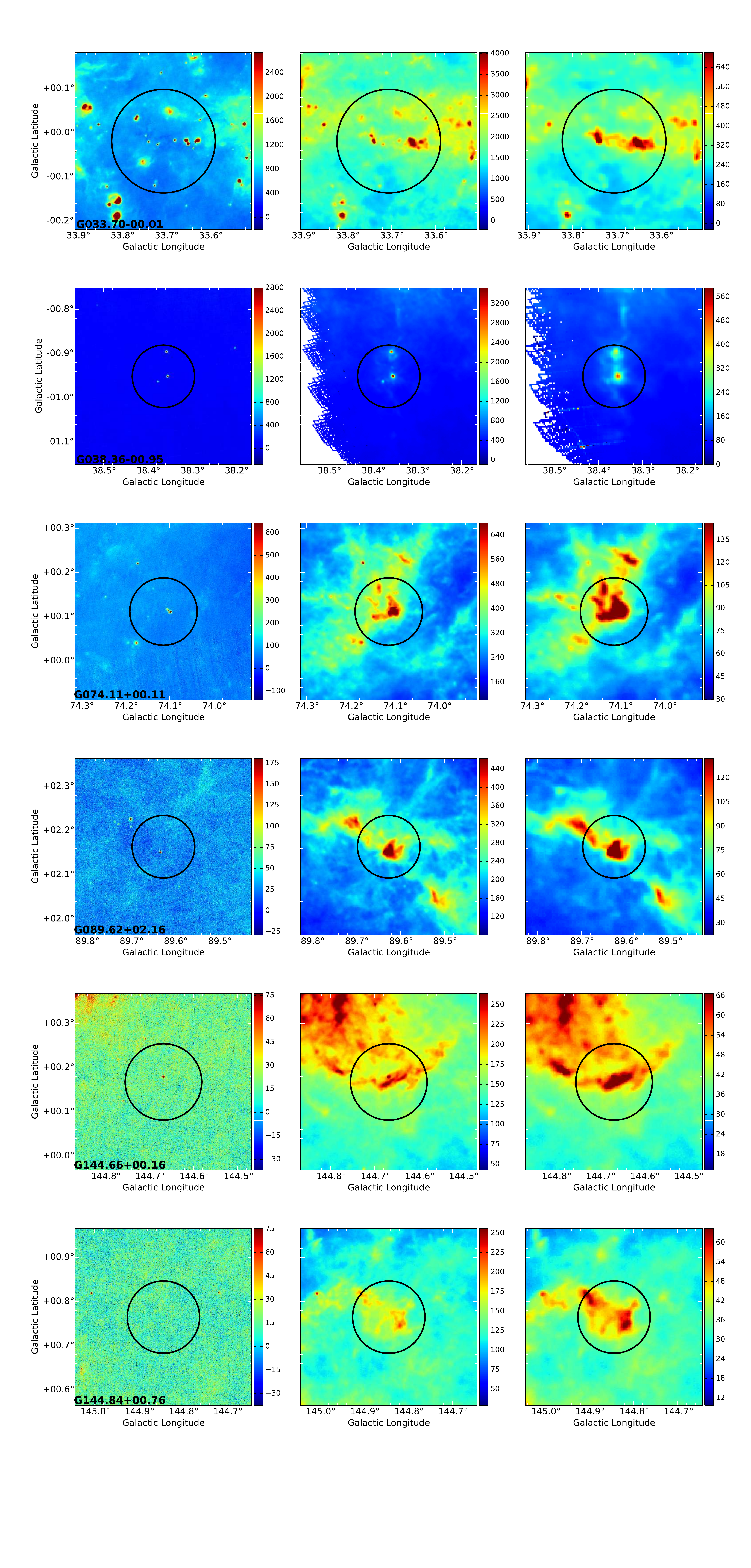}
\caption{Hershel 70, 250 and 500 $\mu$m images in MJy/sr with a resolution of 5$\arcsec$, 18$\arcsec$ and 36$\arcsec$. Black circle shows the Planck clump's position and size based on the major axis.}
      \label{fig:herschel_all_maps_2}
\end{figure}

\begin{figure}[!h]
   \centering
\includegraphics[width=\columnwidth]{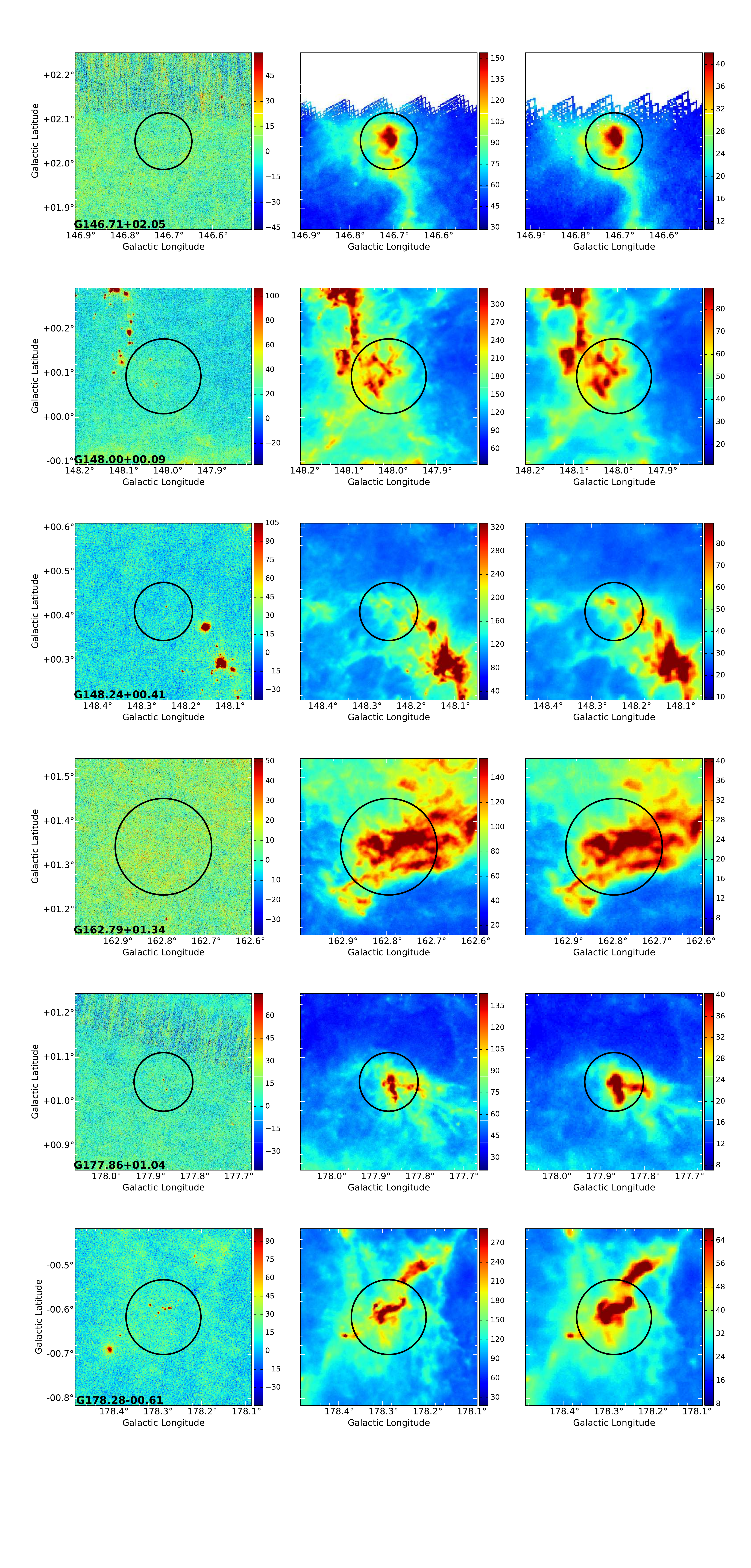}
\caption{Hershel 70, 250 and 500 $\mu$m images in MJy/sr with a resolution of 5$\arcsec$, 18$\arcsec$ and 36$\arcsec$. Black circle shows the Planck clump's position and size based on the major axis.}
      \label{fig:herschel_all_maps_3}
\end{figure}

\begin{figure}[!h]
   \centering
\includegraphics[width=\columnwidth]{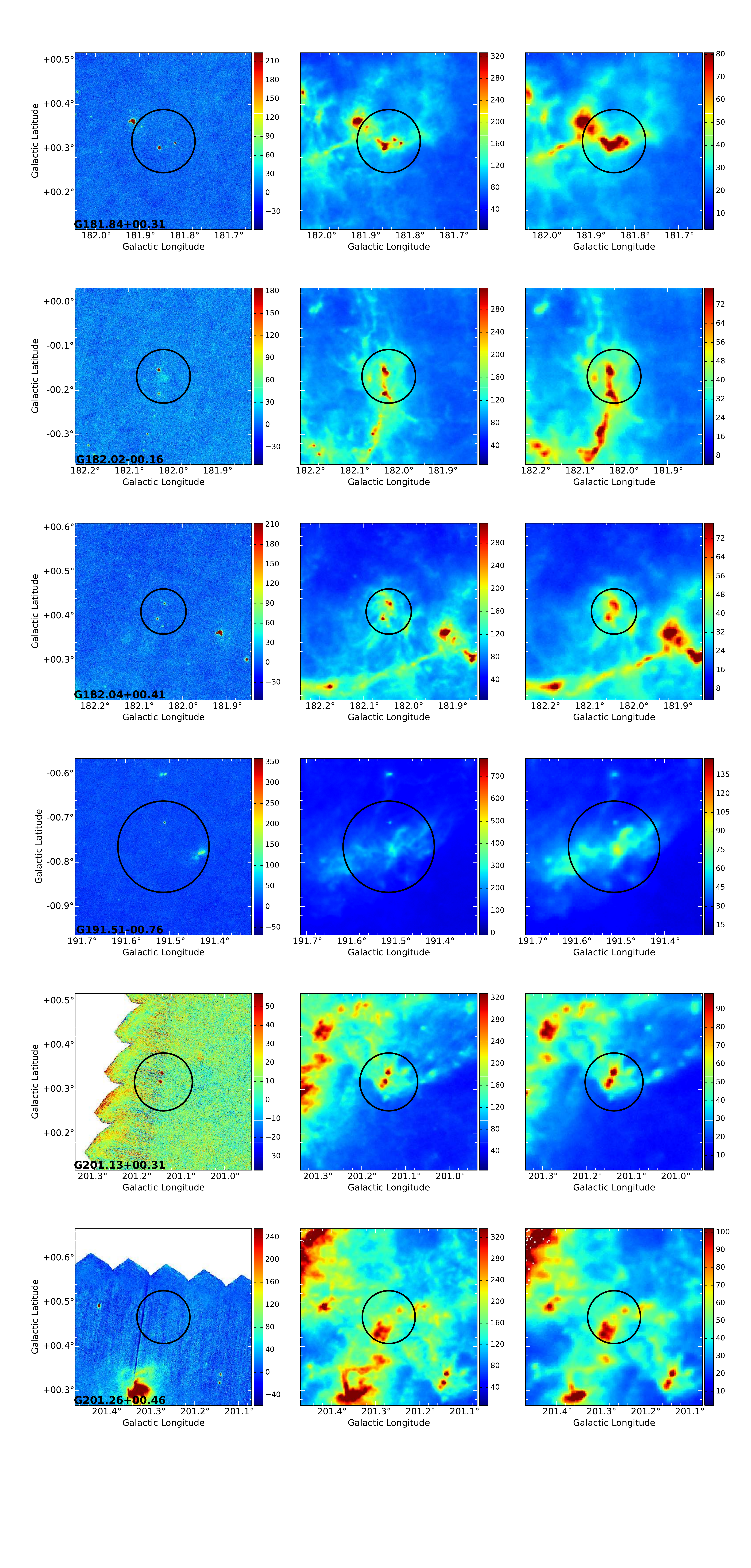}
\caption{Hershel 70, 250 and 500 $\mu$m images in MJy/sr with a resolution of 5$\arcsec$, 18$\arcsec$ and 36$\arcsec$. Black circle shows the Planck clump's position and size based on the major axis.}
      \label{fig:herschel_all_maps_4}
\end{figure}

\begin{figure}[!h]
   \centering
\includegraphics[width=\columnwidth]{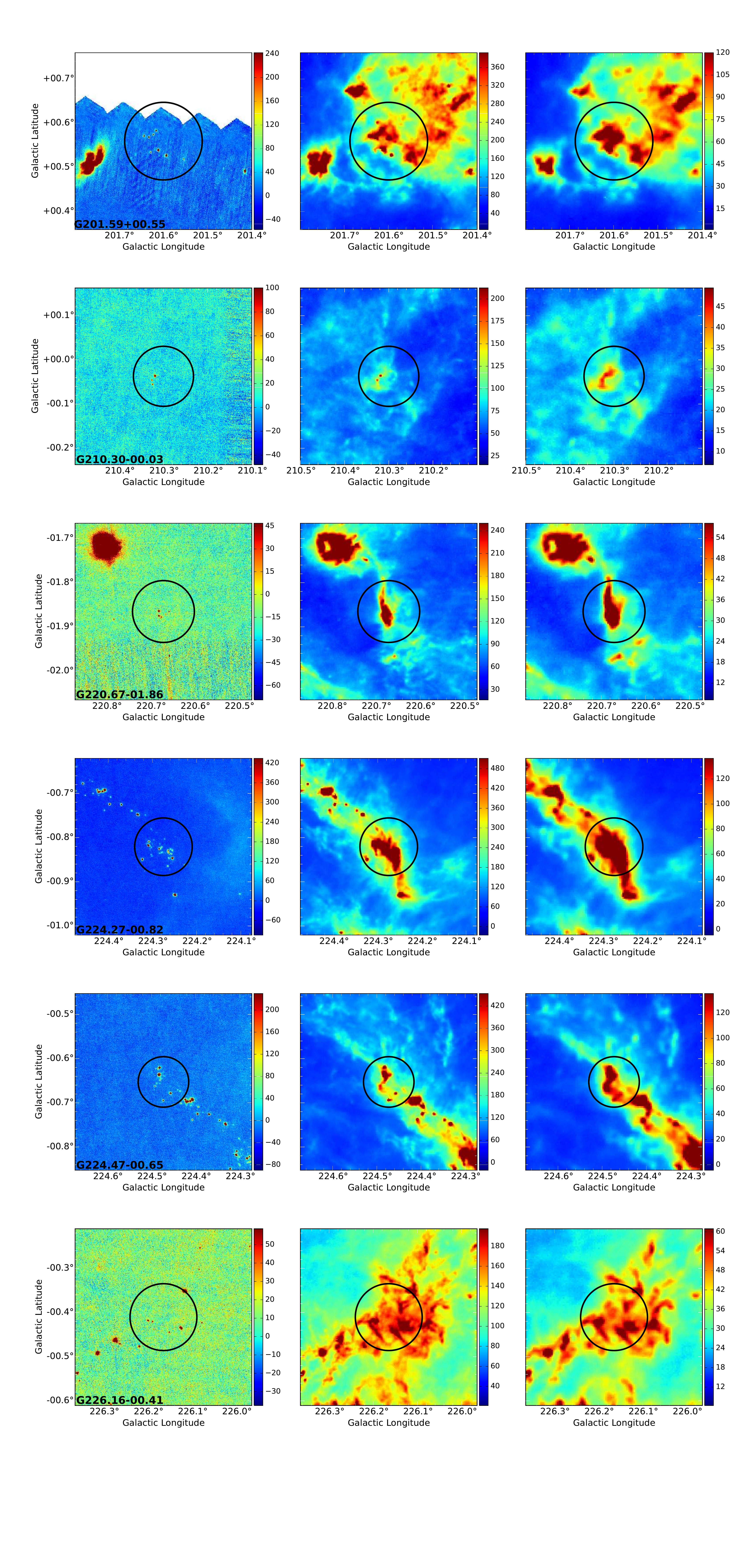}
\caption{Hershel 70, 250 and 500 $\mu$m images in MJy/sr with a resolution of 5$\arcsec$, 18$\arcsec$ and 36$\arcsec$. Black circle shows the Planck clump's position and size based on the major axis.}
      \label{fig:herschel_all_maps_5}
\end{figure}

\begin{figure}[!h]
   \centering
\includegraphics[width=\columnwidth]{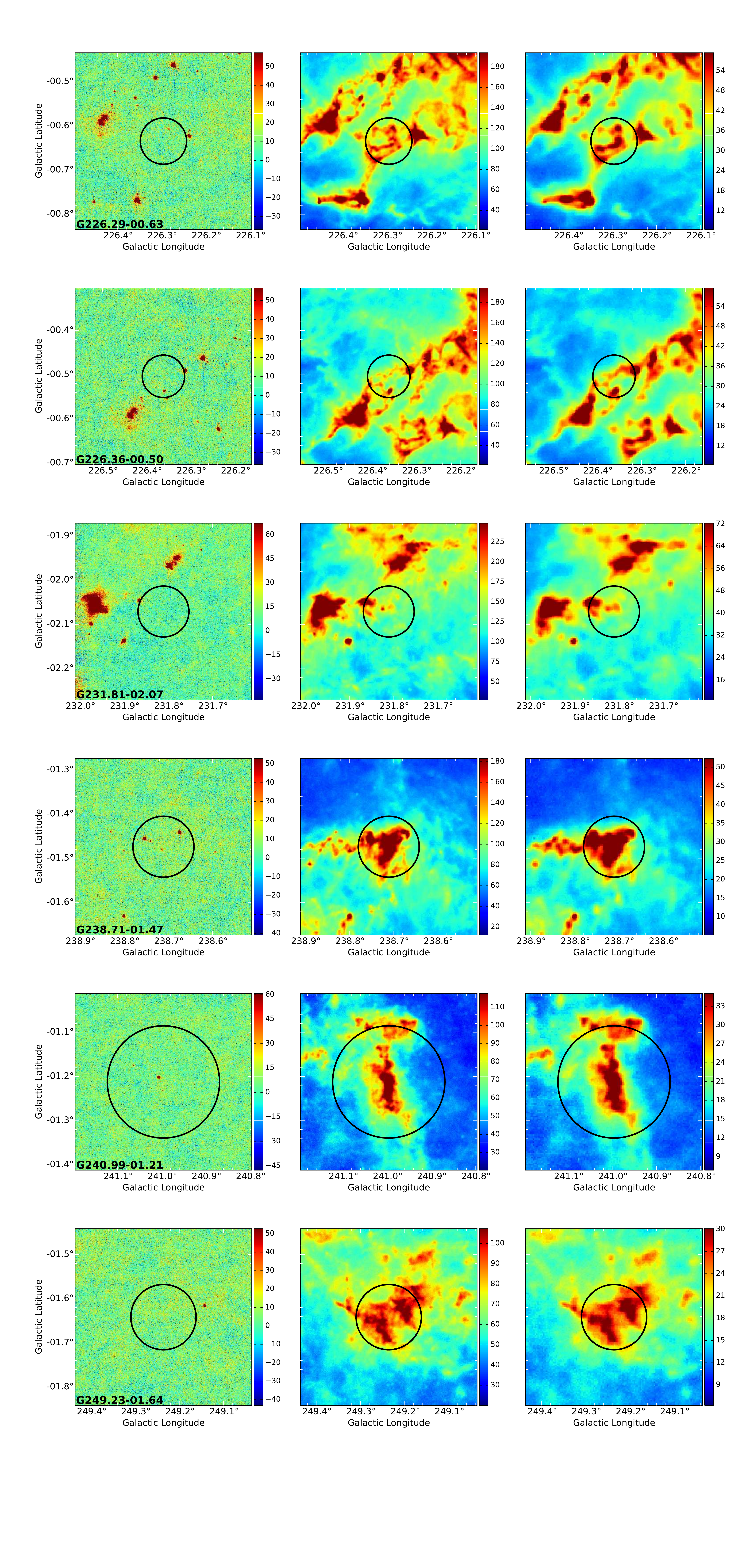}
\caption{Hershel 70, 250 and 500 $\mu$m images in MJy/sr with a resolution of 5$\arcsec$, 18$\arcsec$ and 36$\arcsec$. Black circle shows the Planck clump's position and size based on the major axis.}
      \label{fig:herschel_all_maps_6}
\end{figure}

\begin{figure}[!h]
   \centering
\includegraphics[width=\columnwidth]{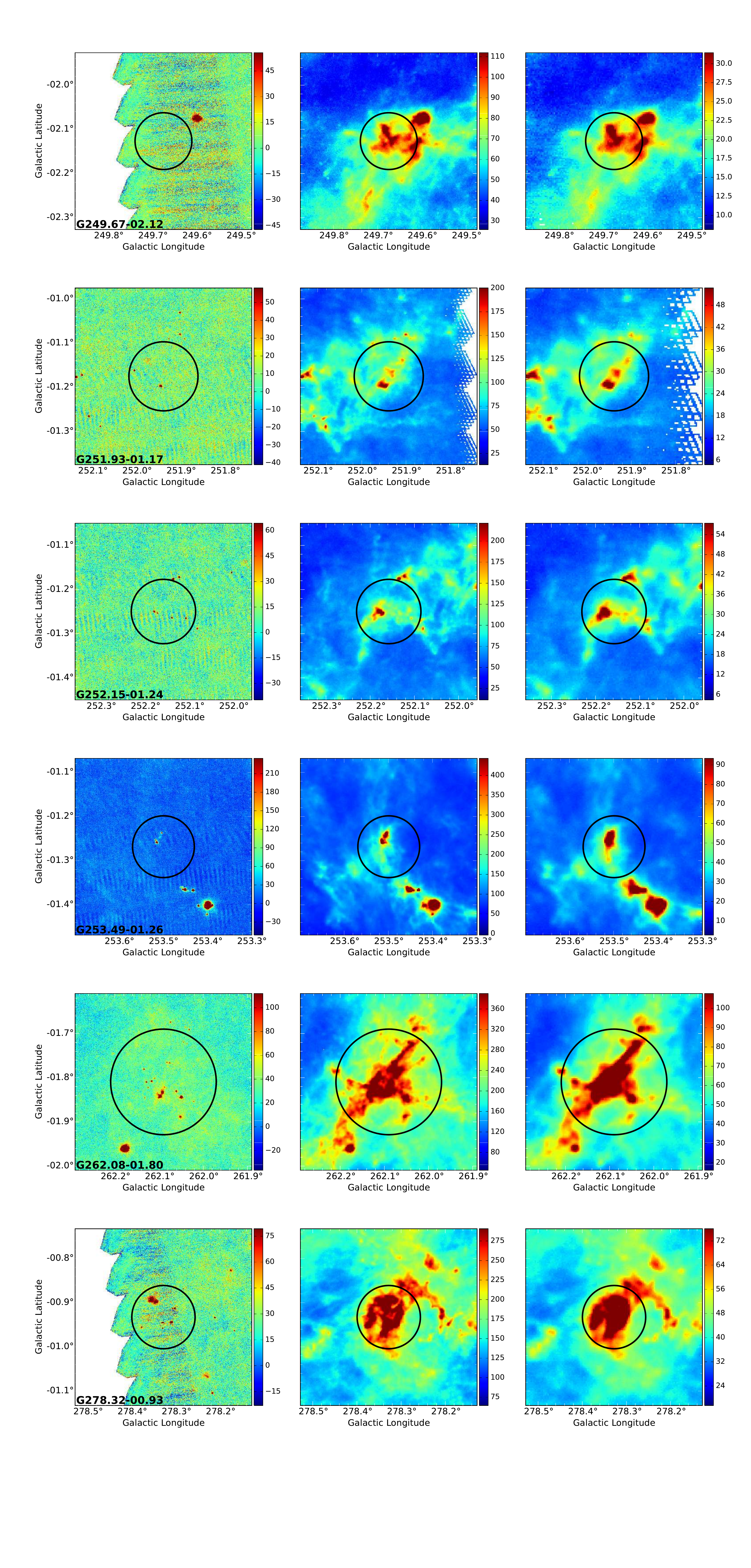}
\caption{Hershel 70, 250 and 500 $\mu$m images in MJy/sr with a resolution of 5$\arcsec$, 18$\arcsec$ and 36$\arcsec$. Black circle shows the Planck clump's position and size based on the major axis.}
      \label{fig:herschel_all_maps_7}
\end{figure}

\begin{figure}[!h]
   \centering
\includegraphics[width=\columnwidth]{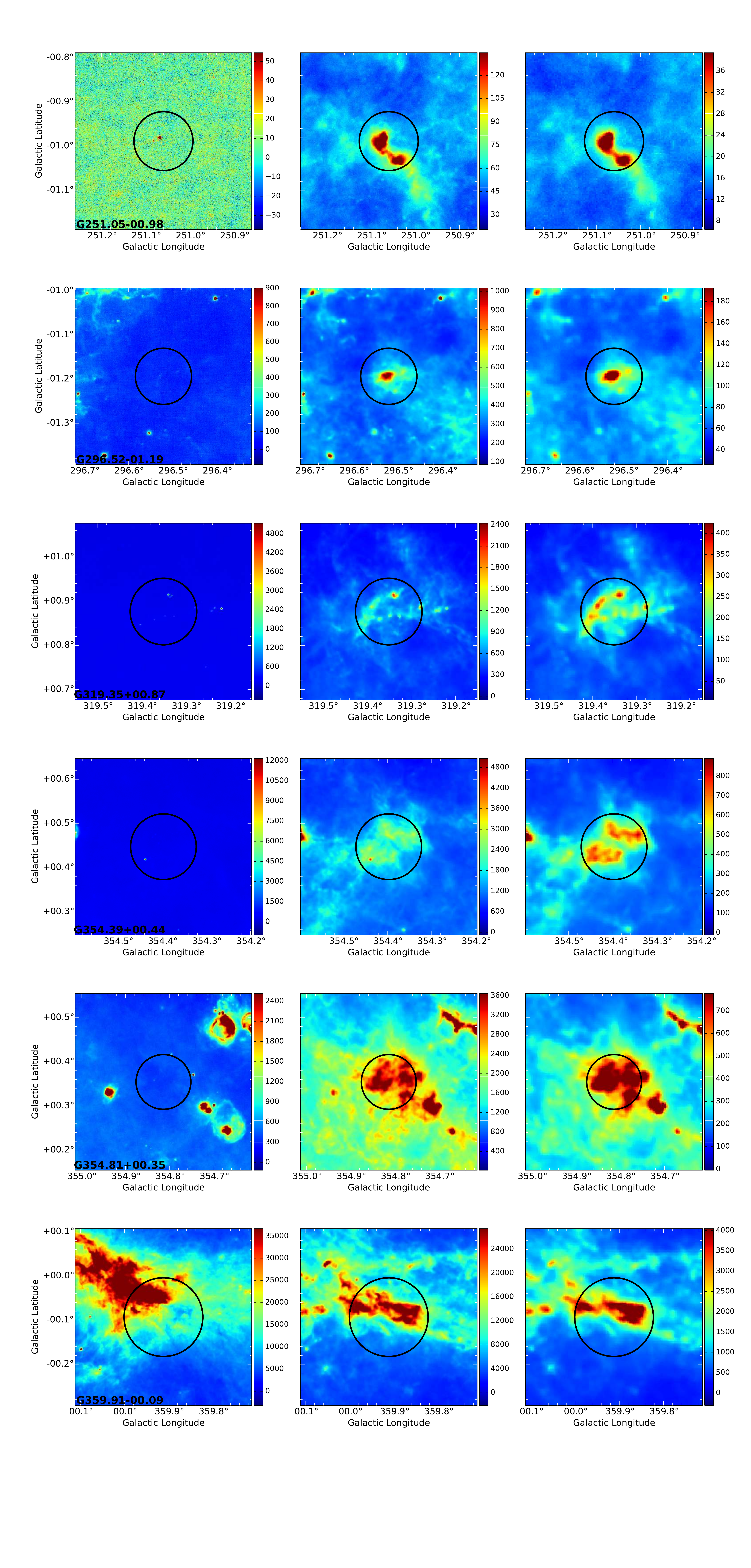}
\caption{Hershel 70, 250 and 500 $\mu$m images in MJy/sr with a resolution of 5$\arcsec$, 18$\arcsec$ and 36$\arcsec$. Black circle shows the Planck clump's position and size based on the major axis.}
      \label{fig:herschel_all_maps_8}
\end{figure}

In Figure \ref{fig:herschel_all_maps}-\ref{fig:herschel_all_maps_8}, we show the Herschel 70, 250 and 500 $\mu$m images of the investigated Planck ECC objects (intensity scale in units of MJy/sr) with a resolution of 5$\arcsec$, 18$\arcsec$ and 36$\arcsec$. Black circles are centered at the Planck clumps' positions and their sizes correspond to the derived major axis of the clumps. 

\section{Available distance estimates for the clumps}
Table \ref{table:distances} shows all the available estimated distances for the ECC objects, see the detailed description in Section 2.3.
\onllongtab{
\begin{landscape}
\begin{longtable}{l | r r | r r | r r | r r | r r | r}
\caption{Estimated distances for the ECC clumps \label{table:distances}}\\
\hline
\hline
	&	\multicolumn{2}{c|}{Wu et al. 2012}			&	\multicolumn{2}{c|}{MALT90}			&	\multicolumn{2}{c|}{APEX}			&	\multicolumn{2}{c|}{CfA}			&	\multicolumn{2}{c|}{IRDC}			&	PGCC 	\\
Name	&	V$_{LSR}$ $^{13}$CO &	D	&	V$_{LSR}$ N$_2$H$^+$	&	D	&	V$_{LSR}$ $^{13}$CO &	D	&	V$_{LSR}$ $^{12}$CO &	D	&	name	&	D	&		\\
	&	[km/s]	&	[kpc]	&	[km/s]	&	[kpc]	&	[km/s]	&	[kpc]	&	[km/s]	&	[kpc]	&		&	[kpc]	&	[kpc]	\\
\hline
\endfirsthead
\caption{Continued.} \\
\hline
	&	\multicolumn{2}{c|}{Wu et al. 2012}			&	\multicolumn{2}{c|}{MALT90}			&	\multicolumn{2}{c|}{APEX}			&	\multicolumn{2}{c|}{CfA}			&	\multicolumn{2}{c|}{IRDC}			&	PGCC	\\
Name	&	V$_{LSR}$ $^{13}$CO &	D	&	V$_{LSR}$ N$_2$H$^+$	&	D	&	V$_{LSR}$ $^{13}$CO &	D	&	V$_{LSR}$ $^{12}$CO &	D	&	name	&	D	&		\\
	&	[km/s]	&	[kpc]	&	[km/s]	&	[kpc]	&	[km/s]	&	[kpc]	&	[km/s]	&	[kpc]	&		&	[kpc]	&	[kpc]	\\
\hline
\endhead
\hline
\endfoot
\hline
\endlastfoot
G000.50+00.00	&		&		&	42.24	&	7.93 / 8.70	&		&		&		&		&		&		&	7.4 (4)	\\
G000.65-00.01	&		&		&	85.76	&	8.09 / 8.57	&		&		&		&		&		&		&		\\
G001.64-00.07	&		&		&	54.25	&	7.39 / 9.19	&		&		&		&		&	G001.62-00.08	&		&		\\
G006.96+00.89	&	41.53	&	4.89 / 11.39	&		&		&		&		&	43	&	4.97 /11.33	&	G006.95+00.88	&		&		\\
G009.79+00.87	&		&		&		&		&		&		&	27.2	&	3.27 / 12.75	&		&		&		\\
G028.56-00.24	&	86.66	&	4.74 / 9.77	&		&		&		&		&		&		&	G028.53-00.25	&	5.7	&	5.3 (4)	\\
G033.70-00.01	&	105.89	&	6.94	&		&		&		&		&		&		&	G033.69-00.01	&	7.1	&	7.1 (1)	\\
G038.36-00.95	&	16.69	&	1.08 / 11.74	&		&		&		&		&	15.8	&	1.02 / 11.8	&	G038.35-00.90	&		&	1.2 (1)	\\
G074.11+00.11	&	-1.42	&	4.4 / 4.6	&		&		&		&		&		&		&		&		&	3.45 (5)	\\
G089.62+02.16	&	-0.34	&	0.06	&		&		&		&		&	-0.8	&	0.05	&		&		&		\\
G144.66+00.16	&	-7.68	&	0.35	&		&		&		&		&	-8.6	&	0.42	&		&		&		\\
G144.84+00.76	&	-30.55	&	2.42	&		&		&		&		&	-31.2	&	2.49	&		&		&		\\
G146.71+02.05	&	2.52	&		&		&		&		&		&	0.8	&		&		&		&	-1.03 (5)	\\
G148.00+00.09	&	-6.40	&	0.26	&		&		&		&		&	-33.8	&	3	&		&		&	0.77 (5)	\\
G148.24+00.41	&	-34.56	&	3.12	&		&		&		&		&	-34.1	&	3.07	&		&		&		\\
G162.79+01.34	&	1.36	&		&		&		&		&		&	0	&		&		&		&	-1.26 (5)	\\
G177.86+01.04	&	-18.36	&		&		&		&		&		&	-18.9	&	1.66	&		&		&	0.3 (2)	\\
G178.28-00.61	&		&		&		&		&		&		&	0.04	&		&		&		&		\\
G181.84+00.31	&	3.30	&		&		&		&		&		&		&		&		&		&		\\
G182.02-00.16	&	3.96	&		&		&		&		&		&		&		&		&		&		\\
G182.04+00.41	&	2.91	&		&		&		&		&		&		&		&		&		&		\\
G191.51-00.76	&	0.13	&	0.41	&		&		&		&		&		&		&		&		&	-0.82 (5)	\\
G201.13+00.31	&	5.08	&	0.76	&		&		&		&		&	7.4	&	1.04	&		&		&		\\
G201.26+00.46	&	5.60	&	0.82	&		&		&		&		&	7.6	&	1.06	&		&		&		\\
G201.59+00.55	&	4.42	&	0.66	&		&		&		&		&	7.5	&	1.03	&		&		&	0.63 (2)	\\
G210.30-00.03	&	36.47	&	4.24	&		&		&		&		&		&		&		&		&	5.44 (5)	\\
G220.67-01.86	&	12.11	&	0.96	&		&		&		&		&	33.4	&	2.9	&		&		&		\\
G224.27-00.82	&	14.39	&	1.11	&		&		&		&		&		&		&		&		&		\\
G224.47-00.65	&	15.34	&	1.18	&		&		&		&		&		&		&		&		&		\\
G226.16-00.41	&	15.95	&	1.22	&		&		&		&		&		&		&		&		&	1.05 (5)	\\
G226.29-00.63	&	17.19	&	1.31	&		&		&		&		&		&		&		&		&		\\
G226.36-00.50	&	14.96	&	1.14	&		&		&		&		&		&		&		&		&	0.77 (5)	\\
G231.81-02.07	&	42.43	&	3.46	&		&		&		&		&		&		&		&		&	3.53 (5)	\\
G240.99-01.21	&		&		&		&		&		&		&		&		&		&		&	6.15 (5)	\\
G249.23-01.64	&		&		&		&		&		&		&	11.84	&	1.08	&		&		&		\\
G249.67-02.12	&		&		&		&		&		&		&	10.7	&	0.99	&		&		&		\\
G251.93-01.17	&		&		&		&		&		&		&	2.6	&	0.19	&		&		&		\\
G252.15-01.24	&		&		&		&		&		&		&	2.4	&	0.16	&		&		&		\\
G262.08-01.80	&		&		&		&		&		&		&	3.7	&	0.5	&		&		&		\\
G278.32-00.93	&		&		&		&		&		&		&	-28.6	&	1.21	&		&		&		\\
G319.35+00.87	&		&		&		&		&	-34.22	&	2.9 / 9.95	&	-40.5	&	2.7 / 10.15	&		&		&	2.18 (5)	\\
G354.39+00.44	&		&		&		&		&	72.39	&	4.43 / 11.76	&		&		&	G354.42+00.46	&		&	1.9 (4)	\\
G354.81+00.35	&		&		&	100.79	&	6.04 / 10.41	&		&		&		&		&	G354.84+00.31	&		&		\\
G359.91-00.09	&		&		&	10.58	&	7.93 / 8.57	&		&		&		&		&	G359.87-00.09	&	8	&		\\
\end{longtable}
\tablefoot{The distance determination available in the PGCC catalogue have been obtained by using different methods. The final number in parenthesis corresponds to: 1) kinematic distance estimates, 2) optical extinction based on SDSS DR7, 3) NIR extinction towards IRDCs, 4) NIR extinction. In case of the NIR extinction methods, negative values indicate the upper limits.}
\end{landscape}
}

\section{Background subtraction}
For the precise determination of clump masses and sizes, we have to subtract the background and foreground emission in the observed images. Different methods are available to perform this task: i) the Fourier Transform (FT) method, which separates the large-scale and small-scale structures \citep{wangke15}; ii) to use a reference, relatively emission-free region to derive a constant background value \citep{2012A&A...541A..12J}; and iii) to perform the Gaussian fitting of the Galactic background along the Galactic latitude \citep{2011A&A...535A.128B}. 

For the background subtraction, we used 2$^\circ\times$2$^\circ$ maps around the central positions of the Planck cold clumps. These maps are much larger than the structures of interest. We smoothed all intensity images to the same resolution of 5' and used the smoothed map as background. In contrast to methods ii) and iii), this technique allows spatial variations in the background values across the maps.
After background subtraction, the derived column densities are 15-20\% lower and the dust temperatures are 1-3 K lower. We compared our method with the FT method introduced by \cite{wangke15}. Our column density values are in good agreement with the FT method, and the difference is on average of only $\sim$10\%. We caution, however, that this may be partly due to the specific properties of the power spectrum of the Hi-GAL images in large tiles, as both works have found; the same may not apply for other images with a different power spectrum characterization.
Calculated dust temperature and column density maps based on the background subtracted maps are shown in Appendix \ref{maps}.

\section{T$_{dust}$ and N(H$_2$) maps \label{maps}}
\begin{figure}[!ht]
    \centering
\includegraphics[width=\columnwidth]{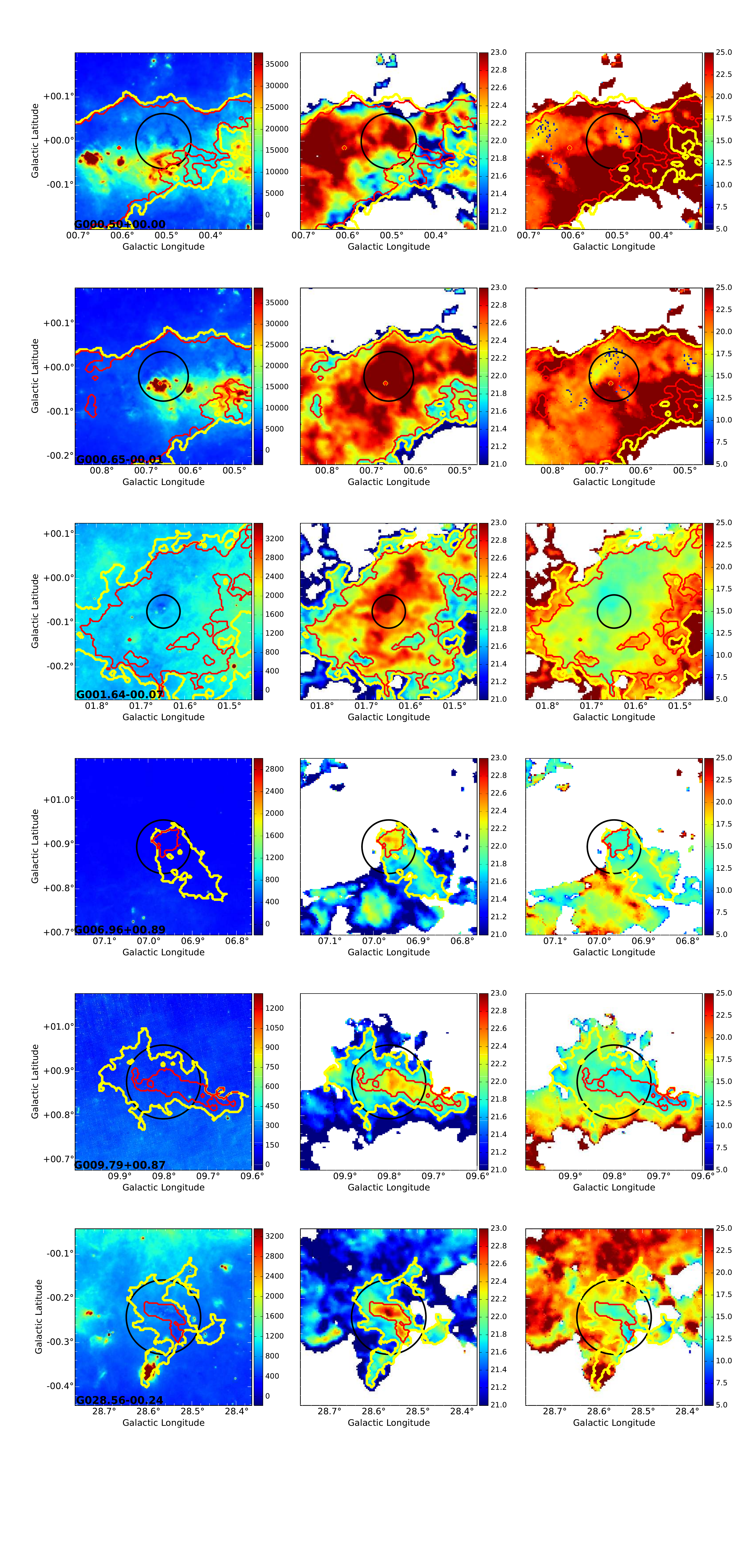}
    \caption{70 $\mu$m images ([MJy/sr], left) and calculated column density ([cm$^{-2}$], middle) in logarithmic scale and dust temperature ([K], right) maps with a resolution of 36$\arcsec$. Yellow and red contour levels are at 3$\times10^{21}$ cm$^{-2}$ and 10$^{22}$ cm$^{-2}$, respectively. Black circles are placed at the Planck clumps' positions and their sizes correspond to the clump's major axis reported in the ECC catalogue.}
    \label{fig:all_calculated_maps}
\end{figure}

\begin{figure}[!ht]
    \centering
\includegraphics[width=\columnwidth]{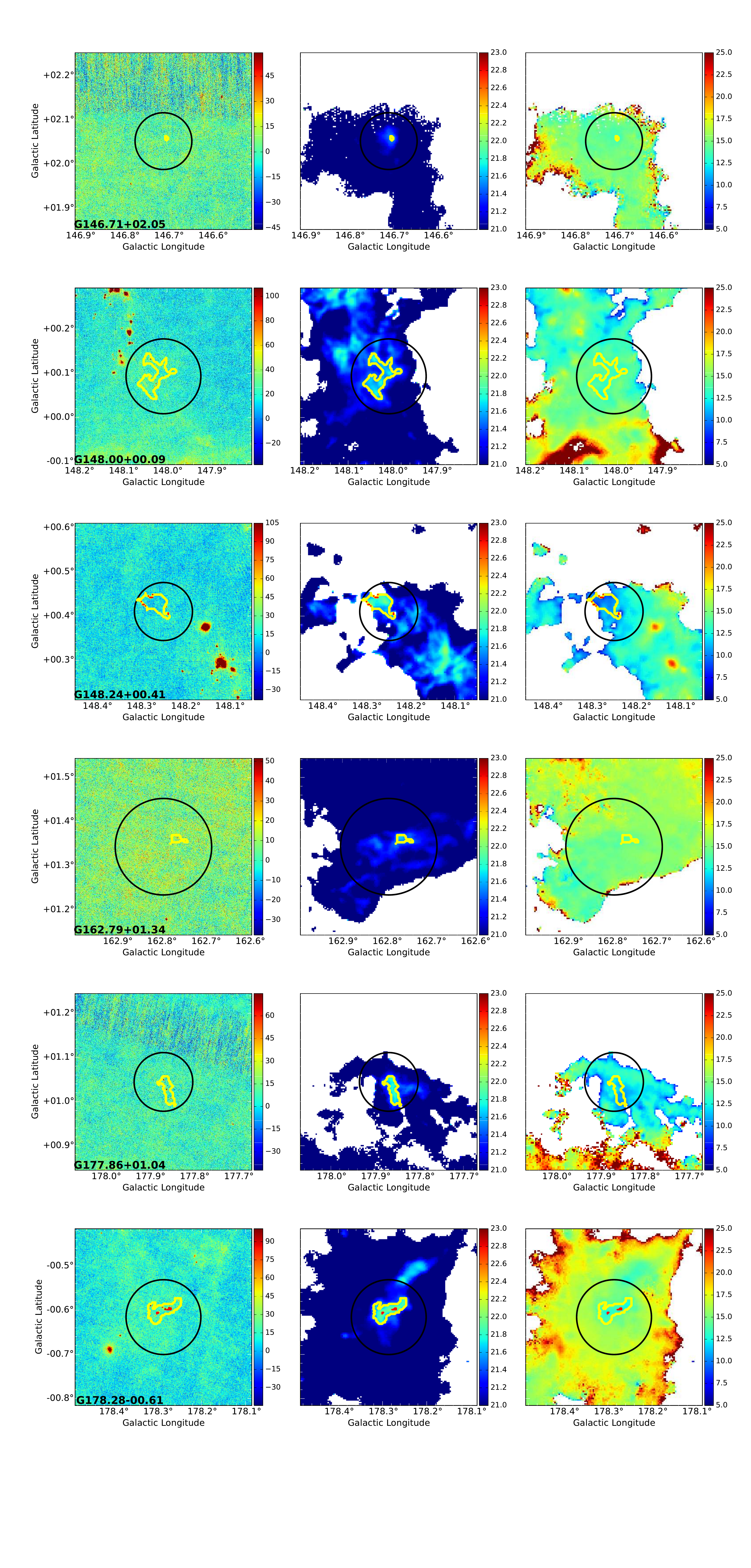}
    \caption{70 $\mu$m images ([MJy/sr], left) and calculated column density ([cm$^{-2}$], middle) in logarithmic scale and dust temperature ([K], right) maps with a resolution of 36$\arcsec$. Yellow and red contour levels are at 3$\times10^{21}$ cm$^{-2}$ and 10$^{22}$ cm$^{-2}$, respectively. Black circle shows the Planck clump's position and size based on the given major axis from the ECC catalogue.}
    \label{fig:all_calculated_maps_2}
\end{figure}

\begin{figure}[!ht]
    \centering\
\includegraphics[width=\columnwidth]{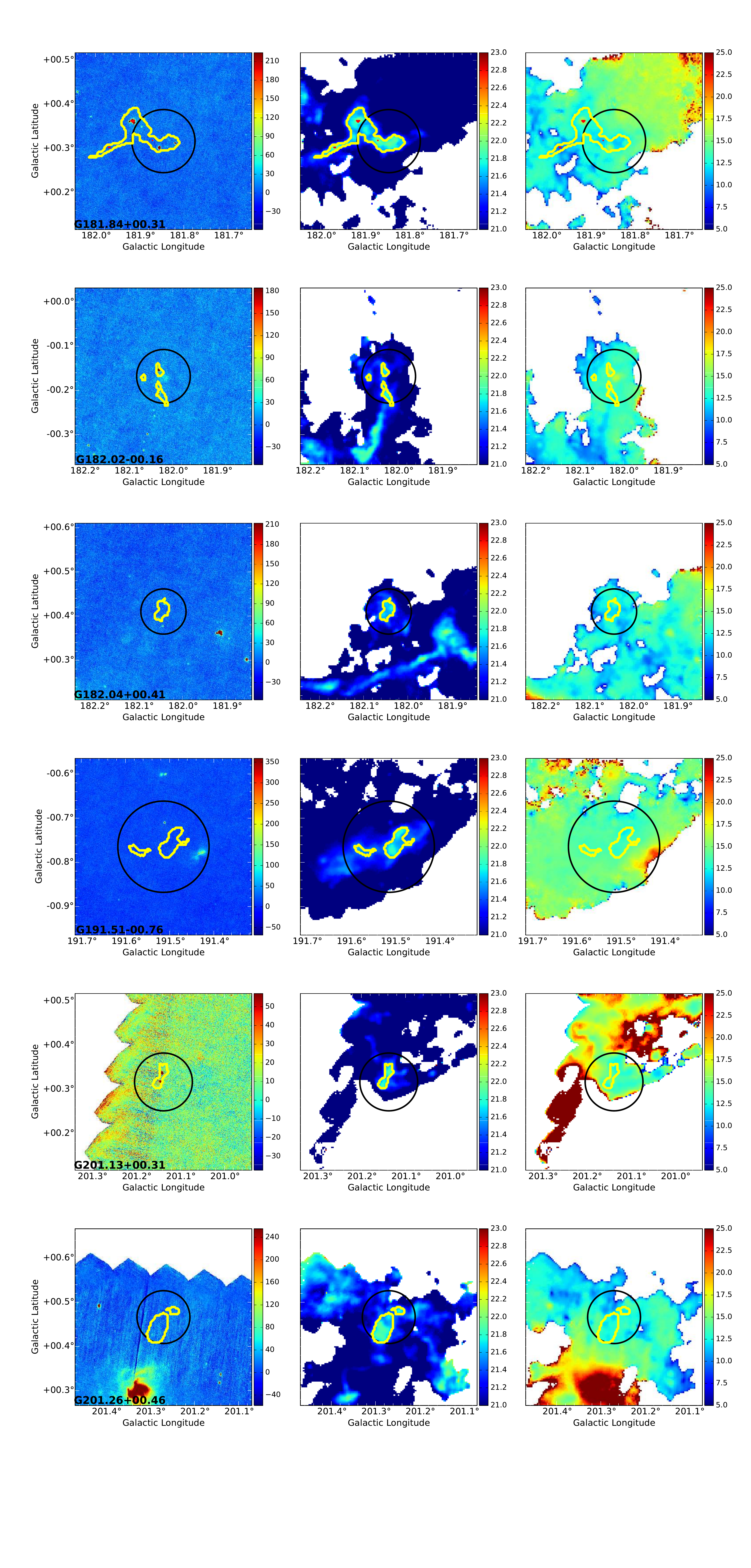}
    \caption{70 $\mu$m images ([MJy/sr], left) and calculated column density ([cm$^{-2}$], middle) in logarithmic scale and dust temperature ([K], right) maps with a resolution of 36$\arcsec$. Yellow and red contour levels are at 3$\times10^{21}$ cm$^{-2}$ and 10$^{22}$ cm$^{-2}$, respectively. Black circle shows the Planck clump's position and size based on the given major axis from the ECC catalogue.}
    \label{fig:all_calculated_maps_3}
\end{figure}

\begin{figure}[!ht]
    \centering
\includegraphics[width=\columnwidth]{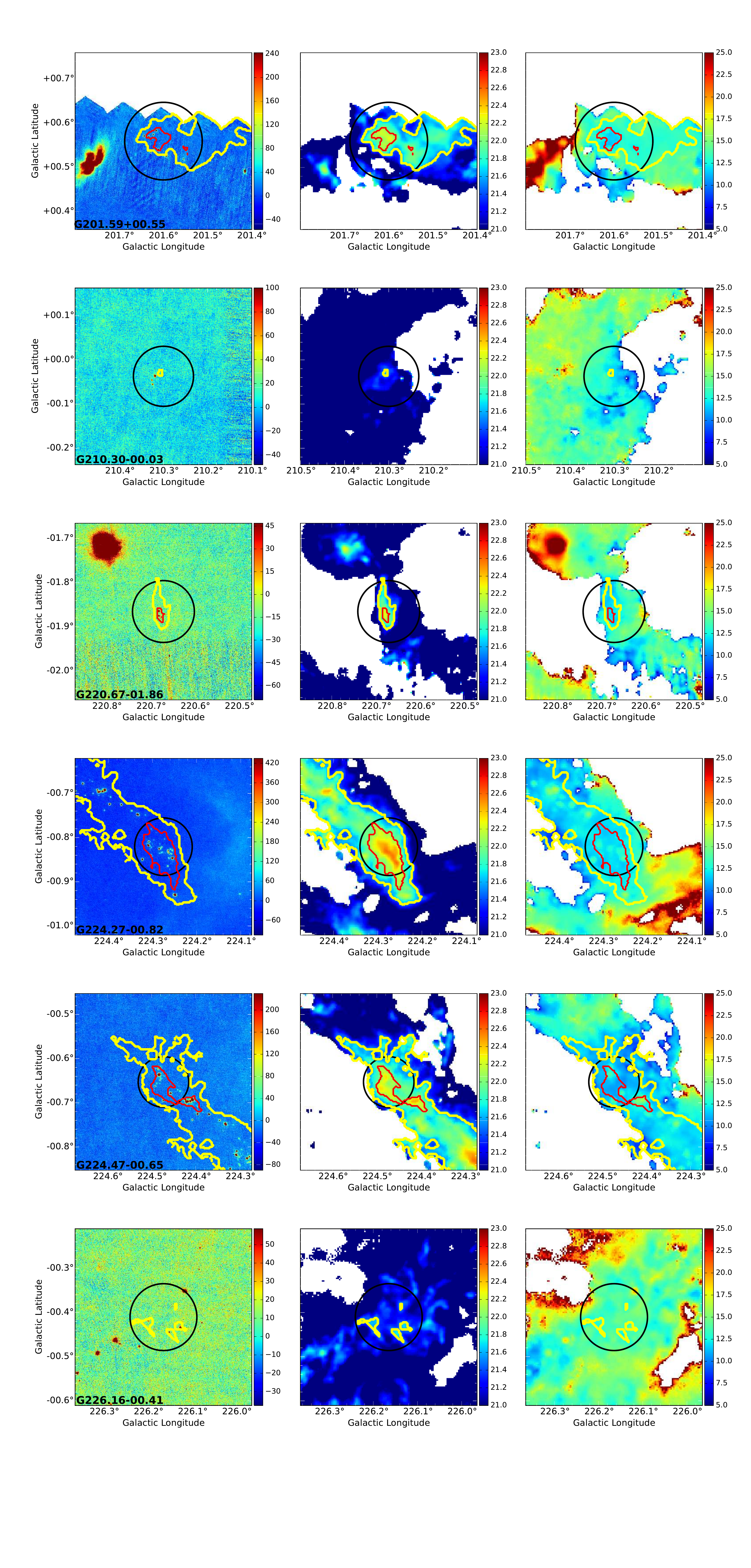}
    \caption{70 $\mu$m images ([MJy/sr], left) and calculated column density ([cm$^{-2}$], middle) in logarithmic scale and dust temperature ([K], right) maps with a resolution of 36$\arcsec$. Yellow and red contour levels are at 3$\times10^{21}$ cm$^{-2}$ and 10$^{22}$ cm$^{-2}$, respectively. Black circle shows the Planck clump's position and size based on the given major axis from the ECC catalogue.}
    \label{fig:all_calculated_maps_4}
\end{figure}

\begin{figure}[!ht]
    \centering
\includegraphics[width=\columnwidth]{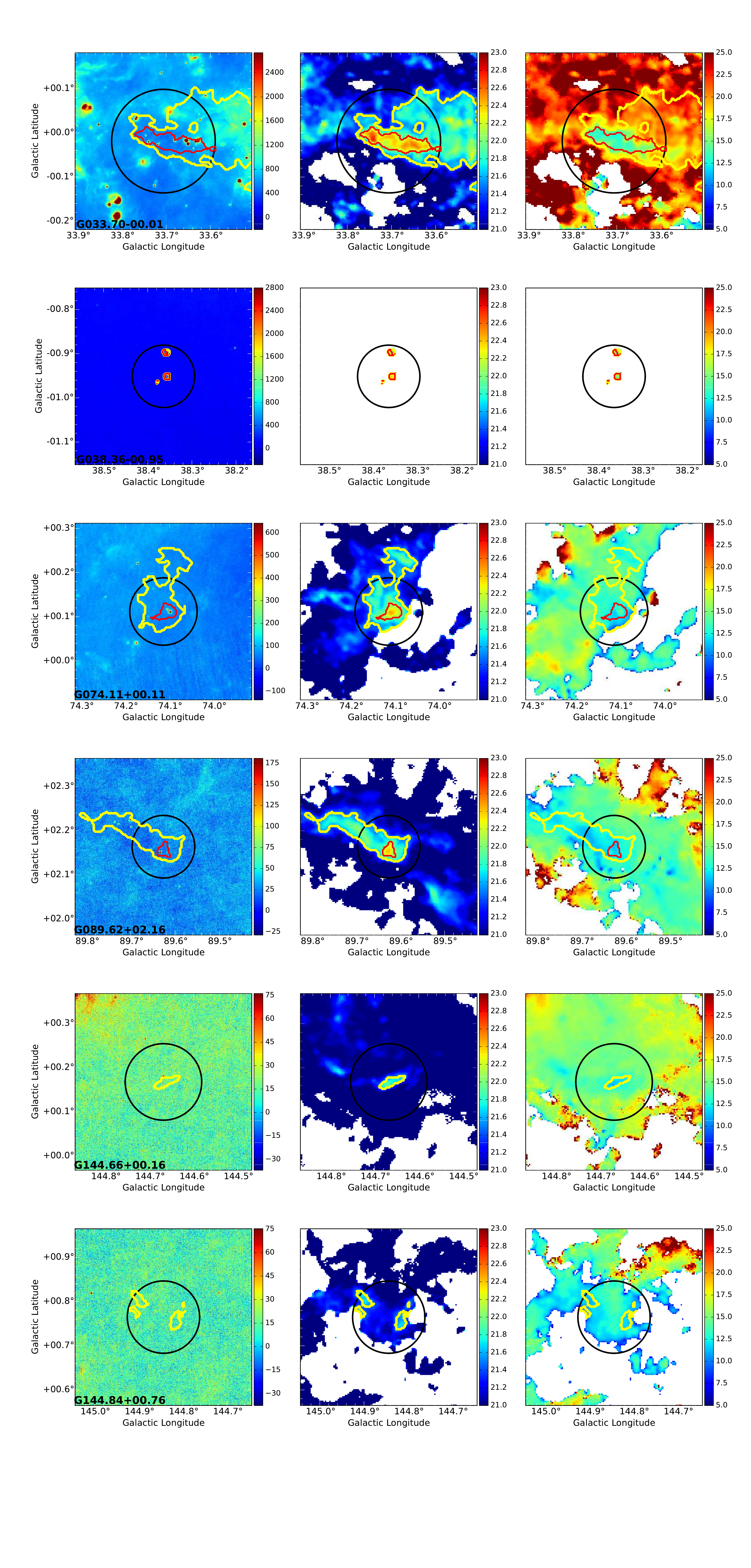}
    \caption{70 $\mu$m images ([MJy/sr], left) and calculated column density ([cm$^{-2}$], middle) in logarithmic scale and dust temperature ([K], right) maps with a resolution of 36$\arcsec$. Yellow and red contour levels are at 3$\times10^{21}$ cm$^{-2}$ and 10$^{22}$ cm$^{-2}$, respectively. Black circle shows the Planck clump's position and size based on the given major axis from the ECC catalogue.}
    \label{fig:all_calculated_maps_5}
\end{figure}

\begin{figure}[!ht]
    \centering
\includegraphics[width=\columnwidth]{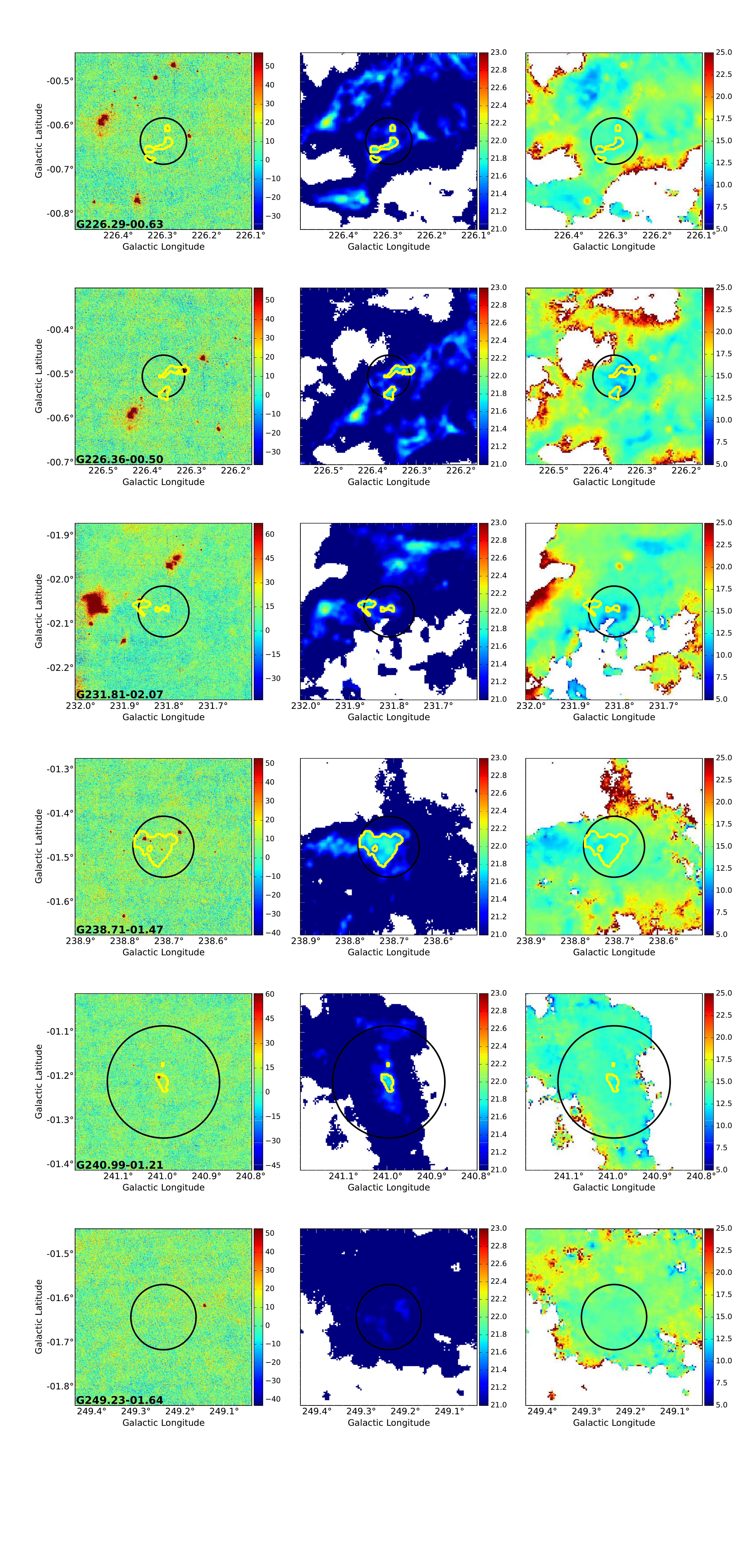}
    \caption{70 $\mu$m images ([MJy/sr], left) and calculated column density ([cm$^{-2}$], middle) in logarithmic scale and dust temperature ([K], right) maps with a resolution of 36$\arcsec$. Yellow and red contour levels are at 3$\times10^{21}$ cm$^{-2}$ and 10$^{22}$ cm$^{-2}$, respectively. Black circle shows the Planck clump's position and size based on the given major axis from the ECC catalogue.}
    \label{fig:all_calculated_maps_6}
\end{figure}

\begin{figure}[!ht]
    \centering
\includegraphics[width=\columnwidth]{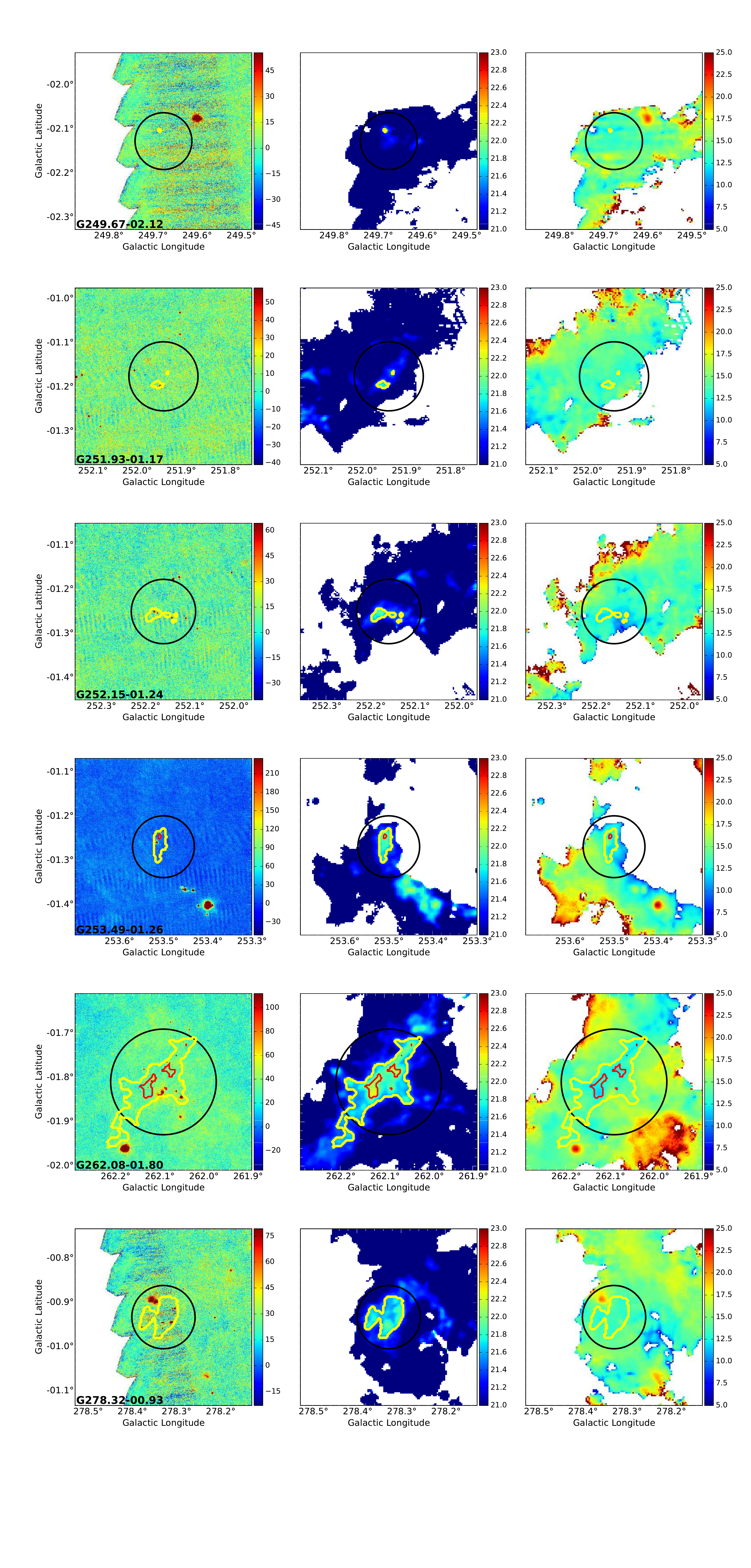}
    \caption{70 $\mu$m images ([MJy/sr], left) and calculated column density ([cm$^{-2}$], middle) in logarithmic scale and dust temperature ([K], right) maps with a resolution of 36$\arcsec$. Yellow and red contour levels are at 3$\times10^{21}$ cm$^{-2}$ and 10$^{22}$ cm$^{-2}$, respectively. Black circle shows the Planck clump's position and size based on the given major axis from the ECC catalogue.}
    \label{fig:all_calculated_maps_7}
\end{figure}

\begin{figure}[!ht]
    \centering
\includegraphics[width=\columnwidth]{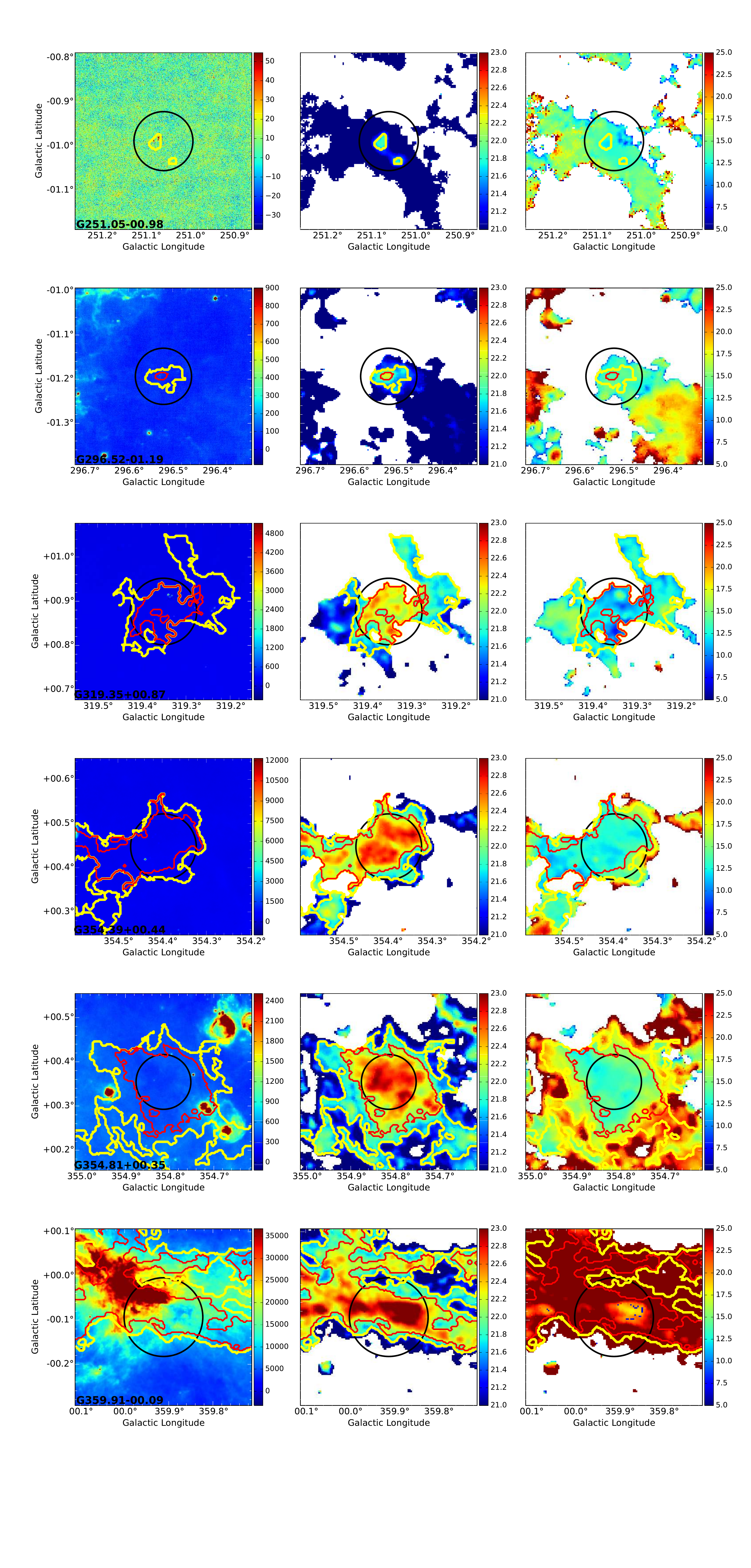}
    \caption{70 $\mu$m images ([MJy/sr], left) and calculated column density ([cm$^{-2}$], middle) in logarithmic scale and dust temperature ([K], right) maps with a resolution of 36$\arcsec$. Yellow and red contour levels are at 3$\times10^{21}$ cm$^{-2}$ and 10$^{22}$ cm$^{-2}$, respectively. Black circle shows the Planck clump's position and size based on the given major axis from the ECC catalogue.}
    \label{fig:all_calculated_maps_8}
\end{figure}

Figure \ref{fig:all_calculated_maps}-\ref{fig:all_calculated_maps_8} shows the Herschel 70 $\mu$m images ([MJy/sr], left panels), the calculated column density ([cm$^{-2}$], middle panels) in logarithmic scale and the dust temperature maps ([K], right panels) for the ECC sources with a resolution of 36$\arcsec$. The calculation was based on the background subtracted Herschel 160 - 500 $\mu$m images. Black circles show the location (central position) and size (major axis given in the ECC catalogue) of the Planck clumps. Yellow and red contour levels refer to the H$_2$ column density thresholds of 3$\times$10$^{21}$ and 10$^{22}$ cm$^{-2}$, respectively.

\end{appendix}

\end{document}